%% file: intro.tex
\documentclass[11pt, a4paper]{book}		
\usepackage[latin1]{inputenc}
\usepackage[english]{babel}
\usepackage{graphicx}				
\usepackage{epsf}
\usepackage{amsfonts}
\usepackage{slashed}
\usepackage{amssymb}
\usepackage{amsmath}
\usepackage{beton}
\begin{document}
\pagestyle{empty}

\input cover1.tex

\newpage

\input cover2.tex

\newpage

\pagenumbering{roman}

\input preface.tex

\newpage

\input contributions.tex

\newpage
\phantom{Tyhjaa}
\newpage

\pagenumbering{arabic}
\tableofcontents
\pagestyle{plain}

\input{foreword.tex}

\input{chap1.tex}

\input{NLOdis.tex}

\input{DYNLO.tex}

\input{hadronprod.tex}
\input{solveDGLAP.tex}
\input{global.tex}

\input{refs.tex}
\end{document}

%% file: cover1.tex
\vspace*{10mm}

\centerline{DEPARTMENT OF PHYSICS, UNIVERSITY OF JYV\"ASKYL\"A}
\centerline{RESEARCH REPORT No. 4/2009}

\vspace{25mm} 


\centerline{\bf GLOBAL ANALYSIS OF NUCLEAR PARTON}
\centerline{\bf DISTRIBUTION FUNCTIONS AT LEADING AND}
\centerline{\bf NEXT-TO-LEADING ORDER PERTURBATIVE QCD}

\vspace{13mm}

\centerline{\bf BY}
\centerline{\bf HANNU PAUKKUNEN}

\vspace{13mm}

\centerline{Academic Dissertation}
\centerline{for the Degree of}
\centerline{Doctor of Philosophy}

\vspace{13mm}

\vspace{20mm}

\begin{figure}[!h]
\center
\includegraphics[scale=0.12]{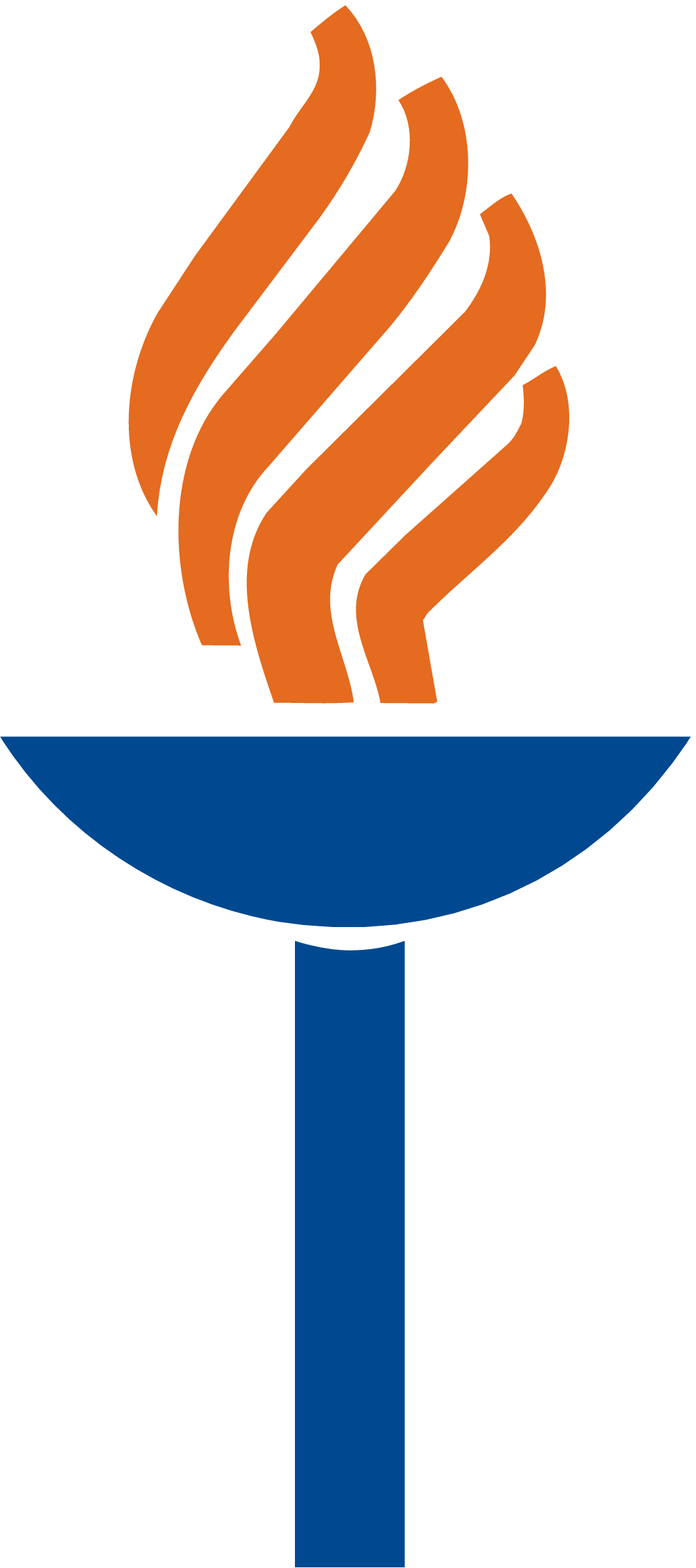}
\end{figure}

\centerline{Jyv\"askyl\"a, Finland}
\centerline{June, 2009}

\pagebreak

%% file: cover2.tex

\centerline{DEPARTMENT OF PHYSICS, UNIVERSITY OF JYV\"ASKYL\"A}
\centerline{RESEARCH REPORT No. 4/2009}

\vspace{25mm} 


\centerline{\bf GLOBAL ANALYSIS OF NUCLEAR PARTON}
\centerline{\bf DISTRIBUTION FUNCTIONS AT LEADING AND}
\centerline{\bf NEXT-TO-LEADING ORDER PERTURBATIVE QCD}

\vspace{13mm}

\centerline{\bf BY}
\centerline{\bf HANNU PAUKKUNEN	}

\vspace{13mm}

\centerline{Academic Dissertation}
\centerline{for the Degree of}
\centerline{Doctor of Philosophy}

\vspace{13mm}

\centerline{To be presented, by permission of the}
\centerline{Faculty of Mathematics and Natural Sciences}
\centerline{of the University of Jyv\"askyl\"a,}
\centerline{for public examination in Auditorium FYS 1 of the}
\centerline{University of Jyv\"askyl\"a on June 26, 2009,}
\centerline{at 12 o'clock noon}

\vspace{20mm}

\begin{figure}[!h]
\center
\includegraphics[scale=0.12]{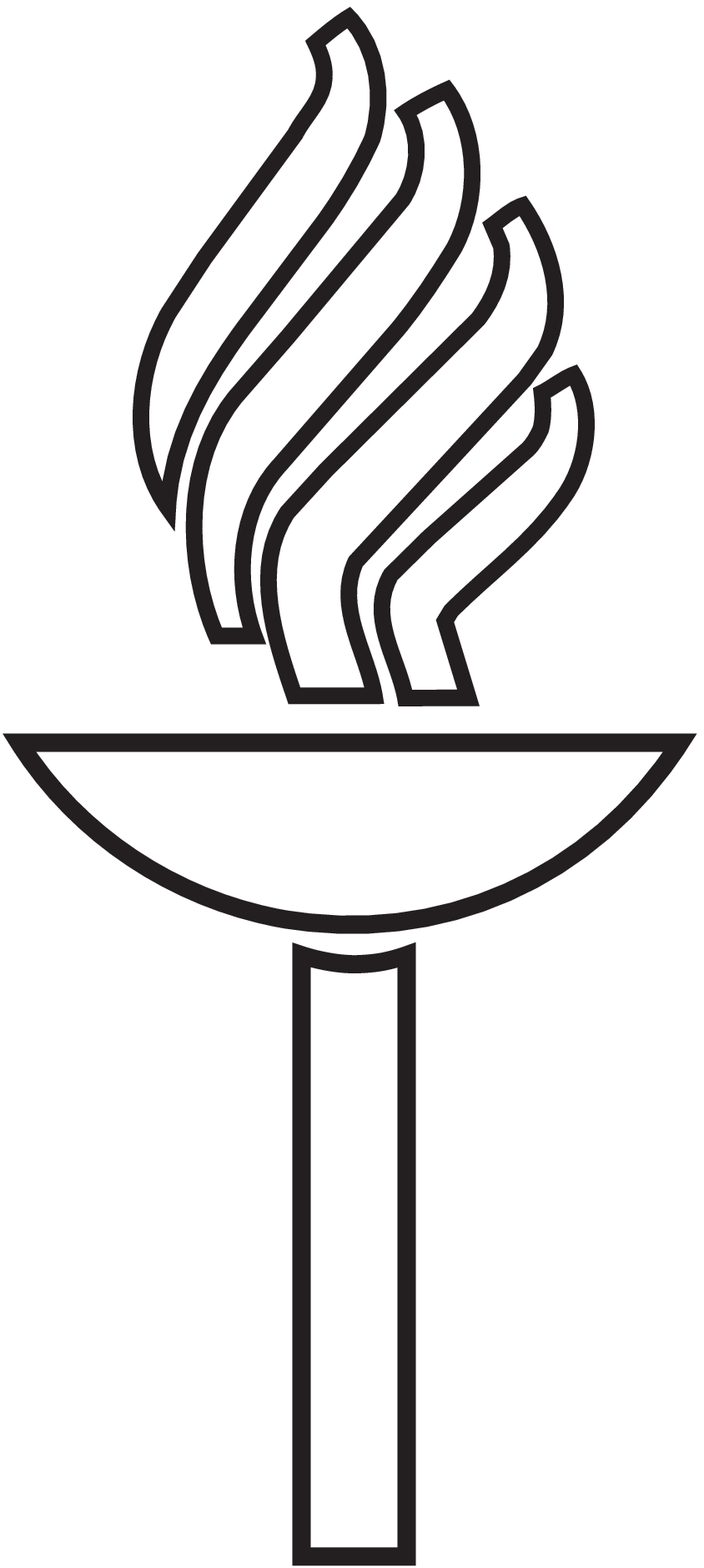}
\end{figure}


\centerline{Jyv\" askyl\" a, Finland}
\centerline{June 2009}

\pagebreak

%% file: preface.tex
\chapter*{Preface}

This thesis summarizes years of work carried out under supervision of Prof. Kari J. Eskola at the University of  Jyv\"askyl\"a, Department of Physics. The financial support from the Graduate School of Particle and Nuclear Physics, the Helsinki Institute of Physics URHIC \& LNCPM Theory Projects, and the Academy of Finland, the projects 206024 and 115262, are gratefully acknowledged.

\vspace{0.3cm}
My special thanks go to my supervisor Prof. Kari J. Eskola for his guidance along these years, Dr. Carlos Salgado and Dr. Vesa Kolhinen as collaborators, and Prof. Paul Hoyer and Prof. Kari Rummukainen for reviewing the original manuscript of this thesis. I also want to acknowledge the friendly atmosphere at the Department of Physics.

\vspace{0.3cm}
Finally, I wish to thank my family for the constant support on this long-distance journey, called life.

\vspace{0.3cm}
Jyv\"askyl\"a, June 2009 \\
Hannu Paukkunen

%% file: contributions.tex

\chapter*{List of Publications}

This thesis consists of an introductory part and of the following
publications:

\begin{itemize}
\item[{\bf I}]
{\bf NuTeV $\mathbf{\sin^2 \theta_{\rm W}}$ anomaly and nuclear parton distributions revisited,} \\
K.~J.~Eskola and H.~Paukkunen, \\
JHEP {\bf 0606} (2006) 008 [arXiv:hep-ph/0603155].

\item[{\bf II}]
{\bf A global reanalysis of nuclear parton distribution functions,} \\
K.~J.~Eskola, V.~J.~Kolhinen, H.~Paukkunen and C.~A.~Salgado, \\
JHEP {\bf 0705} (2007) 002 [arXiv:hep-ph/0703104].

\item[{\bf III}]
{\bf An improved global analysis of nuclear parton distribution functions including RHIC data,} \\
K.~J.~Eskola, H.~Paukkunen and C.~A.~Salgado, \\
JHEP {\bf 0807} (2008) 102 [arXiv:0802.0139].

\item[{\bf IV}]
{\bf EPS09 - a New Generation of NLO and LO Nuclear Parton Distribution Functions,} \\
K.~J.~Eskola, H.~Paukkunen and C.~A.~Salgado, \\
JHEP {\bf 0904} (2009) 065 [arXiv:0902.4154].

\end{itemize}

The author has worked on the calculations and participated in the planning and writing of the first [I] paper. For the second [II] publication, the author contributed by numerical downward parton evolution and inclusive hadron production calculations, as well as participating in the preparation of the article.

\vspace{0.2cm}
The author has been in a significant role in extending the $\chi^2$-analysis technique for the third [III] and fourth [IV] publication, and developing the error propagation procedure employed in the fourth [IV] publication. All the numerical results presented in these articles have been obtained by the fast parton evolution and cross-section programs constructed or largely modified (hadron production) by the author. The author also wrote the original draft versions for both of these publications.

%% file: foreword.tex
\chapter{Foreword}

The Quantum Chromo-dynamics (QCD) is a theory of strong interactions --- interactions between hadrons and, in particular, between their inner constituents. In QCD, the fundamental building blocks are quarks and gluons whose interactions are ultimately defined by the Lagrangian density
\begin{equation}
\mathcal{L}_{\rm QCD} = -\frac{1}{4}F^a_{\mu\nu}F^{a,\mu\nu} + \overline{\Psi}^k\left(i\gamma^\mu D_\mu - m_k \right) \Psi^k,
\end{equation}
where $\Psi^k$ denote the quark fields, $\gamma^\mu$s the standard Dirac matrices, and
\begin{eqnarray}
D_\mu & \equiv & \partial_\mu - ig_s t^a A_\mu^a \\
F^a_{\mu\nu} & \equiv & \partial_\mu A^a_\nu - \partial_\nu A^a_\mu + g_s f^{abc} A^b_\mu A^c_\nu,
\end{eqnarray}
where $A_\mu^a$ are the gluon fields and $g_s$ denotes the strong coupling constant. The matrices $t^a$ are the SU(3) generators and $f^{abc}$ are the corresponding structure constants. The physics content of $\mathcal{L}_{\rm QCD}$ has turned out to be very rich, yet challenging to work out. The most rigorous approaches to probe the inner workings of QCD are the lattice simulations, which have demonstrated encouraging results e.g. for confinement and predicting the mass-hierarchy of light hadrons. The lattice-QCD, however, quickly meets its limitations when the size of the studied system increases and it comes to describing scattering experiments. To apply QCD in such situation, perturbative methods to treat quarks and gluons are to be employed. The ultimate justification for the use of perturbative QCD (pQCD) tools lies in the fact that QCD enjoys what is known as \emph{asymptotic freedom} --- the strong interactions becoming effectively weaker when the inherent momentum scale of the process is large, $Q^2 \gg 1 \, {\rm GeV}^2$, or equivalently, when the probed distances are much smaller than the size of the hadron. As the strong interactions nevertheless bind the quarks and gluons, \emph{partons}, together to make a hadron, the exact way they are distributed inside the hadrons cannot be neglected when applying pQCD to hadronic collisions. Intuitively, the structure of the hadron should not, however, have anything to do with the collision, but is rather something that is inherent for the hadron itself. From the pQCD point of view, such property is known as \emph{factorization}, and the relevant structure of the hadrons is encoded in \emph{parton distribution functions} (PDFs) which are process-independent. In principle, the PDFs should be computable from $\mathcal{L}_{\rm QCD}$ but such task is far from being realized in practice any time soon. Instead, they must be inferred from various experiments with the help of pQCD --- from global analyses.

\vspace{0.5cm}
The role of the proton PFDs becomes emphasized in a hadron-hadron collider like the CERN-LHC where the backrounds are often huge and the expected physics signals relatively weak. Interpreting the experimental measurements in a situation like this, requires reliable knowledge of the PDFs. Similarly, the detailed knowledge of the quark-gluon content of the bound nucleons is of vital importance in precision studies on the properties of the strongly interacting matter expected to be produced in ultrarelativistic Pb+Pb collisions at the LHC and e.g. Au+Au collisions at the BNL-RHIC.

\vspace{0.5cm}
This thesis consists of two parts, the separate introductory part and the published four articles. The introduction begins by a technically detailed description of the DGLAP evolution --- the pQCD-physics behind the global QCD analyses --- as I understand it. I also discuss the fast numerical solving method for the DGLAP equations, which has been used in the numerical works of the published articles of this thesis. A write-up of the next-to-leading order (NLO) calculations for the deeply inelastic scattering (DIS) and the Drell-Yan (DY) dilepton production cross-sections, which are the data types that comprise most of the experimental input employed in the articles of this thesis, is also included. The formalism of the inclusive single hadron production at NLO, the third type of experimental data utilized in these articles, is described as well, although less rigorously. The introductory part ends with a discussion of the global QCD analyses in general, with a special attention paid to the major work of this thesis \cite{EPS09-paperi}, the NLO analysis of nuclear parton densities and their uncertainties. I have tried to avoid unnecessary overlap between the introductory part and the published articles, but yet keep the introductory part such that it is logical and self-contained, without leaning too much on the published articles. The necessary background for understanding what is presented in this thesis is the basic knowledge of Quantum Field Theory and elementary phenomenology of High Energy Physics.

%% file: chap1.tex
\chapter{DGLAP evolution}
\label{DGLAPevolution}

In this chapter, I will discuss the physics of parton evolution. Instead of only sketching general guidelines, I will take a somewhat more detailed point of view, hoping this thesis would also serve as an elementary introduction to the subject. Much of what I present here can be learned from works of Dokshitzer et al. \cite{Dokshitzer:1978hw,Dokshitzer:1991wu} and Altarelli \cite{Altarelli:1981ax}.

\section{Deeply inelastic scattering}
\label{Deeplyinelasticscattering}

\begin{figure}[h]
\centering
 \includegraphics[scale=0.35]{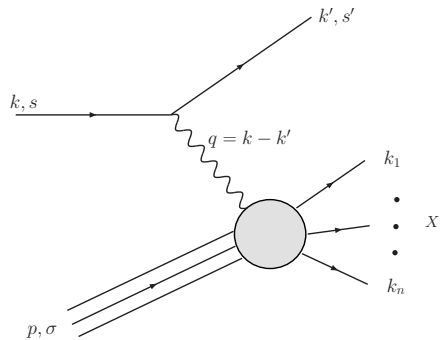}
\caption{Schematic picture of the deeply inelastic scattering. The variables $k$ and $P$ denote the incoming momenta, whereas $k'$ and $k_1,\ldots,k_n$ are all outgoing. The spin states are marked by $s$, $s'$ and $\sigma$.}
\label{Fig:DIS}
\end{figure}
In a deeply inelastic scattering (DIS) a lepton projectile hits a target nucleon breaking it apart to a state $X$ consisting of a plethora of various particles with invariant mass $M_X^2 \gg M^2$, where $M$ denotes the rest mass of the nucleon. In the simplest case the lepton is an electron or muon and the interaction is dominantly mediated by exchanging a virtual photon, as illustrated in Fig.~\ref{Fig:DIS}

In the target rest frame, the four-momenta of the particles can be chosen as
\begin{equation}
\begin{array}{cccl}
 k  & = & (E,   {\bf k}) & = (E,0,0,E) \\
 k' & = & (E',  {\bf k'}) & = (E',E' \sin \theta \cos \phi,E' \sin \theta \sin \phi,E' \cos \theta) \\
 P  & = & (P^0,   {\bf P}) & = (M,0,0,0) \\
 q  & = & (\nu,   {\bf q}) & = (E-E',{\bf k}-{\bf k}),
\end{array}
\end{equation}
where I have neglected the lepton mass. The standard invariant DIS-variables are 
\begin{eqnarray}
Q^2 & \equiv & -q^2 \; \; \, = 4EE' \sin^2 \left( {\theta}/{2} \right) \nonumber \\
x   & \equiv & \frac{Q^2}{2 P \cdot q} = \frac{Q^2}{2M\nu}  \\
y   & \equiv & \frac{P \cdot q}{P \cdot k} \,= \frac{\nu}{E} \nonumber,
\end{eqnarray}
where the latter equalities refer to the target rest frame. The differential, spin-independent cross-section for this process can be written as
\begin{equation}
 \sigma = \frac{\pi M e^4}{|k \cdot P|q^4} \frac{d^3{\bf k'}}{(2\pi)^3 2E'} L^{\mu\nu} W_{\mu\nu},
 \label{eq:DIS1}
\end{equation}
where $e$ is the electric coupling constant and
\begin{eqnarray}
L^{\mu\nu} & \equiv & \frac{1}{2} {\rm Tr}[\slashed{k'} \gamma^\mu \slashed{k} \gamma^\nu] \\
4\pi M W_{\mu\nu} & \equiv & \frac{1}{2} \sum_n \sum_\sigma \prod_{i=1}^{n} \frac{d^3{\bf k_i}}{(2\pi)^3 2k_i^0} (2\pi)^4 \delta^{(4)} ( P+q-\sum_{j=1}^{n} k_j) \label{eq:def_of_hadronic_tensor} \\
& \phantom{\equiv} & \langle n,{\rm out}|\hat{J}_\mu(0)|(P,\sigma), {\rm in} \rangle \langle (P,\sigma),{\rm in}|\hat{J}^\dagger_\nu(0)|n, {\rm out} \rangle \nonumber
\end{eqnarray}
are the leptonic and hadronic tensors. In contrast to the leptonic tensor $L^{\mu\nu}$, the non-perturbative nature of QCD makes it impossible to compute $W_{\mu\nu}$ directly but its general form can nevertheless be written down without much further input. 
Indeed, since $L^{\mu\nu}$ is symmetric under interchange of indices, the relevant part of the hadronic tensor should also satisfy $W_{\mu\nu}=W_{\nu\mu}$, and together with the definition (\ref{eq:def_of_hadronic_tensor}) this implies $W_{\mu\nu}^*=W_{\mu\nu}$. A further restriction is provided by the current conservation $q^\mu W_{\mu\nu}=q^\nu W_{\mu\nu}=0$. The general expression satisfying these conditions can be written as
\begin{equation}
 W_{\mu\nu} = -W_1 \left( g_{\mu\nu} - \frac{q_\mu q_\nu}{q^2} \right) + \frac{W_2}{M^2} \left(P_\mu - \frac{P \cdot q}{q^2}q_\mu \right) \left(P_\nu - \frac{P \cdot q}{q^2}q_\nu \right),
\label{eq:Hadronic tensor}
\end{equation}
where $W_1$ and $W_2$ are, \emph{a priori} unknown coefficients. It is traditional to define dimensionless structure functions
\begin{equation}
 F_1(x,Q^2) \equiv M W_1 \qquad F_2(x,Q^2) \equiv \nu W_2,
\end{equation}
which, in the $M^2 \ll Q^2$ limit, can be projected from the hadronic tensor as
\begin{eqnarray}
 \frac{F_2}{x} & = & \left( -g^{\mu\nu} + \frac{12x^2}{Q^2} P^\mu P^\nu \right) M W_{\mu\nu} \\
{F_1} & = & \left( - \frac{1}{2}g^{\mu\nu} + \frac{2x^2}{Q^2} P^\mu P^\nu \right) M W_{\mu\nu} = \frac{F_2}{2x} - \left(\frac{4x^2}{Q^2} P^\mu P^\nu \right) M W_{\mu\nu}. \nonumber
\end{eqnarray}
In terms of the structure functions $F_{1}$ and $F_{2}$ the cross-section in Eq.~(\ref{eq:DIS1}) can expressed in an invariant way
\begin{equation}
 \frac{d^2\sigma}{dxdQ^2} = \frac{4\pi \alpha^2_{\rm em}}{Q^4} \frac{1}{x} \left[ xy^2 F_1   + F_2 \left(1-y-\frac{xyM^2}{s-M^2} \right)\right],
\label{eq:General_DIS_crosssection}
\end{equation}
where $s \equiv (P+k)^2$ denotes the center-of-mass energy, and $\alpha_{\rm em} \equiv e^2/4\pi$ stands for the fine-structure constant.

\subsubsection{Parton model}

The parton model \cite{Feynman:1969ej,Bjorken:1969ja} can be motivated by considering the DIS not in the target-rest-frame but in the electron-proton center-of-mass system. In such a frame, the nucleon appears Lorentz contracted, and the time dilatation slows down the intrinsic interaction rate of the fundamental constituents of the nucleon, the partons. During the short period it takes for the electron to traverse across the nucleon, the state of the nucleon wave function can thus be envisioned as being frozen to a superposition of free partons collinear with the nucleon. Mathematically, the parton model is defined by the relation
\begin{equation}
d\sigma = \sum_q \int_0^1 d\xi {d\hat{\sigma}^q_0(\xi P)} f_q(\xi),
\label{eq:PartonModel}
\end{equation}
where $\hat{\sigma}^q_0(\xi P)$ is the leading order (Born) cross-section for the electron-parton scattering, with the parton carrying a momentum $p = \xi P$. The functions $f_q(\xi)$ are called \emph{parton distributions}, and represent the number density of partons of \emph{flavor} $q$ in the nucleon. 
\begin{figure}[h]
\centering
 \includegraphics[scale=0.4]{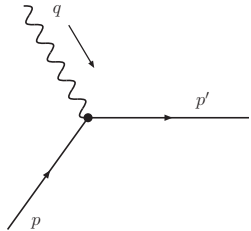}
\caption{The leading-order diagram for photon-quark interaction.}
\label{Fig:LOquark}
\end{figure}

\begin{figure}[h]
\centering
 \includegraphics[scale=0.6]{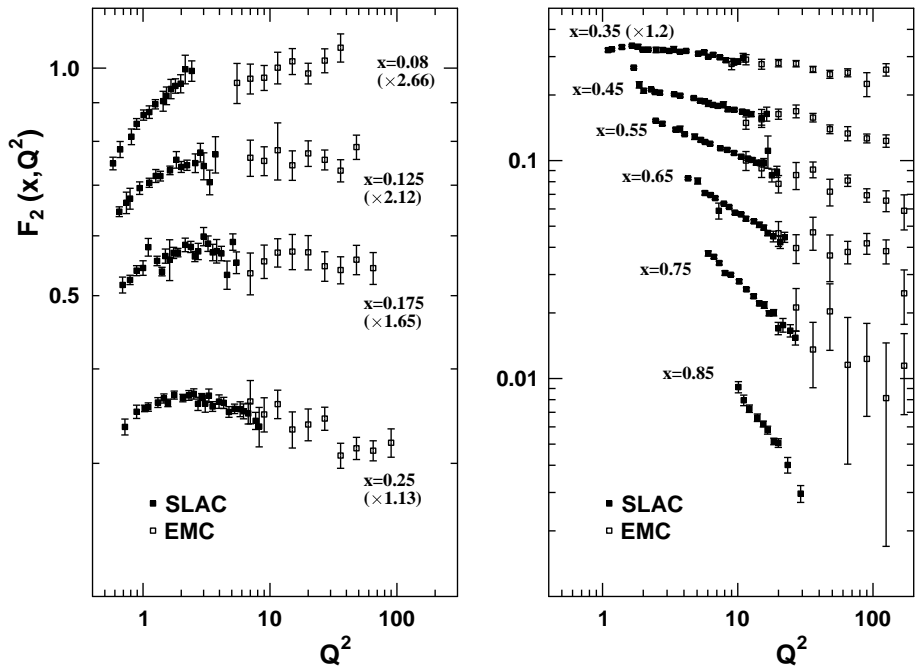}
\caption{Experimental data for proton structure function $F_2$ from SLAC \cite{Whitlow:1991uw} and CERN-EMC \cite{Aubert:1985fx} experiments.}
\label{Fig:DISdata}
\end{figure}

In QCD, only quarks carry an electric charge $e_q$ and the definition (\ref{eq:PartonModel}) with Eq.~(\ref{eq:DIS1}) implies that the hadronic tensor $W_{\mu \nu}$ can be written as
\begin{equation}
W_{\mu \nu} = \sum_q \int_0^1 \frac{d\xi}{\xi} \hat{W}_{\mu \nu}^q f_q(\xi),
\end{equation}
where its partonic counterpart $\hat{W}_{\mu\nu}^q$ is essentially the square of the diagram in Fig.~(\ref{Fig:LOquark})
\begin{eqnarray}
 4\pi M \hat{W}_{\mu\nu}^q & = & \frac{e_q^2}{2} \sum_\sigma \frac{d^3{\bf p'}}{(2\pi)^3 2p'^0} (2\pi)^4 \delta^{(4)} (p+q-p') \, {\rm Tr}[\slashed{p'} \gamma^\mu \slashed{p} \gamma^\nu] \nonumber \\
& = & \frac{e_q^2}{2} \frac{2\pi x}{Q^2} \, {\rm Tr}[\slashed{p'} \gamma^\mu \slashed{p} \gamma^\nu] \delta(\xi-x).
\label{eq:partonicstr}
\end{eqnarray}
Neglecting the nucleon mass $M$ compared to the photon virtuality $Q^2$,
\begin{equation}
 g_{\mu\nu} \, {\rm Tr}[\slashed{p'} \gamma^\mu \slashed{p} \gamma^\nu] = -4Q^2 \qquad p_{\mu}p_{\nu} \, {\rm Tr}[\slashed{p'} \gamma^\mu \slashed{p} \gamma^\nu] = 0,
\end{equation}
we find
\begin{equation}
-g_{\mu\nu} (M \hat{W}^q_{\mu \nu}) = e_q^2 x \delta(\xi-x) \qquad p^{\mu}p^{\nu} (M\hat{W}^q_{\mu \nu}) = 0.
\end{equation}
Consequently, the parton model predictions for the structure functions reduce to an electric-charge-weighted sum of the quark distributions,
\begin{equation} 
2xF_1(x) = F_2(x) = \sum_q e_q^2 x f_q(x),
\end{equation}
and the cross-section in Eq.~(\ref{eq:General_DIS_crosssection}) can be written as
\begin{equation}
\frac{d^2\sigma}{dxdQ^2} = \frac{d^2\hat{\sigma_{0}}}{dxdQ^2} \sum_q e_q^2 f_q(x),
\label{eq:bornDIS}
\end{equation}
where $\hat{\sigma_{0}}$ denotes the partonic Born cross-section
\begin{equation}
\frac{d^2\hat{\sigma}_{0}}{dxdQ^2}  \equiv  \frac{4\pi \alpha^2}{Q^4} \left[\frac{y^2}{2} + \left(1-y-\frac{xyM^2}{s-M^2} \right)\right].
\end{equation}
It is a prediction of the parton model that the structure functions $F_{1,2}$ are only functions of $x$, and should not depend on $Q^2$ in the $Q^2 \gg M^2$ limit. This phenomenon, termed as \emph{Bjorken-scaling}, was indeed observed in the early SLAC experiments providing direct evidence about the inner constituents of the nucleon. Later experiments which have covered a larger domain in the $(x,Q^2)$-plane have revealed, however, that the $Q^2$-independence of the structure functions $F_{1,2}$, although a good first approximation, was not exactly true. Such deviations are clear e.g. in Fig.~\ref{Fig:DISdata}, which shows some experimental data for the proton structure function $F_2$. The \emph{scaling violations}, as they are nowadays called, can however be fully explained by the QCD dynamics --- by so-called DGLAP equations, to be derived shortly. Together with the asymptotic freedom, these equations with factorization theorem constitute the main pillars of perturbative QCD.


\section{Initial state radiation}
\label{Isradiation}

\subsection{Origin of the scaling violations}

\begin{figure}[h]
\begin{center}
\includegraphics[scale=0.35]{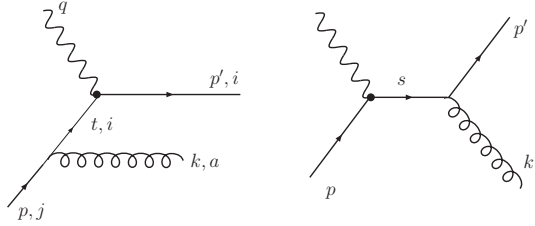}
\caption{Real gluon radiation. The letter $i,j,a$ are the color indices.}
\label{fig:pic2}
\end{center}
\end{figure}

Due to the inclusive nature of the deeply inelastic scattering nothing forbids having additional QCD particles in the final state. First such corrections to the Born-level matrix element originate from a radiation of a real gluon as shown in Fig.~\ref{fig:pic2}. 
Both of these diagrams are divergent as the intermediate quark propagators are close to being on-shell:
\begin{eqnarray}
  \left( p - k\right)^2 & = & -2p^0k^0 \left(1 - \cos \theta \right) \, \,\, \rightarrow 0, \nonumber \\
  \left( p' + k\right)^2 & = & -2p'^0k^0 \left(1 - \cos \theta' \right) \rightarrow 0 \nonumber.
\end{eqnarray}
This can happen either if the momentum of the emitted particle goes to zero, $k^0,{p'}^0 \rightarrow 0$, or if the emission is in the direction of the incoming or the outgoing quark $\theta,\theta' \rightarrow 0$. These are archetypes of \emph{infrared} and \emph{collinear} singularities correspondingly. There are also same kind of divergences stemming from the virtual corrections, and it turns out that all but the collinear divergence related to the gluon radiation from the \emph{incoming quark} will eventually cancel. In what follows, I will show how to extract these divergences and how their resummation gives rise to the parton DGLAP evolution --- the $Q^2$-dependence of the parton distributions observed in the experiments.

\subsection{One gluon emission}
\label{Onegluonemission}
Rather than drawing graphs for matrix elements, for the rest of this Chapter, I will draw the graphs in the cut diagram notation (see e.g. \cite{Brock:1993sz}) directly for the cross-sections. The square of the diagrams in Fig.~\ref{fig:pic2} look as
\begin{figure}[h]
\begin{center}
\includegraphics[scale=0.45]{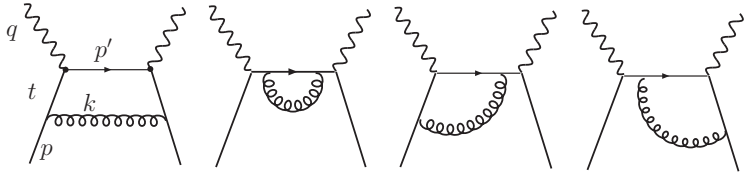}
\caption{Diagrams representing the $\gamma^*q \rightarrow qg$ process. It should be understood that the parton lines in the middle are real, on-shell particles.}
\label{fig:pic5}
\end{center}
\end{figure}

Although the such squared matrix element is certainly gauge invariant, the contribution of an individual graph depends on the choice of gauge. However, as I already mentioned it is only the collinear $\left( p - k\right)^2 \rightarrow 0$ singularity which turns out to be the relevant one. In the Feynman gauge, all but the second of the diagrams in Fig.~\ref{fig:pic5} will contribute in this kinematical limit, but, as it turns out, in the axial gauge it is the first diagram alone that is responsible for the divergent behaviour. For obvious reasons I call it a \emph{ladder diagram}.

\subsubsection{Axial gauge}

The class of axial gauges is specified by a gauge-fixing term $-\frac{1}{2\xi}\left( n \cdot G \right)^2$ in the QCD Lagrangian where $G$ denotes the gluon field, $n$ is an arbitrary four-vector and $\xi$ is the gauge parameter. The gluon propagator in this gauge is
\begin{eqnarray}
D_{\mu\nu}(k) & = & \frac{-i}{k^2} \left[ g_{\mu\nu} - \frac{k_\mu n_\nu + k_\nu n_\mu}{k \cdot n} + \frac{\xi k^2 + n^2 }{(k \cdot n)^2}k_\mu k_\nu\right]. \label{eq:gluonpropagator}
\end{eqnarray}
The sum over the two physical polarization states $\epsilon^{\lambda_{1,2}}(k)$ ($k^2=0$), obeying $k \cdot \epsilon^{\lambda_i}(k) = 0$ and $n \cdot \epsilon^{\lambda_i}(k) = 0$, normalized by $\epsilon^{\lambda_{i}}(k) \cdot \epsilon^{\lambda_{i}}(k)= -1$, reads
\begin{eqnarray}
\sum_{\lambda} \epsilon_\mu^\lambda(k) \epsilon^{*\lambda}_\nu(k) & = & -g_{\mu \nu} + \frac{k_\mu n_\nu + k_\nu n_\mu}{(k \cdot n)} - \frac{n^2 k_\mu k_\nu}{(k \cdot n)^2}. \label{eq:polsum}
\end{eqnarray}

Usually, it is convenient to choose $\xi = 0$ and $n^2 = 0$ which specifies the \emph{light-cone gauge}. The axial gauges are sometimes called \emph{physical gauges}: the reason for this is most distinct in the light-cone gauge as any propagator in a Feynman diagram can be replaced by the polarization sum over the physical states:
\begin{equation}
 D_{\mu\nu}(k) = \frac{i}{k^2}\sum_{\lambda}\epsilon_\mu^\lambda(k)\epsilon^{*\lambda}_\nu(k).
\end{equation}

A convenient choice for the light-like axial vector in the present problem is
\begin{equation}
 n \equiv q + \eta p, \qquad \eta \equiv \frac{-q^2}{2p \cdot q}.
\end{equation}

\subsubsection{Sudakov decomposition}

In extracting the dominant part of the squared matrix elements, it is convenient to parametrize the momenta of the outgoing partons by \cite{Sudakov:1954sw}
\begin{equation}
k = (1-z)p + \beta n + k_\perp, \qquad \beta = \frac{-k_\perp^2}{2(1-z) p \cdot q},
\label{eq:Sudakov1}
\end{equation}
where $k_\perp$ is a space-like 4-vector orthogonal to $n$ and $p$: $k_T^2 < 0$, $n \cdot k_T = p \cdot k_T = 0$. For example, in the center-of-mass frame of $p$ and $n$,
\begin{eqnarray}
p & = & (P,0,0,P)  \nonumber \\
n & = & (P,0,0,-P) \\
k_\perp & = & (0,{\bf k_\perp},0) \nonumber \\
k & = & \left( (1-z)P + \frac{-k_\perp^2}{4(1-z)P}, {\bf k_\perp}, (1-z)P - \frac{ -k_\perp^2}{4(1-z)P} \right) \nonumber
\end{eqnarray}
where $P$ is some reference momentum. In such a frame the interpretation of $k_\perp^2 = -{\bf k}_\perp^2$ as the transverse momentum is evident. Furthermore,
$$t^2 = (p-k)^2 = k_\perp^2/(1-z),$$
and we see that the collinear $t \rightarrow 0$ divergence should be found by extracting the $1/k_\perp^2$-pole.

\subsubsection{The calculation}

The squared matrix element corresponding to the first diagram in Fig.~\ref{fig:pic5} reads
\begin{equation}
|\mathcal{M}^{\rm Ladder}_{\gamma*q \rightarrow qg}|^2_{\mu\nu} = 
C_F g_s^2 \frac{e_q^2}{2} \sum_{\rm pol} \frac{1}{t^4} {\rm Tr} \left[ \slashed{p'} \gamma_\mu \slashed{t} \slashed{\epsilon} \slashed{p} \slashed{\epsilon^*} \slashed{t} \gamma_\nu \right], \quad t=p-k
\end{equation}
where the color factor $C_F=4/3$ arises from (see Fig.~\ref{fig:pic2})
$$
\frac{1}{3} \sum_{i,j,a} t^a_{ij}(t^a_{ij})^* = \frac{1}{3} \sum_{a} {\rm Tr} (t^at^a) = \frac{4}{3} = C_F.
$$
Using the polarization sum Eq.~(\ref{eq:polsum}) one finds
\begin{equation}
 \sum_{\rm pol} \slashed{\epsilon} \slashed{p} \slashed{\epsilon^*} = \frac{2}{1-z} \left( \slashed{k} + \beta \slashed{n} \right),
\end{equation}
and after a short calculation
\begin{equation}
\slashed{t} \left( \slashed{k} + \beta \slashed{n} \right) \slashed{t} = \left(\frac{1+z^2}{1-z}\right) (-k_\perp^2) \, \slashed{p} + \mathcal{O}(\slashed{k}_\perp k_\perp^2),
\end{equation}
where the remaining terms are higher order in $k_\perp$ and will not contribute to the collinear divergence. In total,
\begin{equation}
|\mathcal{M}^{\rm Ladder}_{\gamma*q \rightarrow qg}|^2_{\mu\nu} = g_s^2 C_F \frac{2(1-z)}{-k_\perp^2} \left(\frac{1+z^2}{1-z}\right) \times \frac{e_q^2}{2} {\rm Tr} \left[ \slashed{p'} \gamma_\mu \slashed{p} \gamma_\nu \right] + \cdots.
\label{eq:Mq_to_qg}
\end{equation}
It is essential that the last combination of terms is nothing but the squared matrix element in the Born approximation. Supplying the phase-space element in the Sudakov variables
\begin{equation}
\frac{d^3{\bf k}}{(2\pi)^3 2k^0} = \frac{1}{16\pi^2} \frac{dz}{1-z} d{\bf k_\perp^2},
\end{equation}
one obtains
\begin{equation}
\frac{d^3{\bf k}}{(2\pi)^3 2k^0} |\mathcal{M}^{\rm Ladder}_{\gamma*q \rightarrow qg}|^2_{\mu\nu} = \frac{d{\bf k_\perp^2}}{{\bf k_\perp^2}} dz \left(\frac{\alpha_s}{2\pi}\right) P_{qq}(z) \times \frac{e_q^2}{2} {\rm Tr} \left[ \slashed{p'} \gamma_\mu \slashed{p} \gamma_\nu \right] + \cdots,
\label{eq:1gluonMsquared}
\end{equation}
where
\begin{equation}
P_{qq}(z) \equiv C_F \left(\frac{1+z^2}{1-z}\right)
\label{eq:Pqqnonreg}
\end{equation}
is the so-called \emph{Altarelli-Parisi splitting function} associated with the unpolarized quark $\rightarrow$ quark transition. In the collinear limit, the variable $z$ is readily interpreted as the momentum fraction of the quark left after the gluon emission. The contribution to the quark tensor $\hat{W}_{\mu\nu}^q$ is
\begin{eqnarray}
 4\pi M \hat{W}_{\mu\nu}^q & = & \int \frac{d^3{\bf p'}}{(2\pi)^3 2p'^0} \int \frac{d^3{\bf k}}{(2\pi)^3 2k^0}  |\mathcal{M}^{\rm Ladder}_{\gamma*q \rightarrow qg}|^2_{\mu\nu} (2\pi)^4 \delta^{(4)} (p+q-k-p') \nonumber \\
& = &  2\pi \int \frac{d^3{\bf k}}{(2\pi)^3 2k^0} \, |\mathcal{M}^{\rm Ladder}_{\gamma*q \rightarrow qg}|^2_{\mu\nu}  \, \delta(p'^2) \theta(p'^0). \nonumber
\end{eqnarray}
Neglecting all $\mathcal{O}(k_\perp^2)$ terms which would cancel the collinear singularity in Eq.~(\ref{eq:1gluonMsquared}),
$$
p'^2=2z(p \cdot q) + q^2 + \mathcal{O}(k_\perp^2) \approx Q^2 \left( \frac{\xi z}{x} -1 \right)
$$
and
\begin{equation}
 g_{\mu\nu} \, {\rm Tr}[\slashed{p'} \gamma^\mu \slashed{p} \gamma^\nu] \approx -4Q^2 \frac{\xi}{x}  \qquad p_{\mu}p_{\nu} \, {\rm Tr}[\slashed{p'} \gamma^\mu \slashed{p} \gamma^\nu] = 0.
\end{equation}
Thus, the dominant $\mathcal{O}(\alpha_s)$ piece in the quark tensor is
\begin{equation}
-g^{\mu\nu} M\hat{W}_{\mu\nu}^q  =  e_q^2 \left[ \left(\frac{\alpha_s}{2\pi}\right) \int \frac{d{\bf k_\perp^2}}{{\bf k_\perp^2}} \int \frac{dz}{z} P_{qq}(z) \right] \xi \delta \left( \xi - \frac{x}{z} \right) + \cdots,
\end{equation}
which contributes to the hadronic tensor by
\begin{equation}
-g^{\mu\nu} M\hat{W}_{\mu\nu}   =   \sum_q e_q^2 \left[ \left(\frac{\alpha_s}{2\pi}\right) \int \frac{d{\bf k_\perp^2}}{{\bf k_\perp^2}} \int_x^1 \frac{dz}{z} P_{qq}(z) \right] f_q\left( \frac{x}{z} \right) + \cdots.
\end{equation}
As anticipated, the collinear divergence manifests itself in the $\int {d{\bf k_\perp^2}}/{{\bf k_\perp^2}}$ integral. The upper bound for this integral is proportional to $Q^2$ but the lower limit remains zero for massless quarks. Even if the quark had a small regulating mass $m$, the resulting logarithm $\log(Q^2/m^2)$ would not be \emph{infrared safe}: the resulting cross-section would be sensitive to the value of $m^2$ in the large-$Q^2$ limit. The solution to this problem will require resumming a whole tower of such logarithms. For the time being, however, I add this divergent piece to the DIS cross-section
\begin{equation}
\frac{d^2\sigma}{dxdQ^2} \stackrel{\rm LL}{=} \frac{d^2\hat{\sigma_{0}}}{dxdQ^2} \sum_q e_q^2 \left[1 + \left(\frac{\alpha_s}{2\pi}\right) \log \left( \frac{Q^2}{m^2} \right) P_{qq}  \right] \otimes f_q,
\label{eq:1gluonfinal}
\end{equation}
where the designation LL means that I have kept only the leading logarithmic contribution, and the shorthand notation $\otimes$ stands for the convolution
\begin{eqnarray}
P_{qq} \otimes f_q & \equiv & \,\, \int_x^1 \frac{dz}{z}P_{qq}(z) f_q \left( \frac{x}{z} \right) \,\,\,  = f_q  \otimes P_{qq}, \\
1 \otimes f_q & \equiv & \int_x^1 \frac{dz}{z} \delta(1-z) f_q \left( \frac{x}{z} \right) = f_q(x) \nonumber.
\end{eqnarray}
Since the left-hand side of Eq.~(\ref{eq:1gluonfinal}) is a measurable, finite, quantity the non-perturbative parton density $f_q$ is inevitably intertwined with the arbitrary cut-off scale $m^2$ such that the cross-section is finite. 

\vspace{0.5cm}
I still need to prove my claim that in the axial gauge this is the only collinear logarithm related to the initial state gluon radiation. This is actually quite a simple task: writing down the cross term
\begin{figure}[h]
\hspace{0.6cm}
\includegraphics[scale=0.25]{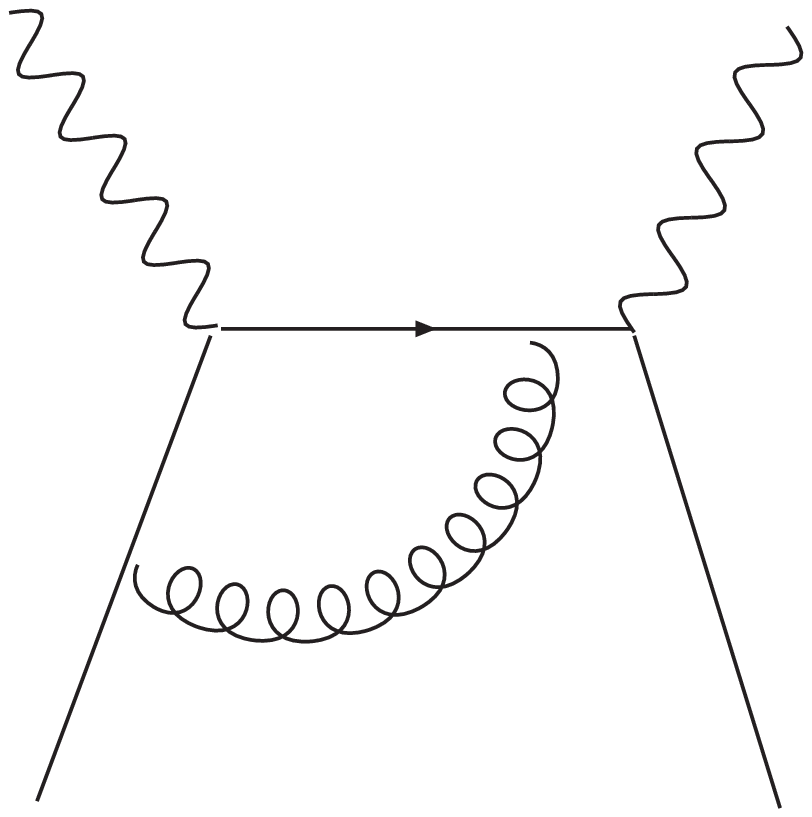}
\label{fig:pic8}
\end{figure}
\vspace{-2.1cm}
$$
\sim \frac{1}{{k_\perp^2}} \sum_{\rm pol} {\rm Tr} \left[ \slashed{p'} \gamma_\mu (\slashed{p}-\slashed{k}) \slashed{\epsilon} \slashed{p} \gamma_\nu (\slashed{q}+\slashed{p}) \slashed{\epsilon^*} \right],
$$

\vspace{0.8cm}
one realizes that if the trace is independent of $k_\perp$ there will be a similar collinear logarithm as found above. However, noting that
$$
(\slashed{p}-\slashed{k}) \slashed{\epsilon} \slashed{p} \approx z \, \slashed{p} \slashed{\epsilon} \slashed{p} =  2z \left(p \cdot \epsilon \right),
$$
the polarization sum reveals the structure
\begin{equation}
p_\mu \left( -g^{\mu \nu} + \frac{k^\mu n^\nu + k^\nu n^\mu}{(k \cdot n)} \right) \sim k_\perp^\nu,
\end{equation}
demonstrating that no $k_\perp$-independent term exists and the proof is complete\footnote{In the Feynman gauge with $\sum_{\lambda} \epsilon_\mu^\lambda(k) \epsilon^{*\lambda}_\nu(k) =  -g_{\mu \nu}$ this last step would not be true.}. Thus, in the collinear limit and in the axial gauge, there is no interference with the outgoing quark and, in the spirit of parton model, the factor
$$
\left(\frac{\alpha_s}{2\pi}\right) \log \left( \frac{Q^2}{m^2} \right) P_{qq}(z)
$$
can be interpreted as a probability density for the quark to radiate a gluon carrying a fraction $1-z$ of the quark momentum, before getting struck by the photon. One should note that due to the $1/(1-z)$-pole in $P_{qq}(z)$, this probability diverges in the infrared $z\rightarrow 1$ limit, making the convolution integrals apparently ill-defined. However, the probability that the quark re-absorbs the emitted gluon diverges similarly and will wash out the $z=1$ singularity, as will be discussed soon.

\subsection{Multiple gluon emissions}

\begin{figure}[h]
\begin{center}
\includegraphics[scale=0.4]{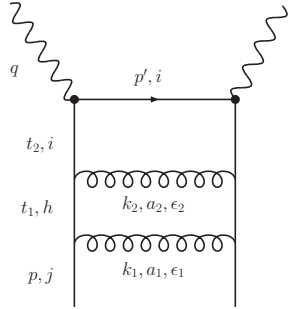}
\end{center}
\caption{Ladder graph for two-gluon emission. Momenta of the produced gluons with polarizations $\epsilon_1,\epsilon_2$ are denoted by $k_1,k_2$, and the intermediate quark momenta are $t_1 \equiv p-k_1$, $t_2 \equiv p-k_1-k_2$. The color indices are denoted by $i,j,h,a_1,a_2$.}
\label{fig:2gluons}
\end{figure}

Based on the previous section, it is natural to expect to find two similar collinear divergences as in Eq.~(\ref{eq:1gluonfinal}) if double gluon emission, shown in Fig.~\ref{fig:2gluons}, is considered. This is indeed the case and employing the method introduced earlier one can extract an $\alpha_s^2 \log^2(Q^2/m^2)$ contribution to cross-section Eq.~(\ref{eq:1gluonfinal}). This is how it goes.

The squared and spin-summed matrix element for the ladder diagram in Fig.~\ref{fig:2gluons} reads
\begin{eqnarray}
|\mathcal{M}^{\rm Ladder}_{\gamma*q \rightarrow q,2g}|^2_{\mu\nu} & = & 
g_s^4 \frac{e_q^2}{2} C_F^2 \sum_{\rm pol} \frac{1}{t_1^4t_2^4}
{\rm Tr} 
\left[ \slashed{p'} \gamma_\mu \slashed{t_2} \slashed{\epsilon_2} \slashed{t_1} \slashed{\epsilon_1} \slashed{p} \slashed{\epsilon_1^*} \slashed{t_1} \slashed{\epsilon_2^*} \slashed{t_2} \gamma_\nu \right]
\label{eq:2gluonsm1}
\end{eqnarray}
where the color factor $C_F^2=(4/3)^2$ arises in the following way (summation over all indices is implicit):
\begin{eqnarray}
& & \frac{1}{3} (t_{ih}^{a_2} t_{hj}^{a_1}) (t_{is}^{a_2} t_{sj}^{a_1})^* = 
\frac{1}{3} (t_{ih}^{a_2} t_{hj}^{a_1}) (t_{si}^{a_2} t_{js}^{a_1}) \nonumber \\
& = & \frac{1}{3} (t^{a_1}t^{a_1})_{hs} (t^{a_2}t^{a_2})_{sh}  \\
& = & \frac{1}{3} {\rm Tr}(t^{a_1}t^{a_1}) \frac{1}{3} {\rm Tr}(t^{a_2}t^{a_2}) = \left(\frac{4}{3} \right) \left( \frac{4}{3} \right) = C_F^2, \nonumber 
\end{eqnarray}
where I used $(t^{a}t^{a})_{hs} = (1/3) \, \delta_{hs}$. Introducing the Sudakov decomposition for the lower gluon momentum
\begin{eqnarray}
k_1 & = & (1-z_1)p + \beta_1 n + k_{1\perp}, \qquad \hspace{0.35cm} \beta_1 = \frac{-k_{1\perp}^2}{2(1-z_1) p \cdot q},
\label{eq:Sudakov2}
\end{eqnarray}
one immediately obtains, reading from the preceding calculation, that 
\begin{equation}
\sum_{\rm pol_1} \slashed{t_1} \slashed{\epsilon_1} \slashed{p} \slashed{\epsilon_1^*} \slashed{t_1} = \frac{2}{1-z_1} \left(\frac{1+z_1^2}{1-z_1}\right) ({-k_{1\perp}^2}) \slashed{p} + \ldots,
\end{equation}
where I have again omitted the terms higher order in $k_{1\perp}$. In the same way, writing the Sudakov decomposition for the upper gluon momentum as
\begin{equation}
k_2  =  z_1(1-z_2)p + \beta_2 n + k_{2\perp}, \qquad \beta_2 = \frac{-k_{2\perp}^2}{2z_1(1-z_2) p \cdot q},
\label{eq:Sudakov3}
\end{equation}
and dropping terms higher order in $k_{1\perp}$ and $k_{2\perp}$, one finds the leading contribution
\begin{equation}
\sum_{\rm pol_2} \slashed{t_2} \slashed{\epsilon_2} \slashed{p} \slashed{\epsilon_2^*} \slashed{t_2} = \frac{2}{1-z_2} \left(\frac{1+z_2^2}{1-z_2}\right) ({-k_{2\perp}^2}) \slashed{p} + \ldots.
\end{equation}
Thus, the squared matrix element (\ref{eq:2gluonsm1}) acquires a form
\begin{eqnarray}
|\mathcal{M}^{\rm Ladder}_{\gamma*q \rightarrow q,2g}|^2_{\mu\nu} & = & 
g_s^4 \frac{-k_{1\perp}^2}{t_1^4} \left[ \frac{2P_{qq}(z_1)}{1-z_1}   \right] \frac{-k_{2\perp}^2}{t_2^4} \left[ \frac{2P_{qq}(z_2)}{1-z_2}   \right] \nonumber \\
 & & \frac{e_q^2}{2} {\rm Tr} \left[ \slashed{p'} \gamma_\mu \slashed{p} \gamma_\nu \right] + \cdots,
\label{eq:eq:2gluonsm2}
\end{eqnarray}
where the last factor is again the Born matrix-element that has penetrated through the calculation. If there were not the factors $t_{1,2}^4$ in the denominator, the leading factors for both emitted gluons would be identical. However,
\begin{eqnarray}
t_1^2 & = & \frac{k_{1\perp}^2}{1-z_1} \nonumber \\
t_2^2 & = & \frac{k_{2\perp}^2}{1-z_2} + \frac{1-z_1(1-z_2)}{1-z_1} k_{1\perp}^2 + k_{1\perp} \cdot k_{2\perp},
\end{eqnarray}
where the latter one looks bad. In the region of phase space where $-k_{1\perp}^2 < -k_{2\perp}^2$ one can power expand Eq.~(\ref{eq:eq:2gluonsm2}) in $k_{1\perp}^2/k_{2\perp}^2$, schematically
\begin{equation}
|\mathcal{M}^{\rm Ladder}_{\gamma*q \rightarrow q,2g}|^2_{\mu\nu} \propto \frac{1}{k_{1\perp}^2} \frac{1}{k_{2\perp}^2} \left[1 + A \left( \frac{k_{1\perp}^2}{k_{2\perp}^2}\right) + B \left( \frac{k_{1\perp}^2}{k_{2\perp}^2} \right)^2 + \ldots \right]
\end{equation}
where the odd powers of $k_{1\perp}$ are absent as they would vanish upon integration. Whereas the integration over the first term gives the leading double logarithm,
\begin{equation}
\int^{Q^2}_{m^2} \frac{d{\bf k_{2\perp}^2}}{{\bf k_{2\perp}^2}} \int^{{\bf k_{2\perp}^2}}_{m^2} \frac{d{\bf k_{1\perp}^2}}{{\bf k_{1\perp}^2}} = \frac{1}{2!} \log^2\left( \frac{Q^2}{m^2} \right),
\label{eq:doublelog}
\end{equation}
the rest can give only a single logarithm. In the opposite transverse momentum ordering $-k_{1\perp}^2 > -k_{2\perp}^2$, 
one again obtains only single logarithms. 
\begin{figure}[h]
\begin{center}
\includegraphics[scale=0.35]{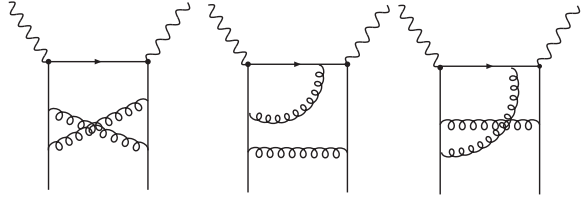}
\end{center}
\caption{Diagrams for two-gluon emission that do not contain double-logarithms.}
\label{fig:2gluons_nonleading}
\end{figure}
Thus, the leading contribution stems from the transverse momentum ordering $-k_{1\perp}^2 \ll -k_{2\perp}^2$ for the emitted gluons, and
\begin{eqnarray}
|\mathcal{M}^{\rm Ladder}_{\gamma*q \rightarrow q,2g}|^2_{\mu\nu} & \stackrel{\rm LL}{=} & 
\frac{1-z_1}{{-k_{1\perp}^2}} \left[ 2g_s^2P_{qq}(z_1) \right] \, \frac{1-z_2}{{-k_{2\perp}^2}} \left[ 2g_s^2P_{qq}(z_2) \right]  \\
 & & \frac{e_q^2}{2} {\rm Tr} \left[ \slashed{p'} \gamma_\mu \slashed{p} \gamma_\nu \right], \nonumber
\end{eqnarray}
Following the same steps as earlier, we find
\begin{eqnarray}
-g^{\mu\nu} M\hat{W}_{\mu\nu}^{q \rightarrow q,2g}  & =e_q^2 & \frac{1}{2} \left[ \frac{\alpha_s}{2\pi} \log \left( \frac{Q^2}{m^2} \right)\right]^2 
\int \frac{dz_2}{z_2} P_{qq}(z_2) \int \frac{dz_1}{z_1} P_{qq}(z_1) \nonumber \\
& & \xi \delta \left( \xi - \frac{x}{z_1z_2} \right),
\end{eqnarray}
which contributes to the hadronic tensor by
\begin{eqnarray}
-g^{\mu\nu} M\hat{W}_{\mu\nu} & = 
& \sum_q e_q^2 \frac{1}{2} \left[ \frac{\alpha_s}{2\pi} \log \left( \frac{Q^2}{m^2} \right)\right]^2 P_{qq} \otimes P_{qq} \otimes f_q.
\end{eqnarray}
The convolution between three objects above is defined by
\begin{equation}
 P_{qq} \otimes P_{qq} \otimes f_q = \int_x^1 \frac{dz_2}{z_2} P_{qq}(z_2) \int_{x/z_2}^1 \frac{dz_1}{z_1} P_{qq}(z_1) f_q \left( \frac{x}{z_1z_2} \right),
\end{equation}
with obvious extension to convolutions between an arbitrary number of functions. Thus, to $\mathcal{O}(\alpha_s^2)$, the leading logarithms organize themselves as
\begin{eqnarray}
\frac{d^2\sigma}{dxdQ^2} \stackrel{\rm LL} {=}  \frac{d^2\hat{\sigma_{0}}}{dxdQ^2} \sum_q  e_q^2 & & \hspace{-0.6cm} \left[1 + \left(\frac{\alpha_s}{2\pi}\right)  \log \left( \frac{Q^2}{m^2} \right) P_{qq} \right.  \\
& & \hspace{-0.8cm} + \left. \frac{1}{2} \left(\frac{\alpha_s}{2\pi}\right)^2  \log^2 \left( \frac{Q^2}{m^2} \right) P_{qq}\otimes P_{qq} \right]  \otimes f_q. \nonumber
\label{eq:2gluonfinal}
\end{eqnarray}

Based on a similar reasoning as in the end of the previous subsection, the diagrams like those in Fig.~\ref{fig:2gluons_nonleading} cannot contain $\mathcal{O}(\alpha_s^2 \log^2 (Q^2/m^2))$ terms in the axial gauge --- it is the ladder diagram in Fig.~\ref{fig:2gluons} alone that gives the leading logarithmic singularity.

The generalization to an arbitrary number of collinear gluon emissions from the initial quark is now quite straightforward: For $n$ emitted gluons the leading logarithms originate from the region of the phase space where the transverse momenta are \emph{strongly ordered}
$$-k_{1\perp}^2 \ll -k_{2\perp}^2 \ll \cdots \ll -k_{n-1\perp}^2 \ll -k_{n\perp}^2 \ll Q^2,$$
and the contribution to the DIS cross-section is 
\begin{figure}[h!]
\begin{center}
\includegraphics[scale=0.4]{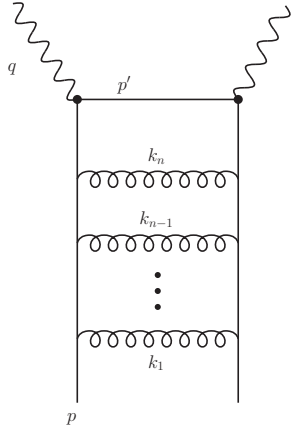}
\end{center}
\caption{Ladder graph for n-gluon emission.}
\label{fig:Ngluons}
\end{figure}
\begin{eqnarray}
\frac{d^2\hat{\sigma_{0}}}{dxdQ^2} \sum_q  e_q^2 \frac{1}{n!} \left(\frac{\alpha_s}{2\pi}\right)^n  \log^n \left( \frac{Q^2}{m^2} \right) \underbrace{P_{qq} \otimes  P_{qq} \otimes \cdots \otimes P_{qq}}_{n \, {\rm times}} \otimes f_q
\label{eq:Ngluonfinal}.
\end{eqnarray}
Thus, the leading logarithm contributions to the DIS cross-section constitute a series which is formally an exponential 
\begin{eqnarray}
\frac{d^2\sigma}{dxdQ^2} \stackrel{\rm LL} {=}  \frac{d^2\hat{\sigma_0}}{dxdQ^2} \sum_q  e_q^2  \exp \left[ \frac{\alpha_s}{2\pi} \log \left( \frac{Q^2}{m^2} \right) P_{qq} \right] \otimes f_q.
\label{eq:allgluonfinal} 
\end{eqnarray}
Comparing this expression to the corresponding parton model prediction, given in Eq.~(\ref{eq:bornDIS}), one can see that the resummation of the leading logarithms is equivalent to replacing the $Q^2$-independent parton distribution function by
\begin{equation}
 f_q(x) \rightarrow f_q(x,Q^2) \equiv \exp \left[ \frac{\alpha_s}{2\pi} \log \left( \frac{Q^2}{m^2} \right) P_{qq} \right] \otimes f_q.
\label{eq:simplestQ2quark}
\end{equation}
Taking the $Q^2$-derivative we see that $f_q(x,Q^2)$ satisfies the following integro-differential equation
\begin{equation}
\label{eq:1stDGLAP}
Q^2 \frac{\partial}{\partial Q^2}f_q(x,Q^2) = \frac{\alpha_s}{2\pi} P_{qq} \otimes f_q(x,Q^2),
\end{equation}
which is an archetype of the \emph{Dokshitzer-Gribov-Lipatov-Altarelli-Parisi evolution equations} \cite{Dokshitzer:sg,Gribov:ri,Gribov:rt,Altarelli:1977zs}, or DGLAP equations in brief.

\subsection{More splitting functions}

\begin{figure}[!h]
\begin{center}
\includegraphics[scale=0.4]{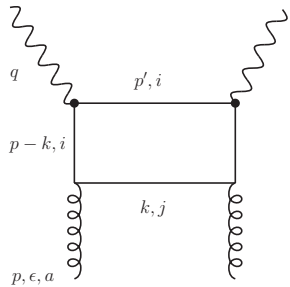}
\end{center}
\caption{A gluon-initiated ladder diagram.}
\label{fig:Gluetoqq}
\end{figure}

The gluon emission discussed above is, of course, only one possibility among other QCD-interactions. For example, from $\mathcal{O}(\alpha_s)$ onwards, also gluon-initiated subprocesses contribute to the deeply inelastic cross-section. The simplest such diagram is shown in Fig.~\ref{fig:Gluetoqq}. 
Similar to the gluon radiation graphs discussed in the preceding sections, also this diagram --- and in the axial gauge this ladder-type diagram only --- gives a collinear divergence. Extracting this divergence is rather straightforward having already carefully done the groundworks during the previous sections. Starting from the squared matrix element
\begin{equation}
|\mathcal{M}^{\rm Ladder}_{\gamma*G \rightarrow q\overline{q}}|^2_{\mu\nu} = 
g_s^2 T_R e_q^2 \frac{1}{2} \sum_{\rm pol} \frac{1}{t^4} {\rm Tr} \left[ \slashed{p'} \gamma_\mu \slashed{t} \slashed{\epsilon} \slashed{k} \slashed{\epsilon^*} \slashed{t} \gamma_\nu \right], \quad t=p-k
\end{equation}
where
$$ T_R \equiv \frac{1}{8} \, t^a_{ij}(t^a_{ij})^* = \frac{1}{8} {\rm Tr} (t^a t^a) = \frac{4}{8} = \frac{1}{2}$$
is the appropriate color factor, and introducing the Sudakov decomposition (\ref{eq:Sudakov1}) for the outgoing quark momentum $k$, the polarization sum gives
\begin{equation}
 \sum_{\rm pol} \slashed{\epsilon} \slashed{k} \slashed{\epsilon^*} = 2\left[(1-z)\slashed{p} + \beta \slashed{n} \right].
\end{equation}
To leading power in $k_\perp^2$ we find
\begin{equation}
\slashed{t} \left[(1-z)\slashed{p} + \beta \slashed{n} \right]  \slashed{t} = \frac{-k_\perp^2}{1-z} \left[ (1-z)^2 + z^2 \right] \slashed{p},
\end{equation}
and consequently
\begin{equation}
|\mathcal{M}^{\rm Ladder}_{\gamma*G \rightarrow q\overline{q}}|^2_{\mu\nu} \stackrel{\rm LL}{=} g_s^2 \frac{2(1-z)}{-k_\perp^2} \, T_R \left[ (1-z)^2 + z^2 \right] \times \frac{e_q^2}{2} {\rm Tr} \left[ \slashed{p'} \gamma_\mu \slashed{p} \gamma_\nu \right].
\end{equation}
Comparing to Eq.~(\ref{eq:Mq_to_qg}) we realize that the leading contribution to the deeply inelastic cross-section from the graph \ref{fig:Gluetoqq} becomes
\begin{equation}
\frac{d^2\hat{\sigma_{0}}}{dxdQ^2} \sum_q e_q^2 \, \left(\frac{\alpha_s}{2\pi}\right) \log \left( \frac{Q^2}{m^2} \right) P_{qg}  \otimes f_g,
\label{eq:2quarkfinal}
\end{equation}
where
\begin{equation}
 P_{qg}(z) \equiv T_R \left[ z^2 + (1-z)^2 \right]
\end{equation}
is the splitting function for a gluon$\rightarrow$quark transition and $f_g$ is the parton distribution function for the gluons.

\vspace{0.5cm}
Having now considered two different ladder vertices, we can also pile them on top of each other to form a parton ladder like the one in Fig.~\ref{fig:Gluetoqqg} below.
\begin{figure}[!h]
\begin{center}
\includegraphics[scale=0.4]{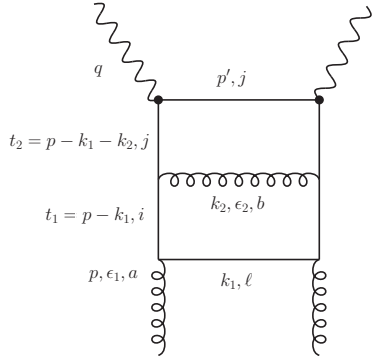}
\end{center}
\caption{Another gluon-initiated ladder diagram.}
\label{fig:Gluetoqqg}
\end{figure}
The corresponding squared matrix element reads
\begin{equation}
|\mathcal{M}^{\rm Ladder}_{\gamma*G \rightarrow q\overline{q}g}|^2_{\mu\nu} = 
g_s^2 T_R C_F \frac{e_q^2}{2} \sum_{\rm pol} \frac{1}{t_1^4 t_2^4} {\rm Tr} \left[ \slashed{p'} \gamma_\mu \slashed{t_2} \slashed{\epsilon_2^*} \slashed{t_1} \slashed{\epsilon_1} \slashed{k_1} \slashed{\epsilon_1^*} \slashed{t_1} \slashed{\epsilon_2} \slashed{t_2} \gamma_\nu \right],
\end{equation}
where the color factor arises as
\begin{eqnarray}
\frac{1}{8} (t_{ji}^{b} t_{i\ell}^{a})(t_{jn}^{b} t_{n\ell}^{a})^*  & = &
\frac{1}{8} (t^{a}t^{a})_{in} (t^{b}t^{b})_{ni} \nonumber \\
& = & \frac{1}{8} {\rm Tr}(t^{a}t^{a}) \frac{1}{3}{\rm Tr}(t^{b}t^{b}) \nonumber \\
& = & \frac{1}{2} \, \frac{4}{3} = T_R C_F \nonumber
\end{eqnarray}
Using the momentum decomposition (\ref{eq:Sudakov2}), the inner part of the trace above gives the leading factor
\begin{equation}
\sum_{\rm pol_1} \slashed{t_1} \slashed{\epsilon} \slashed{k} \slashed{\epsilon^*} \slashed{t_1} = 2\frac{-k_{1\perp}^2}{1-z_1} \left[ (1-z_1)^2 + z_1 \right] \slashed{p} + \ldots,
\end{equation}
and what is left inside the trace is, using (\ref{eq:Sudakov3}),
\begin{equation}
\sum_{\rm pol_2} \slashed{t_2} \slashed{\epsilon_2^*} \slashed{p} \slashed{\epsilon_2} \slashed{t_2} = 2\frac{({-k_{2\perp}^2})}{1-z_2} \left(\frac{1+z_2^2}{1-z_2}\right) \slashed{p} + \ldots.
\end{equation}
The leading region is again the one with $-k_{1\perp}^2 \le -k_{2\perp}^2$, and therefore
\begin{eqnarray}
|\mathcal{M}^{\rm Ladder}_{\gamma*q \rightarrow q,2g}|^2_{\mu\nu} & \stackrel{\rm LL}{=} & 
\frac{1-z_2}{{-k_{2\perp}^2}} \left[ 2g_s^2P_{qq}(z_2) \right] \,
\frac{1-z_1}{{-k_{1\perp}^2}} \left[ 2g_s^2P_{qg}(z_1) \right] \\
 & & \frac{e_q^2}{2} {\rm Tr} \left[ \slashed{p'} \gamma_\mu \slashed{p} \gamma_\nu \right], \nonumber
\end{eqnarray}
which leads to a contribution
\begin{eqnarray}
\frac{d^2\hat{\sigma_{0}}}{dxdQ^2} \sum_q  e_q^2 \, \frac{1}{2} \left(\frac{\alpha_s}{2\pi}\right)^2  \log^2 \left( \frac{Q^2}{m^2} \right) P_{qq}\otimes P_{qg} \otimes f_g \nonumber
\end{eqnarray}
to the deeply inelastic cross-section.
 
\begin{figure}[!h]
\begin{center}
\includegraphics[scale=0.4]{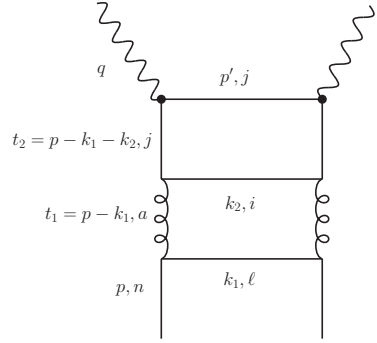}
\end{center}
\caption{Ladder diagram with gluon as a vertical line.}
\label{fig:Quarktoqq}
\end{figure}

\vspace{0.5cm}
Thus far we have only considered parton ladders with quarks at vertical lines. To illustrate how the calculation works when there are vertical gluon lines, let us consider the diagram shown in Fig.~\ref{fig:Quarktoqq}.
The squared matrix element is in this case
\begin{eqnarray}
|\mathcal{M}^{\rm Ladder}_{\gamma*q' \rightarrow q'\overline{q}q}|^2_{\mu\nu} & = &  
g_s^2 C_F T_R \frac{e_q^2}{2} \sum_{\rm pol} \frac{1}{t_1^4 t_2^4} {\rm Tr} \left[ \slashed{p'} \gamma_\mu \slashed{t_2} \gamma_\alpha \slashed{k_2} \gamma_\eta \slashed{t_2} \gamma_\nu \right] {\rm Tr} \left[ \slashed{k_1} \gamma_\beta \slashed{p} \gamma_\phi \right] \nonumber \\
& & \left[ -g^{\alpha\beta} + \frac{t_1^\alpha n^\beta + t_1^\beta n^\alpha}{t_1 \cdot n} \right] \left[ -g^{\eta\phi} + \frac{t_1^\eta n^\phi + t_1^\phi n^\eta}{t_1 \cdot n} \right],
\end{eqnarray}
where the color factor comes from
\begin{eqnarray}
\frac{1}{3} (t_{ji}^{a} t_{\ell n}^{a}) (t_{ji}^{b} t_{\ell n}^{b})^* & = &
\frac{1}{3} {\rm Tr}(t^{a}t^{b}) {\rm Tr}(t^{a}t^{b}) \nonumber \\
& = & \frac{1}{3} {\rm Tr}(t^{a}t^{a}) \frac{1}{8}{\rm Tr}(t^{b}t^{b}) \nonumber \\
& = & \frac{4}{3} \, \frac{1}{2} = C_F T_R  \nonumber.
\end{eqnarray}
Applying the Sudakov decomposition (\ref{eq:Sudakov2}), we find
\begin{eqnarray}
& & {\rm Tr} \left[ \slashed{k_1} \gamma_\beta \slashed{p} \gamma_\phi \right] \left[ -g^{\alpha\beta} + \frac{t_1^\alpha n^\beta + t_1^\beta n^\alpha}{t_1 \cdot n} \right] \left[ -g^{\eta\phi} + \frac{t_1^\eta n^\phi + t_1^\phi n^\eta}{t_1 \cdot n} \right] \nonumber \\
& & = \frac{8}{z_1^2} k_{1\perp}^\alpha k_{1\perp}^\eta + 2\frac{-k_{1\perp}^2}{1-z_1}\left[ -g^{\alpha\eta} + \frac{p^\alpha n^\eta + p^\eta n^\alpha}{p \cdot n} \right] + \cdots
\end{eqnarray}
Whereas the second term in the lower line explicitly contains the sum over the vertical gluon polarization states, the first term looks a bit puzzling. The trick is to notice that under the integration over the transverse momentum this term becomes
\begin{equation}
 \int d^2{\bf k_{\perp}} \frac{k_{\perp}^\alpha k_{\perp}^\eta}{D(k_{\perp}^2)} = \frac{1}{2} \left[ -g^{\alpha\eta} + \frac{p^\alpha n^\eta + p^\eta n^\alpha}{p \cdot n} \right] \int d^2{\bf k_{\perp}} \frac{-k_{\perp}^2}{D(k_{\perp}^2)},
\end{equation}
where $D$ is a function that depends only on $k_{\perp}^2$ and not on the transverse components separately. Thus, we may write
\begin{eqnarray}
& & {\rm Tr} \left[ \slashed{k_1} \gamma_\beta \slashed{p} \gamma_\phi \right] \left[ -g^{\alpha\beta} + \frac{t_1^\alpha n^\beta + t_1^\beta n^\alpha}{t_1 \cdot n} \right] \left[ -g^{\eta\phi} + \frac{t_1^\eta n^\phi + t_1^\phi n^\eta}{t_1 \cdot n} \right] \nonumber \\
& & = \frac{-2k_{1\perp}^2}{z_1(1-z_1)} \left[ \frac{1+(1-z)^2}{z_1} \right] \left( -g^{\alpha\eta} + \frac{p^\alpha n^\eta + p^\eta n^\alpha}{p \cdot n} \right).
\end{eqnarray}
Applying again the Sudakov decomposition (\ref{eq:Sudakov3}) we have
\begin{equation}
\slashed{t_2} \gamma_\alpha \slashed{k_2} \gamma_\eta \slashed{t_2}\left( -g^{\alpha\eta} + \frac{p^\alpha n^\eta + p^\eta n^\alpha}{p \cdot n} \right) = 2\frac{-k_{2\perp}^2}{1-z_2} \left[ (1-z_2)^2 + z_2 \right] z_1 \slashed{p} + \ldots \nonumber,
\end{equation}
and in the leading region
\begin{eqnarray}
|\mathcal{M}^{\rm Ladder}_{\gamma*q' \rightarrow q'\overline{q}q}|^2_{\mu\nu} & \stackrel{\rm LL}{=} & 
\frac{1-z_2}{{-k_{2\perp}^2}} \left[ 2g_s^2P_{qg}(z_2) \right] \,
\frac{1-z_1}{{-k_{1\perp}^2}} \left[ 2g_s^2P_{gq}(z_1) \right] \\
 & & \frac{e_q^2}{2} {\rm Tr} \left[ \slashed{p'} \gamma_\mu \slashed{p} \gamma_\nu \right], \nonumber
\end{eqnarray}
where
\begin{equation}
 P_{gq}(z) \equiv C_F \left[ \frac{1+(1-z)^2}{z} \right]
\end{equation}
is the splitting function for the quark$\rightarrow$gluon transition. The squared matrix element above now leads to a term
\begin{eqnarray}
\frac{d^2\hat{\sigma_{0}}}{dxdQ^2} \sum_{q,q'}  e_q^2 \, \frac{1}{2} \left(\frac{\alpha_s}{2\pi}\right)^2  \log^2 \left( \frac{Q^2}{m^2} \right) P_{qg}\otimes P_{gq'} \otimes f_{q'} \nonumber
\end{eqnarray}
in the deeply inelastic cross-section.

\vspace{0.5cm}
\begin{figure}[!h]
\begin{center}
\includegraphics[scale=0.4]{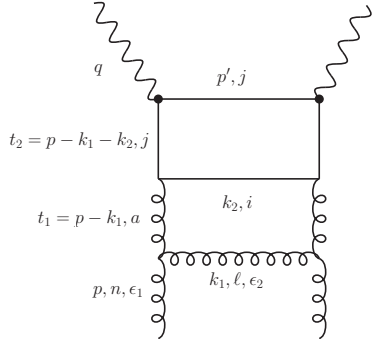}
\end{center}
\caption{A ladder diagram from which one can compute sthe plitting function for gluon$\rightarrow$gluon transition.}
\label{fig:ggg}
\end{figure}
The remaining splitting function to be calculated is $P_{gg}$ corresponding to the gluon$\rightarrow$gluon transition. This can be computed from the ladder diagram depicted in Fig.~\ref{fig:ggg} which corresponds to the squared matrix element
\begin{eqnarray}
|\mathcal{M}^{\rm Ladder}_{\gamma*g \rightarrow g\overline{q}q}|^2_{\mu\nu} & = &  
g_s^2 C_G T_R \frac{e_q^2}{2} \sum_{\rm pol} \frac{1}{t_1^4 t_2^4} {\rm Tr} \left[ \slashed{p'} \gamma_\mu \slashed{t_2} \gamma_\alpha \slashed{k_2} \gamma_\delta \slashed{t_2} \gamma_\nu \right] \\
& & \hspace{-1cm} \left[ -g^{\beta\eta}(t_1+p)^\phi + g^{\eta\phi}(p+k_1)^\beta +g^{\phi\beta}(t_1-k_1)^\eta \right] \nonumber \\
& & \hspace{-1cm} \left[ -g^{\rho\xi}(t_1+p)^\chi \, + g^{\xi\chi}(p+k_1)^\rho \, + g^{\chi\rho}(t_1-k_1)^\xi \right] \nonumber \\
& & \hspace{-1cm} \left[ -g^{\alpha\beta} + \frac{t_1^\alpha n^\beta + t_1^\beta n^\alpha}{t_1 \cdot n} \right] \left[ -g^{\delta \rho} + \frac{t_1^\delta n^\rho + t_1^\rho n^\delta}{t_1 \cdot n} \right] \epsilon_{1\eta} \epsilon_{1\xi}^* \epsilon_{2\chi} \epsilon_{2\phi}^*. \nonumber
\end{eqnarray}
The evaluation of this matrix element is somewhat more tedious than the previous ones, yet straighforward. The resulting leading logarithmic contribution to the deeply inelastic cross-section is
\begin{eqnarray}
\frac{d^2\hat{\sigma_{0}}}{dxdQ^2} \sum_{q}  e_q^2 \, \frac{1}{2} \left(\frac{\alpha_s}{2\pi}\right)^2  \log^2 \left( \frac{Q^2}{m^2} \right) P_{qg}\otimes P_{gg} \otimes f_{g} \nonumber,
\end{eqnarray}
with
\begin{equation}
 P_{gg}(z) \equiv 2C_G \left[ \frac{1-z}{z} + \frac{z}{1-z} + z(1-z)\right], \qquad C_G=3.
\end{equation}

\begin{figure}[!h]
\begin{center}
\includegraphics[scale=0.4]{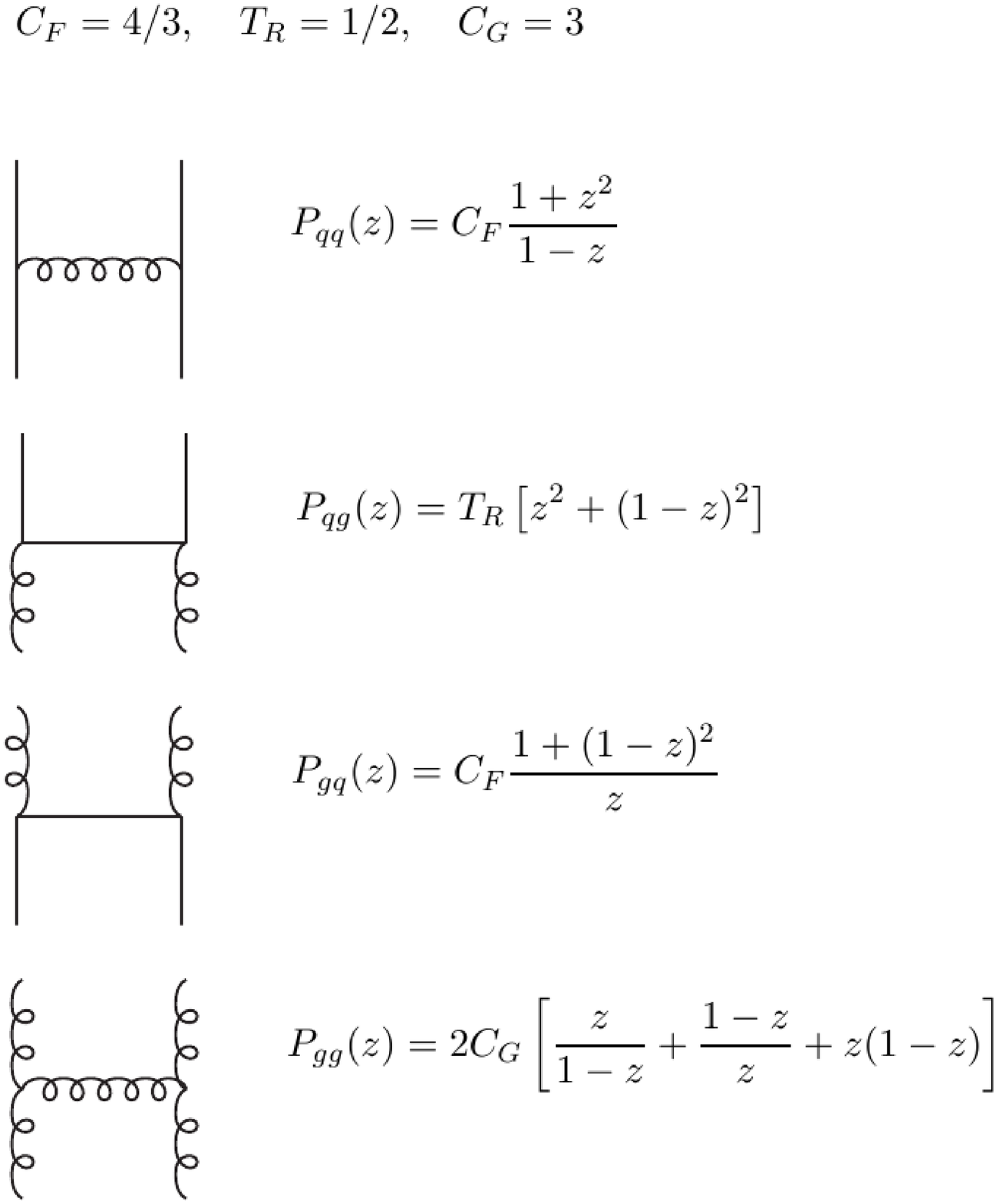}
\end{center}
\caption{Unpolarized splitting functions.}
\label{fig:Splittings1}
\end{figure}

\vspace{0.5cm}
From now on, one can pretty much see how this goes on: each additional ladder-compartment in which parton of flavor $i$ transforms to $j$, effectively just increments the power of ${\alpha_s}\log {Q^2}/{m^2}$ by one unit and adds the corresponding splitting function $P_{ij}$ to the convolution integral. The possible building blocks for constructing the ladders are displayed in Fig.~\ref{fig:Splittings1} together with the characteristic splitting functions. 
Clearly, we should account for all possible parton ladders --- also the gluon-triggered ones --- when defining the scale-dependent quark densities. Therefore, we define the scale-dependent parton distributions as a sum of all possible ladders that end up with the specific parton. The appropriate generalization of the definition (\ref{eq:simplestQ2quark}) can be neatly written down as a matrix equation
\begin{equation}
\left(
\begin{array}{c}
f_{q_i}(x,Q^2) \\
f_g(x,Q^2)
\end{array}
\right)
\equiv
\exp \left[ \frac{\alpha_s}{2\pi} \log \left( \frac{Q^2}{m^2} \right)
\left(
\begin{array}{cc}
P_{q_i q_j} & P_{q_i g} \\
P_{g q_j} & P_{g g}
\end{array}
\right)
\right]
\otimes
\left(
\begin{array}{c}
f_{q_j} \\
f_g
\end{array}
\right), 
\end{equation}
where $q_j$ should be understood as being a vector with different quark flavors as its components and the splitting functions $P_{q_i q_j}$, $P_{q_i g}$, $P_{g q_j}$ as matrices with the appropriate dimension. In the leading logarithm approximation, the splitting functions are flavor-blind and we can write, explicitly
\begin{equation}
\left(
\begin{array}{cc}
P_{q_i q_j} & P_{q_i g} \\
P_{g q_j} & P_{g g}
\end{array}
\right)
=
\left(
\begin{array}{ccccc}
P_{qq}  & 0       & 0       & \cdots & P_{qg}  \\
0       & P_{qq}  & 0       & \cdots & P_{qg}  \\
0       & 0       & P_{qq}  & \cdots & P_{qg}  \\
\vdots  & \vdots  & \vdots  & \ddots & \vdots  \\
P_{gq}  & P_{gq}  & P_{gq}  & \cdots & P_{g g}
\end{array}
\right).
\end{equation}
The complete set of DGLAP evolution equations follow by taking the $Q^2$-derivative:
\begin{equation}
Q^2 \frac{\partial }{\partial Q^2}
\left(
\begin{array}{c}
f_{q_i}(x,Q^2) \\
f_g(x,Q^2)
\end{array}
\right)
\equiv
\frac{\alpha_s}{2\pi}
\left(
\begin{array}{cc}
P_{q_i q_j} & P_{q_i g} \\
P_{g q_j} & P_{g g}
\end{array}
\right)
\otimes
\left(
\begin{array}{c}
f_{q_j}(Q^2) \\
f_g(Q^2)
\end{array}
\right).
\label{eq:completeDGLAP1}
\end{equation}



In summary, the leading collinear singularities in the perturbative Feynman-diagram expansion can be factored to the scale-dependent parton distributions $f_i(x,Q^2)$ such that the parton model prediction for the DIS cross-section stays formally intact, but the parton densities no longer respect the Bjorken-scaling but are $Q^2$-dependent. Also the interpretation of the parton distributions as simple number densities upgrades to being number densities with transverse momentum up to $Q^2$. This and further extensions to the simple parton model are often referred to as \emph{pQCD-improved parton model}.


\vspace{0.5cm}
There are still, however, two serious deficiencies in the equations (\ref{eq:completeDGLAP1}):
\begin{itemize}
\item
The splitting functions $P_{qq}(z)$ and $P_{gg}(z)$ diverge as $z \rightarrow 1$ due to the $1/(1-z)$-poles, making the convolution integrals thereby meaningless.
\item
The argument of the strong coupling constant $\alpha_s$ remains undefined.
\end{itemize}
In order to fill these gaps, we need to discuss also the virtual corrections to the parton ladder on a same footing with the real parton radiation. 

\section{Virtual corrections}

\subsection{Quark self-energy}

The Sudakov decomposition of the momentum provided a useful tool for extracting the collinear limits of the parton radiation diagrams. This is also true in the case of loop-integrations, but the parametrization must be slightly modified to account for the non-zero virtuality of the loop momentum $k$. Also, the virtuality $p^2$ of the parton (as it may lie in the middle of the ladder) should be kept arbitrary. As an appropriate extension to (\ref{eq:Sudakov1}), I will decompose the loop momentum as\footnote{An explicit definition of the axial vector $n$ is not needed here.}
\begin{equation}
k = zp + \beta n + k_\perp, \qquad \beta = \frac{k^2 - k_\perp^2 - z^2 p^2}{2z (p \cdot n)}.
\end{equation}

\begin{figure}[h]
\begin{center}
\includegraphics[scale=0.4]{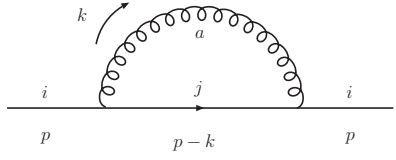}
\end{center}
\caption{Quark self-energy correction at 1-loop. The momentum of the parent quark is $p$ while the loop momentum is denoted by $k$. The letters $i,j,a$ stand for the colors indices.}
\label{fig:QuarkSelf}
\end{figure}

\begin{figure}[h]
\begin{center}
\includegraphics[scale=0.5]{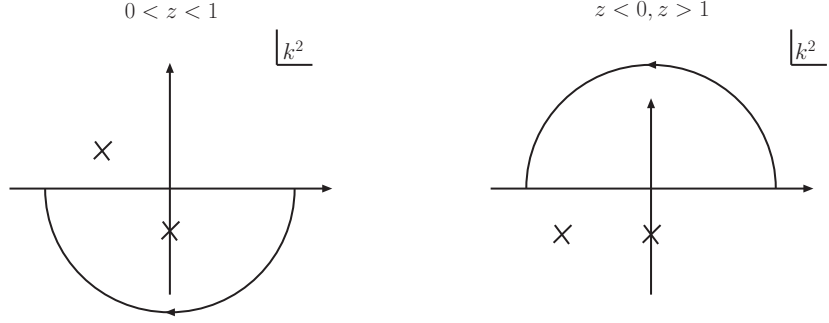}
\end{center}
\caption{The integration contours and the pole structure for $0<z<1$ (left) and $z<0, \, z>1$ (right).}
\label{fig:Integcontours}
\end{figure}

The axial-gauge expression for the quark self-energy diagram shown in Fig.~\ref{fig:QuarkSelf} reads
\begin{eqnarray}
\Sigma(p) & = & -g_s^2 C_F \int \frac{d^4k}{(2\pi)^4} \frac{ \gamma_\mu (\slashed{p} - \slashed{k}) \gamma_\nu}{[(p-k)^2 + i\epsilon][k^2 + i \epsilon]} \left[ g^{\mu \nu} - \frac{k^\mu n^\nu + k^\nu n^\mu}{k \cdot n} \right] \nonumber \\
& = & \Sigma_{\rm Feynman}(p) + \Sigma_{\rm Axial}(p).
\end{eqnarray}
I will now carefully show how to extract the collinear divergence starting with the Feynman gauge contribution
\begin{equation}
\Sigma_{\rm Feynman}(p) = -g_s^2 C_F \int \frac{d^4k}{(2\pi)^4} \frac{ \gamma_\mu (\slashed{p} - \slashed{k}) \gamma^\mu}{[(p-k)^2 + i\epsilon][k^2 + i \epsilon]}. 
\end{equation}
First, in the numerator
\begin{eqnarray}
 \gamma_\mu (\slashed{p} - \slashed{k}) \gamma^\mu & = & -2(\slashed{p} - \slashed{k}) \nonumber  \\
 & = & -2 \left[(1-z)\slashed{p} - \beta \slashed{n} - \slashed{k}_\perp \right] \\
 & = & -2 \left[(1-z)\slashed{p} - \beta \slashed{n} \right] \nonumber, 
\end{eqnarray}
where I dropped the $\slashed{k}_\perp$-term as its contribution will vanish as an odd integral. The denominator from the quark propagator is
\begin{equation}
 (p-k)^2 + i\epsilon= -\frac{1-z}{z} \left[ k^2 - \frac{k_\perp^2}{1-z} - z p^2 - i {\epsilon}' \right],
\end{equation}
where I defined
\begin{equation}
\epsilon' \equiv \frac{z}{1-z} \epsilon = \left\{
\begin{array}{l}
 < 0 \,\,\, {\rm if} \,\,\, z < 0 \,\,\, {\rm or} \,\,\, z > 1 \\
 > 0 \,\,\, {\rm if} \,\,\, 0 < z < 1.
\end{array}
\right.
\end{equation}
The sign of $\epsilon'$ is essential in defining the locations of poles in the $k^2$-plane.
\begin{eqnarray}
\Sigma_{\rm Feynman}(p) = - \frac{\alpha_s}{2\pi} C_F \frac{1}{2\pi} \int dz d {\bf k_\perp^2} dk^2 \frac{z}{|z|} \frac{1}{1-z}  \nonumber \\ \frac{(1-z)\slashed{p} - \beta \slashed{n} }{[k^2 + i \epsilon]\left[ k^2 - \frac{k_\perp^2}{1-z} - z p^2 - i {\epsilon}' \right]} 
\end{eqnarray}
Except for the $k^2$-terms in the numerator which will not contribute to the $k_\perp^2 \rightarrow 0$ divergence, the $k^2$-integral can be evaluated as a contour integral. Depending on the value of $z$, the locations of the $k^2$-poles and the convenient integration contours are shown in Fig.~(\ref{fig:Integcontours}). 
If $0<z<1$, the integration contour can be closed in the lower half-plane enclosing $k^2 = -i\epsilon$ pole
\begin{equation}
 \int_{-\infty}^{+\infty} \frac{dk^2}{[k^2 + i \epsilon]\left[ k^2 - \frac{k_\perp^2}{1-z} - z p^2 - i {\epsilon}' \right]} = -2 \pi i \frac{1-z}{{\bf k_\perp^2} -z(1-z)p^2},
\end{equation}
while for other values of $z$ the integral contour can be closed in the upper half-plane where there are no poles and integral gives zero. Therefore,
\begin{eqnarray}
\Sigma_{\rm Feynman}(p) & \approx & i \left(\frac{\alpha_s}{2\pi}\right) C_F \int_0^1 dz  \int_0^{\Lambda^2} \frac{d{\bf k_\perp^2}}{{\bf k_\perp^2} -z(1-z)p^2} \\
& & \left[ (1-z)\slashed{p} + \frac{-{\bf k_\perp^2} + z^2p^2}{2z (p \cdot n)} \slashed{n} \right] \nonumber \\
& \approx & i \slashed{p} \left(\frac{\alpha_s}{2\pi}\right) C_F \log \left( \frac{\Lambda^2}{-p^2} \right) \int_0^1 dz (1-z), \nonumber
\end{eqnarray}
where I have dropped all but the logarithmic contribution diverging as $p^2 \rightarrow 0$, and regulated the UV-divergence by a cut-off $\Lambda^2$. 
The axial contribution 
\begin{equation}
\Sigma_{\rm Axial}(p) = g_s^2 C_F \int \frac{d^4k}{(2\pi)^4} \frac{ \slashed{k}(\slashed{p} - \slashed{k})\slashed{n}  + \slashed{n}(\slashed{p} - \slashed{k})\slashed{k}}{[(p-k)^2 + i\epsilon][k^2 + i \epsilon]} \frac{1}{k \cdot n}
\end{equation}
can dealt with similarly. Applying the Sudakov decomposition plus dropping all $k^2$-terms and those ones odd in $k_\perp$, the numerator simplifies to 
\begin{equation}
 \slashed{k}(\slashed{p} - \slashed{k})\slashed{n}  + \slashed{n}(\slashed{p} - \slashed{k})\slashed{k} \approx 2 \slashed{n} \frac{{\bf k_\perp^2}}{z}.
\end{equation}
Apart from the  factor $k \cdot n = z p \cdot n$, the denominator is identical with the Feynman-gauge case and the $k^2$-integral can be performed in a similar way. The result is
\begin{eqnarray}
\Sigma_{\rm Axial}(p) & \approx & i \frac{\slashed{n}}{p \cdot n} \left(\frac{\alpha_s}{2\pi}\right) C_F \int_0^1 \frac{dz}{z^2}  \int_0^{\Lambda^2} \frac{d{\bf k_\perp^2}{\bf k_\perp^2}}{{\bf k_\perp^2} -z(1-z)p^2} \\
& \approx & i \frac{p^2}{2p \cdot n}\slashed{n} \left(\frac{\alpha_s}{2\pi}\right) C_F \log \left( \frac{\Lambda^2}{-p^2} \right) \int_0^1 dz \frac{2z}{1-z} \nonumber.
\end{eqnarray}
Thus, the total self-energy correction is of the form
$$
\Sigma(p) = \Sigma_{\rm Feynman}(p) + \Sigma_{\rm Axial}(p) = i \left[ A \slashed{p} + B \frac{p^2}{2 p \cdot n} \slashed{n} \right],
$$
and the 1-loop corrected quark propagator becomes
\begin{equation}
 \frac{i \slashed{p}}{p^2} + \frac{i \slashed{p}}{p^2} \Sigma(p) \frac{i \slashed{p}}{p^2} = \frac{i \slashed{p}}{p^2} \left[ 1-A-B \right] +i\frac{B}{2 p \cdot n}\slashed{n}.
\end{equation}
The last term above will lead to a loss of a leading logarithm in the parton ladder and thus the 1-loop correction to the quark propagator can be effectively accounted by a multiplicative renormalization constant 
\begin{equation}
 Z_q(p^2) = 1-A-B = 1 - \left(\frac{\alpha_s}{2\pi}\right) C_F \log \left( \frac{\Lambda^2}{-p^2} \right) \int_0^1 dz \left( \frac{1+z^2}{1-z} \right).
\label{eq:Zquark}
\end{equation}
This is also the residue of the $p^2 \rightarrow 0$ propagator pole needed (through the LSZ-reduction) if the quark is an on-shell final/initial state particle. Although I did not carefully keep track of the UV-divergences, the Eq.~(\ref{eq:Zquark}) is also accurate in this sense --- it is a cut-off regulated version of the result regulated by going to $4-\epsilon$ dimensions\footnote{The loop calculations in the light-cone gauge are also little tricky, see \cite{Pritchard:1978ts,Stirling:1979az}.}
$$
1 - \left(\frac{\alpha_s}{2\pi}\right) \left[\frac{2}{\epsilon} + \log \left( \frac{\mu^2}{-p^2} \right) \right] C_F \int_0^1 dz \left( \frac{1+z^2}{1-z} \right).
$$
 
To understand how this damps the singular behaviour of $P_{qq}(z)$, it is simplest to look at the effect of applying the $\mathcal{O}(\alpha_s)$ external leg correction to the leading order contribution in the DIS cross-section (\ref{eq:1gluonfinal}). 
\begin{eqnarray}
{d\sigma} & \stackrel{\rm LL}{=} & {d\hat{\sigma_{0}}} \sum_q e_q^2
\left[Z_q(-m^2) + \left(\frac{\alpha_s}{2\pi}\right) \log \left( \frac{Q^2}{m^2} \right) P_{qq}  \right] \otimes f_q \nonumber \\
& = & {d\hat{\sigma_{0}}} \sum_q e_q^2 \left[Z_q(-Q^2) + \left(\frac{\alpha_s}{2\pi}\right) \log \left( \frac{Q^2}{m^2} \right) \tilde{P}_{qq}  \right] \otimes f_q, \label{eq:Pqqregulation} 
\end{eqnarray}
where I defined the regulated splitting function  $\tilde{P}_{qq}(z)$ by
\begin{eqnarray}
\tilde{P}_{qq}(z) & \equiv & C_F \left(\frac{1+z^2}{1-z}\right)_+ \\
\left(\frac{1+z^2}{1-z}\right)_+ & \equiv & \left(\frac{1+z^2}{1-z}\right) - \delta(1-z) \int_0^1 dx \left(\frac{1+x^2}{1-x}\right)
\end{eqnarray}
Here, we meet so-called \emph{plus distribution}, which should be understood through integration against sufficiently smooth ``test-function'' $h(x)$:
\begin{equation}
\int_0^1 dz h(z) \left(\frac{1+z^2}{1-z}\right)_+ = \int_0^1 dz \left(\frac{1+z^2}{1-z}\right) \left[h(z)-h(1)\right].
\label{eq:plus1}
\end{equation}
There are two important things to be emphasized: First, the inclusion of the virtual correction serves to regulate the $z \rightarrow 1$ singularity of the splitting function $P_{qq}(z)$, as promised. Second, the virtual piece $Z_q(-m^2)$ gets replaced by $Z_q(-Q^2)$ which no longer contains collinear $m^2 \rightarrow 0$ divergence.


\subsection{Gluon self-energy}

\begin{figure}[h]
\begin{center}
\includegraphics[scale=0.5]{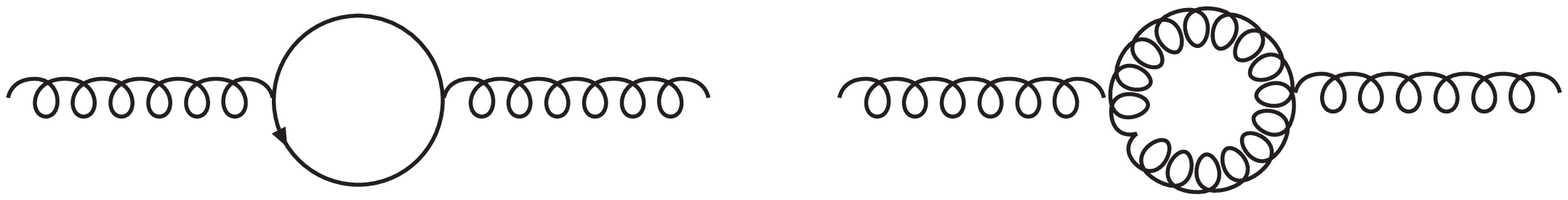}
\end{center}
\caption{Diagrams contributing to the gluon self-energy correction at 1-loop in the axial gauge.}
\label{fig:GluonSelf}
\end{figure}
The gluon-self energy 1-loop correction can be calculated following essentially the same procedure. In the axial gauge, there are only two contributing diagrams, shown in Fig.~(\ref{fig:GluonSelf}).
Evaluating the quark and gluon loops the following logarithmic pieces are found
\begin{eqnarray}
 \Pi^q_{\mu\nu}(p) & = & - i\frac{\alpha_s}{2\pi} p^2 \left(g_{\mu\nu} - \frac{p_\mu p_\nu}{p^2} \right) \log \left( \frac{\Lambda^2}{-p^2} \right) \frac{2}{3} n_f T_f \\
\Pi^g_{\mu\nu}(p) & = & i \frac{\alpha_s}{2\pi} \log \left( \frac{\Lambda^2}{-p^2} \right) C_G \left[ p^2\left( g_{\mu\nu} - \frac{p_\mu p_\nu}{p^2} \right) \left( \frac{11}{6} -2\int_0^1 \frac{dz}{1-z} \right)\right. \nonumber \\
& & + \quad \left. \left(p^\mu - \frac{p^2}{p \cdot n}n^\mu \right) \left(p^\nu - \frac{p^2}{p \cdot n}n^\nu \right) 2 \left( 1- \int_0^1 \frac{dz}{1-z} \right) \right]. \nonumber
\end{eqnarray}
Thus, the result is of the form
$$
 \Pi_{\mu\nu}(p) = i \left[Ap^2 \left(g_{\mu\nu} - \frac{p_\mu p_\nu}{p^2} \right) + B \left(p^\mu - \frac{p^2}{p \cdot n}n^\mu \right) \left(p^\nu - \frac{p^2}{p \cdot n}n^\nu \right) \right],
$$
inducing a correction
\begin{eqnarray}
& & \frac{-i}{k^2} \left( g^{\mu\eta} - \frac{k^\mu n^\eta + k^\eta n^\mu}{k \cdot n} \right) \Pi_{\eta \xi}(p) \frac{-i}{k^2} \left( g^{\xi\nu} - \frac{k^\xi n^\nu + k^\nu n^\xi}{k \cdot n} \right) \nonumber \\
& = & A \, \frac{-i}{k^2} \left( g^{\mu\eta} - \frac{k^\mu n^\eta + k^\eta n^\mu}{k \cdot n} \right) -i B \frac{n^\mu n^\nu}{(k \cdot n)^2} \nonumber 
\end{eqnarray}
to the gluon propagator. Again, the latter part above does not give a leading logarithm, and the loop insertions can again be effectively accounted for by a multiplicative renormalization constant 
\begin{equation}
Z_G(p^2) = 1 + \frac{\alpha_s}{2\pi} \log \left( \frac{\Lambda^2}{-p^2} \right) \left[ C_G \left( \frac{11}{6} -2\int_0^1 \frac{dz}{1-z} \right) - \frac{2}{3} n_f T_R\right].
\label{eq:Z_G}
\end{equation}
As a consequence, the gluon $\rightarrow$ gluon splitting function $P_{gg}(z)$ gets replaced by a regulated one
\begin{equation}
\tilde{P}_{gg}(z) \equiv 2C_G \left[ \frac{1-z}{z} + \frac{z}{(1-z)_+} + z(1-z)\right] + \left[ \frac{11}{6}C_G - \frac{2}{3}n_f T_f \right] \delta(1-z).
\end{equation}

\subsection{Renormalization of the ladder vertex}

The standard field theory text books (e.g. \cite{Field:1989uq,Kaku:1993ym}) relate the running coupling constant $g_s(\mu^2)$ to the bare one $g_{0s}$ e.g. by
\begin{eqnarray}
 g_s(\mu^2) & \equiv & \frac{\sqrt{Z_G(k_A^2)}\sqrt{Z_q(k_B^2)}\sqrt{Z_q(k_C^2)}}{Z_1(k_A^2,k_B^2,k_C^2)} g_{0s},
\label{eq:defstrong}
\end{eqnarray}
with $k_A^2 = k_B^2 = k_C^2 = -\mu^2$,  and where $Z_1$ is the renormalization factor for the qqg-vertex. 
At 1-loop, we have the well-known result
\begin{eqnarray}
 \alpha_s(\mu^2) & = & \frac{\alpha_s^0}{1-\frac{\alpha_s^0}{4\pi} \beta_0 \log \left( \frac{\Lambda^2}{\mu^2} \right)} = \alpha_s^0 \left[ 1 + \frac{\alpha_s^0}{4\pi} \beta_0 \log \left( \frac{\Lambda^2}{\mu^2} \right) + \cdots \right], \nonumber \\
& = & \frac{4\pi}{\beta_0 \log \left( {\mu^2}/{\Lambda_{\rm QCD}^2} \right)}
\end{eqnarray}
where $\beta_0 = \frac{11}{3}C_G - \frac{4}{3} T_R n_f$, $\Lambda^2$ in the first line denotes the cut-off, and $\Lambda_{\rm QCD} \approx 200 \, {\rm MeV}$ is the QCD scale parameter.

\begin{figure}[h]
\begin{center}
\includegraphics[scale=0.5]{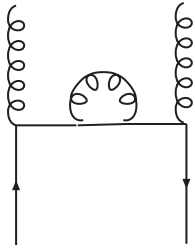}
\end{center}
\caption{Diagram for gluon radiation from the outgoing quark.}
\label{fig:qqg_plus_g}
\end{figure}

\vspace{0.5cm}
In a parton ladder, however, the kinematics appear quite different from that in (\ref{eq:defstrong}) --- the virtualities $k_A^2, k_B^2$ of the partons in the vertical lines are strongly ordered while the horizontal rung $k_C^2$ is on-shell:
\begin{equation}
 -k_A^2, -k_B^2 > 0, \qquad -k_A^2 < -k_B^2, \qquad k_C^2 = 0.
 \label{eq:kinematic_conf}
\end{equation}
In order to discuss what is involved, I will adopt a specific example with quark (A) splitting to a on-shell quark (C) and virtual gluon (B).
\begin{figure}[h]
\begin{center}
\includegraphics[scale=0.5]{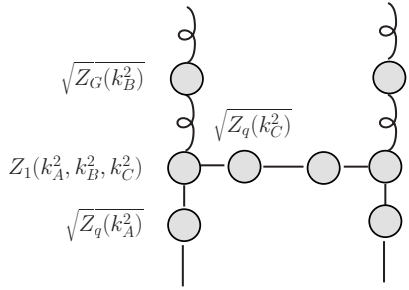}
\end{center}
\caption{Various renormalization factors for a ladder vertex.}
\label{fig:LadderRen}
\end{figure}
The outgoing quark contributes by a factor $Z_q(-m^2)$ with $m^2 \rightarrow 0$. However, to the same order in $\mathcal{O}(\alpha_s)$ we should also consider the contribution from the horizontal quark rung radiating an additional gluon as shown in Fig.~\ref{fig:qqg_plus_g}. 
A calculation along the lines presented in Sec.~\ref{Onegluonemission}, reveals that the leading contribution is effectively a multiplicative factor
\begin{equation}
\left[{\rm Ladder \,\, skeleton}\right] \times C_F \frac{\alpha_s}{2\pi} \int_{m^2}^{- k_{B}^2 \sim {\bf  k_{\perp}^2}} \frac{d{\bf k^{2}_{g\perp}}}{\bf  k_{g\perp}^2} \int_0^1 dz \frac{1+z^2}{1-z},
\end{equation}
where the upper limit ${\bf k_{\perp}^2}$ denotes the aggregate transverse momentum carried by the outgoing quark-gluon system and derives from the requirement not to upset the underlying logarithmic structure of the ladder. 
This is exactly of the same form as the quark renormalization factor $Z_q(-m^2)$, and when combined, the sum is clearly free from collinear divergences. In effect, we may simply make a replacement $Z_q(-m^2) \rightarrow Z_q(k_{B}^2)$. Similarly, the incoming quark line involves a renormalization factor $Z_q(k_A^2)$ which gets replaced by $Z_q(k_{B}^2)$ by the mechanism demonstrated in Eq.~(\ref{eq:Pqqregulation}) when the contribution of gluon radiation is included.

\vspace{0.3cm}
The remaining piece is the vertex part $Z_1(k_{A}^2,k_{B}^2,k_{C}^2)$. However, in the axial gauge it happens that if $k_{B}^2$ is kept finite, $Z_1$ does not contain terms that would be divergent in the $k_{A,C}^2 \rightarrow 0$ limit. In other words, in axial gauge all mass singularities are contained in the self-energy factors and we may safely replace $Z_1(k_{A}^2,k_{B}^2,k_{C}^2)$ by $Z_1(k_{B}^2,k_{B}^2,k_{B}^2)$ without losing large logarithms. 
Thus, the virtual corrections indeed eventually combine to the usual definition of the running coupling,
\begin{eqnarray}
\frac{\sqrt{Z_G(k_A^2)}\sqrt{Z_q(k_B^2)}\sqrt{Z_q(k_C^2)}}{Z_1(k_A^2,k_B^2,k_C^2)} \approx \frac{\sqrt{Z_G(k_B^2)}\sqrt{Z_q(k_B^2)}\sqrt{Z_q(k_B^2)}}{Z_1(k_B^2,k_B^2,k_B^2)} = g_s(-k_B^2). \nonumber
\end{eqnarray}

Incorporation of the running coupling to the resummation of leading logarithms is straightforward: In each ladder vertex we change $\alpha_s \rightarrow \alpha_s({\bf k_{\perp}^2})$, and do the nested transverse momentum integrals like (\ref{eq:doublelog}) by a change of variables
\begin{equation}
 \kappa({\bf k_{\perp}^2}) \equiv \frac{2}{\beta_0} \log \left[ \frac{\alpha_s(m^2)}{\alpha_s({\bf k_{\perp}^2})} \right],
\end{equation}
such that 
\begin{eqnarray}
& & \int^{Q^2}_{m^2} \frac{d{\bf k_{2\perp}^2}}{{\bf k_{2\perp}^2}} \frac{\alpha_s({\bf k_{2\perp}^2})}{2\pi} \int^{{\bf k_{2\perp}^2}}_{m^2} \frac{d{\bf k_{1\perp}^2}}{{\bf k_{1\perp}^2}} \frac{\alpha_s({\bf k_{1\perp}^2})}{2\pi} \nonumber \\
& = & \int^{\kappa(Q^2)}_{0} d\kappa({\bf k_{2\perp}^2})  \int^{\kappa({\bf k_{2\perp}^2})}_{0} d\kappa({\bf k_{1\perp}^2}) \\
& = & \frac{1}{2} \kappa^2 (Q^2). \nonumber
\end{eqnarray}

The correspondingly corrected definition for the scale-dependent parton densities becomes
\begin{equation}
\left(
\begin{array}{c}
f_{q_i}(x,Q^2) \\
f_g(x,Q^2)
\end{array}
\right)
\equiv
\exp \left[ \kappa(Q^2) 
\left(
\begin{array}{cc}
P_{q_i q_j} & P_{q_i g} \\
P_{g q_j} & P_{g g}
\end{array}
\right)
\right]
\otimes
\left(
\begin{array}{c}
f_{q_j}(x) \\
f_g(x)
\end{array}
\right),
\end{equation}
where the splitting functions are understood as being the regulated ones. By differentiating with respect to $\kappa$ and changing variables, we obtain the complete DGLAP evolution equations that resum all leading logarithms:
\begin{eqnarray}
Q^2 \frac{\partial f_{q_i}(x,Q^2)}{\partial Q^2} & = & \frac{\alpha_s(Q^2)}{2\pi} \left[P_{q_iq_j} \otimes f_{q_j}(Q^2) + P_{q_ig} \otimes f_g (Q^2) \right] \nonumber \\
\\
Q^2 \frac{\partial f_{g}(x,Q^2)}{\partial Q^2} & = & \frac{\alpha_s(Q^2)}{2\pi} \left[P_{gg} \otimes f_{g}(Q^2) + P_{gq_j} \otimes f_{q_j} (Q^2) \right]. \nonumber 
\label{eq:completeDGLAP3}
\end{eqnarray}

\section{Higher orders}

In this section I have systematically kept to the leading logarithm approximation, in which the logic is to retain only terms of the form $\alpha_s^n \log^n(Q^2/m^2)$, discarding all contributions which are suppressed by additional powers of $\alpha_s$. However, if one keeps track also of the non-leading contributions 
$$\alpha_s^{n+1} \log^n(Q^2/m^2), \, \alpha_s^{n+2} \log^n(Q^2/m^2), \ldots$$
one finds that the splitting functions $P_{f_if_j}$ actually constitute a power series in $\alpha_s$,
\begin{equation}
P_{f_if_j}(z) = P_{f_if_j}^{(0)}(z) + \frac{\alpha_s}{2\pi} P_{f_if_j}^{(1)}(z) + \left(\frac{\alpha_s}{2\pi}\right)^2 P_{f_if_j}^{(2)}(z) + \ldots.
\end{equation}
Of these, $P^{(1)}$ have been known since 1980's \cite{Furmanski:1980cm,Curci:1980uw} but $P^{(2)}$ have been computed only recently \cite{Vogt:2004mw,Moch:2004pa}. While the kernels $P_{ij}^{(0)}$ are unique, independent of how the collinear divergences are regulated, the higher order splitting functions $P_{ij}^{(n)}$, $n > 0$ are not unique but they depend on the framework.

\section{Factorization in deeply inelastic scattering}

For couple of times during this section, I noted that when the divergent logarithms from the parton ladders were extracted, a multiplicative term that coincided with the leading order cross-section for photon-quark scattering, was found. This property is not, however, a special feature of the leading order cross-section, but order by order in perturbative calculations, the divergent logarithms can be systematically factored apart from the perturbative parton-level pieces
\begin{equation}
d\hat\sigma_i = d\hat\sigma^{(0)}_i + \left(\frac{\alpha_s}{2\pi}\right)d\hat\sigma^{(1)}_i +  \left(\frac{\alpha_s}{2\pi}\right)^2 d\hat\sigma^{(2)}_i + \ldots
\end{equation}
That is, the cross-section for deeply inelastic scattering retains its same simple form
\begin{equation}
\frac{d^2\sigma}{dxdQ^2} = \sum_i \frac{d^2\hat{\sigma}_i(x,Q^2)}{dxdQ^2} \otimes f_i(Q^2),
\label{eq:DISfac1}
\end{equation}
where the parton densities $f_q(Q^2)$ obey the DGLAP equations (\ref{eq:completeDGLAP3}). This is a special case of the pQCD \emph{factorization theorem} \cite{Ellis:1978ty,Collins:1989gx} which states that the collinear singularities can be, order by order, factored to the scale dependent parton densities, and finiteness of (\ref{eq:DISfac1}) is guaranteed. A more complete treatment indicates that the factorization is subject to power corrections $\mathcal{O}\left({\Lambda^2_{\rm QCD}}/{Q^2} \right)^n$ which may become important for small $Q^2$. Such terms appear, for example, if the partons are not exactly collinear with the parent nucleon, but are allowed to carry some ``primordial'' transverse momentum ${\bf k_\perp}$. More generally, such terms arise form multi-parton interactions.
 
\vspace{1cm}
The definition of the parton densities is not unique: starting from parton densities $f_i(x,Q^2)$ we may define another version $f'_i(x,Q^2)$ by
\begin{equation}
f'_i(x,Q^2) \equiv \sum_j C_{ij} \otimes f_j(Q^2),
\end{equation}
where $C_{ij}(z)=1 + \frac{\alpha_s}{2\pi} C^{(1)}_{ij}(z) + \left( \frac{\alpha_s}{2\pi} \right)^2 C^{(2)}_{ij}(z) + \ldots$. In terms of the primed densities, the cross-section can be written as
$$\sigma = \hat \sigma_i \otimes f_i = \hat \sigma_i \otimes C^{-1}_{ij} \otimes f'_j = \hat \sigma'_i \otimes f'_i,$$
where I defined $\hat \sigma'_i \equiv \hat \sigma_i \otimes C^{-1}_{ij}$. The same reshuffling implies that $f'_i$ obey the DGLAP equations with splitting functions
$$P'_{ij} = \left[C \otimes P \otimes C^{-1} - 2\pi \beta(\alpha_s)\frac{dC}{d\alpha_s}C^{-1}\right]_{ij}.$$
In other words, beyond leading order, the perturbative coefficients $\hat \sigma_i$, and the splitting functions $P$ are intertwined with the exact definition for the parton densities. These kind of different ``bases'' for computing are known as \emph{factorization schemes} and, in principle, the predictions for physical, measurable cross-sections should be independent of the chosen scheme. However, as the change from a scheme to another is facilitated by the perturbative coefficients $C_{ij}$ --- often motivated by some particular cross-sections --- the predictions from various schemes are not precisely equal. However, such difference is always one power higher in $\alpha_s$ than to which the computation was performed. 

\vspace{1cm}
There is a similar ambiguity in choosing the argument of $f_q(Q^2)$ in (\ref{eq:DISfac1}). This is because different scales are related perturbatively as
\begin{eqnarray}
f_i(x,Q^2) & = & f_i(x,Q_f^2) + \frac{\alpha_s}{2\pi}\log \left({Q^2}/{Q^2_f} \right) P_{ij} \otimes f_j(Q_f^2) \\
& + & \frac{1}{2} \left(\frac{\alpha_s}{2\pi}\right)^2 \log^2 \left({Q^2}/{Q^2_f} \right) P_{ij} \otimes P_{jk} \otimes f_k(Q_f^2) \nonumber \\
& + & \ldots \nonumber \\
& = & \sum_j D_{ij}(Q^2/Q_f^2) \otimes f_j(Q_f^2),
\end{eqnarray}
and defining $\hat \sigma_j(x,Q^2,Q_f^2) \equiv \hat \sigma_i(Q^2,Q_f^2) \otimes D_{ij}(Q^2/Q_f^2)$, the factorization formula (\ref{eq:DISfac1}) becomes
\begin{equation}
\frac{d^2\sigma}{dxdQ^2} = \sum_i \frac{d^2\hat{\sigma}_i(x,Q^2,Q^2_f)}{dxdQ^2} \otimes f_i(Q_f^2).
\label{eq:DISfac2}
\end{equation}
The scale $Q_f^2$ is called the \emph{factorization scale}, which we are free to choose. A typical choice is $Q_f^2 = cQ^2$, where $c$ is between $0.5$ and $2$.  Again, in all-orders calculation a physical cross-section does not depend on this choice, but in practice, as only the first few terms in the perturbative expansion are known, the prediction retains sensitivity to the adopted choice.

\section{Factorization for other processes}

\begin{figure}[h!]
\begin{center}
\includegraphics[scale=0.3]{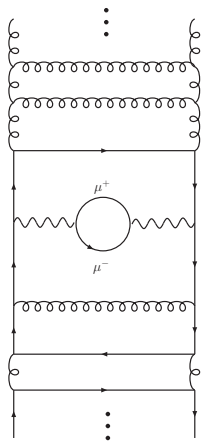}
\end{center}
\caption{Ladder-type diagram that gives rise to a leading logarithmic term.}
\label{fig:DYfac}
\end{figure}

Although I have here considered only the deeply inelastic scattering, the underlying physics is shared in variety of other processes involving hadrons in the initial state --- the structure of the collinear singularities is the same. It follows that the parton densities should be independent of the actual hard process, \emph{universal}. For example in the Drell-Yan production of dileptons in nucleon-nucleon collisions, the leading logarithms originate from diagrams like that in Fig.~\ref{fig:DYfac}.
The formal proofs for factorization are highly technical and mathematically demanding. Therefore, there are only few processes for which such all-order proofs actually exists \cite{Collins:1989gx}, but it is typically assumed that for hadronic interactions that are ``hard enough'' (involve a large invariant scale), the pQCD-improved parton model is applicable. Ultimately, however, it is the comparison with experiments that is of essence.
 
\section{Example of parton densities}
\begin{figure}[h]
\centering
 \includegraphics[scale=0.6]{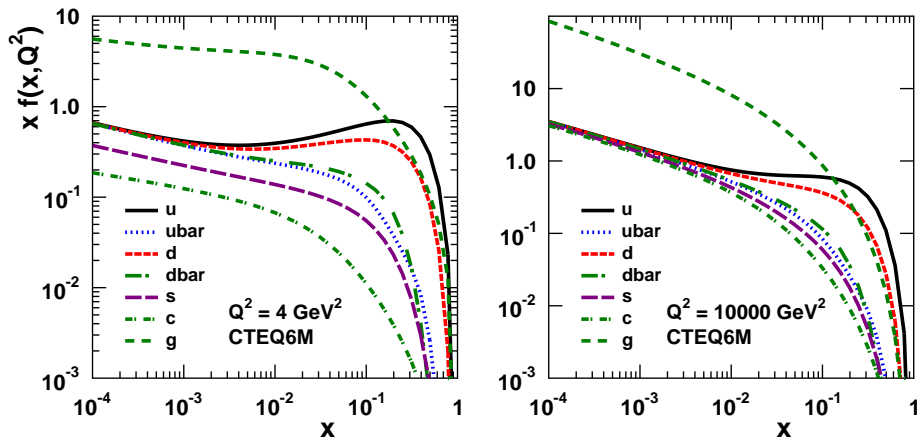}
\caption{The CTEQ6M  partons \cite{Stump:2003yu}.}
\label{Fig:CTEQ6}
\end{figure}

\begin{figure}[h]
\centering
 \includegraphics[scale=0.62]{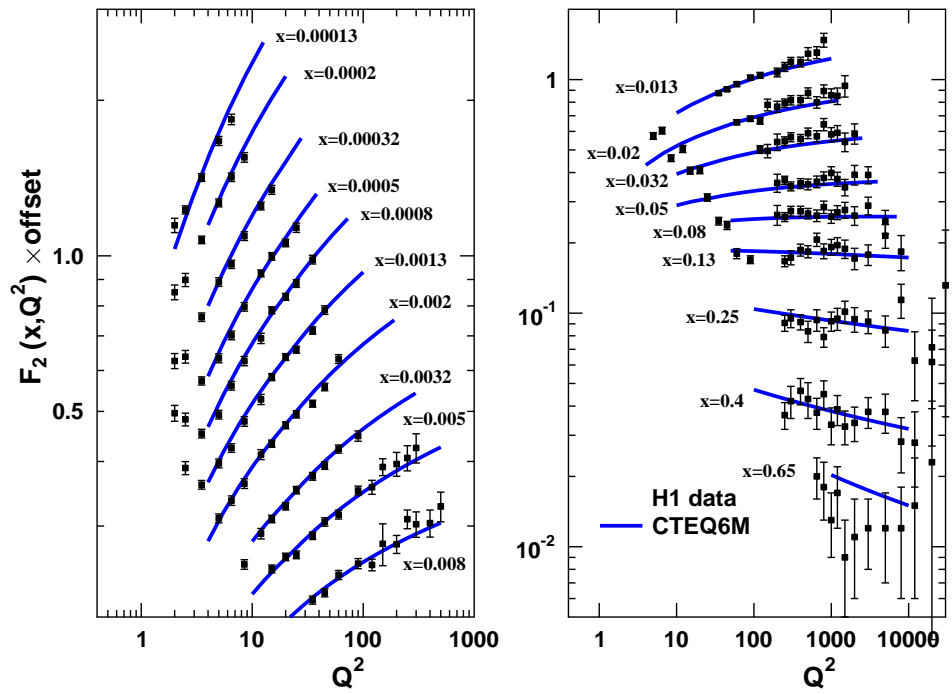}
\caption{Experimental data for proton structure function $F_2$ from H1 experiment \cite{Adloff:1999ah, Adloff:2000qk} compared with the CTEQ6M PDFs \cite{Stump:2003yu}.}
\label{Fig:H1data}
\end{figure}

In their full glory, the factorization and DGLAP equations are exploited and tested in \emph{global QCD-analyses}, to be described in Chapter \ref{AboutGlobal}. In short, their purpose is to extract the $x$-dependence of the parton distributions from the experimental data. The Fig.~\ref{Fig:CTEQ6} displays a typical outcome of such analyses, and Fig.~\ref{Fig:H1data} demonstrates how the $Q^2$-dependence of the experimental $F_2$ data becomes correctly reproduced by the DGLAP evolution.

\vspace{3cm}

%% file: NLOdis.tex
\chapter{Deeply inelastic scattering at NLO}
\label{Deeply_inelastic_scattering_at_NLO}

In this chapter I present the full calculation of the deeply inelastic scattering cross-section at next-to-leading order pQCD. Based on the discussion in Chapter \ref{DGLAPevolution}, there will be three types of divergences: collinear, infrared, and ultraviolet. The most sophisticated and gauge-invariant way to regulate these is to use the dimensional regularization methods \cite{Marciano:1975de} and perform all calculations in $N=4-\epsilon$ space-time dimensions as in \cite{Altarelli:1979ub}, or e.g. \cite{KJE_Lectures}. The divergences will appear as $1/\epsilon^n$-poles but only those which correspond to the collinear divergences will remain. These are removed by absorbing them into the parton densities, after which the $\epsilon \rightarrow 0$ limit can be safely taken.

The generic form of the hadronic tensor $W_{\mu\nu}$ in Eq.~(\ref{eq:Hadronic tensor}) is independent of the space-time dimension, but the projections to the two independent structure functions $F_1$ and $F_2$ receive some $\epsilon$-dependence:
\begin{eqnarray}
 \frac{F_2}{x} & = & \frac{2}{2-\epsilon} \left[ -g^{\mu\nu} + (3-\epsilon)\frac{4x^2}{Q^2} P^\mu P^\nu \right] M W_{\mu\nu} \label{eq:F12DIM} \\
{F_1} & = & \frac{F_2}{2x} - \left(\frac{4x^2}{Q^2} P^\mu P^\nu \right) M W_{\mu\nu}, \nonumber
\end{eqnarray}
as may be directly verified, remembering that $g_{\mu\nu}g^{\mu\nu}=4-\epsilon$.

\section{Leading-order contribution}

For consistency, also the leading-order contribution must be computed in $N=4-\epsilon$ dimensions. The appropriate $N$-dimensional extension of the partonic tensor for the leading-order $\gamma^* q \rightarrow q$ process in Eq.~(\ref{eq:partonicstr}) is
\begin{eqnarray}
 4\pi M \hat{W}_{\mu\nu}^q & = & \frac{e_q^2}{2} \frac{d^{N-1}{\bf p'}}{(2\pi)^{N-1} 2p'^0} (2\pi)^{N} \delta^{(N)} (p+q-p') \, {\rm Tr}[\slashed{p'} \gamma^\mu \slashed{p} \gamma^\nu] \nonumber \\
& = & \frac{e_q^2}{2} \frac{2\pi x}{Q^2} \, {\rm Tr}[\slashed{p'} \gamma^\mu \slashed{p} \gamma^\nu] \delta(\xi-x). \nonumber
\end{eqnarray}
The contractions with $g_{\mu\nu}$ and $p^\mu p^\nu$ give
\begin{equation}
 g_{\mu\nu} \, {\rm Tr}[\slashed{p'} \gamma^\mu \slashed{p} \gamma^\nu] = -2(2-\epsilon)Q^2 \qquad p_{\mu}p_{\nu} \, {\rm Tr}[\slashed{p'} \gamma^\mu \slashed{p} \gamma^\nu] = 0,
\end{equation}
and we find
\begin{equation}
-g_{\mu\nu} (M \hat{W}^q_{\mu \nu}) = e_q^2 z \left(1- \frac{\epsilon}{2} \right) \delta(1-z) \qquad P^{\mu}P^{\nu} (M\hat{W}^q_{\mu \nu}) = 0,
\label{eq:LO_in_D}
\end{equation}
where I have defined the variable $z \equiv x/\xi = \frac{-q^2}{2p \cdot q}$.

\section{Gluon radiation}

\begin{figure}[h]
\centering
 \includegraphics[scale=0.4]{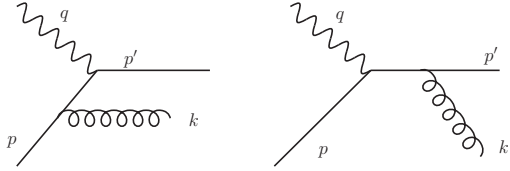}
\caption{The graphs for $\mathcal{O}(\alpha_s)$ real gluon radiation.}
\label{Fig:NLODISq_qg}
\end{figure}

The diagrams contributing to the real-gluon emission process $\gamma^* q \rightarrow qg$ are shown in Fig.~(\ref{Fig:NLODISq_qg}).
The squared, spin-independent matrix element contracted with $-g_{\mu\nu}$ reads
\begin{eqnarray}
 -g^{\mu\nu} \left| \mathcal{M}_{\gamma^* q \rightarrow qg} \right|_{\mu\nu}^2 & = & \frac{e_q^2}{2} g_s^2 C_F (\mu^2)^{\epsilon/2} 8(1-\frac{\epsilon}{2}) \label{eq:q_qg} \\
 & & \left[ (1-\frac{\epsilon}{2}) \left(-\frac{\hat s}{\hat t} - \frac{\hat t}{\hat s}\right) - 2\frac{q^2 \hat u}{\hat t \hat s} + \epsilon + \mathcal{O}(\epsilon^2)\right] \nonumber,
\end{eqnarray}
where $g_s (\mu^2)^{\epsilon/4}$ is the strong coupling in $N=4-\epsilon$ dimensions and where the Mandelstam variables $\hat s,\hat t,\hat u$ are defined as
\begin{eqnarray}
\hat s & \equiv & (p+q)^2  = (k+p')^2 \nonumber \\
\hat t & \equiv & (p-k)^2  = (q-p')^2 \nonumber \\
\hat u & \equiv & (p-p')^2 = (q-k)^2 \nonumber.
\end{eqnarray}
As we are eventually taking the $\epsilon \rightarrow 0$ limit, the last two terms in Eq.~(\ref{eq:q_qg}) will not contribute and I will forget them from now on. The 2-particle phase space in $N$ dimensions can be written as

\begin{eqnarray}
d \Pi^{(2)}  & = & \frac{d^{N-1}{\bf p'}}{(2\pi)^{N-1} 2k^0} \frac{d^{N-1}{\bf k}}{(2\pi)^{N-1} 2k^0} (2\pi)^{N} \delta^{(N)} (p+q-k-p')  \nonumber \\
& = &  \frac{d^{N}{\bf p'}}{(2\pi)^{N-1}} \frac{d^{N-1}{\bf k}}{(2\pi)^{N-1}2k^0} (2\pi)^{N} \delta^{(N)} (p+q-k-p') \theta(p'^0)\delta(p'^2) \nonumber \\
& = & (2\pi)^{2-N} \frac{d^{N-1}{\bf k}}{2|{\bf k}|} \, \delta_+((p+q-k)^2) \label{eq:phase1}
\end{eqnarray}
In the center-of-mass frame of $q$ and $p$, we may choose the momenta as
\begin{eqnarray}
p  & = & (|{\bf p}|, 0, 0, \ldots,  |{\bf p}|) \nonumber \\
q  & = & (q^0, 0, 0, \ldots, -|{\bf p}|) \nonumber \\
k  & = & (|{\bf k}|, \ldots, |{\bf k}|\cos \theta) \nonumber \\
p' & = & (|p'^0|,    \ldots, -|{\bf k}|\cos \theta) \nonumber.
\end{eqnarray}
The squared matrix element does not depend on the momenta which were left unspecified above and we may perform the redundant angular integrations
\begin{eqnarray}
d^{N-1}{\bf k} & = & |{\bf k}|^{N-2} d|{\bf k}| \Omega_{N-1} \\
& = & |{\bf k}|^{N-2} \sin^{N-3} \theta \, d|{\bf k}| d\theta d \Omega_{N-2} \nonumber \\
\int d \Omega_{N-2} & = & \frac{2\pi^{\left( \frac{N-2}{2} \right)}}{\Gamma \left(\frac{N-2}{2}\right)} \nonumber,
\end{eqnarray}
where $\int d \Omega_{n}$ is the surface area of an $n$-dimensional euclidean unit-sphere. The phase space (\ref{eq:phase1}) thus becomes
\begin{equation}
d \Pi^{(2)} = \frac{1}{4\pi} \frac{(4\pi)^{\epsilon/2}}{\Gamma(1-\frac{\epsilon}{2})} d|{\bf k}| |{\bf k}|^{1-\epsilon} d(\cos\theta) \left(1-\cos^2\theta \right)^{-\epsilon/2} \delta ( \hat s - 2|{\bf k}| \sqrt{\hat s}).
\end{equation} 
The $|{\bf k}|$-integration eliminates the remaining $\delta$-function
\begin{equation}
\int d|{\bf k}| |{\bf k}|^{1-\epsilon} \delta ( \hat s - 2|{\bf k}| \sqrt{\hat s}) = \frac{1}{4} \left( \frac{4}{\hat s} \right)^{\epsilon/2},
\end{equation}
and after introducing an angular variable $y \equiv \frac{1}{2}(1+\cos\theta)$, the 2-particle phase space becomes
\begin{equation}
d \Pi^{(2)} = \frac{1}{8\pi} \frac{1}{\Gamma(1-\frac{\epsilon}{2})} \left( \frac{4\pi}{\hat s} \right)^{\epsilon/2} \int_0^1 dy \left[y(1-y)\right]^{-\epsilon/2}.
\label{eq:2plePH}
\end{equation}
In terms of $y$, $Q^2$, and  $z = {-q^2}/{2p \cdot q}$, the Mandelstam invariants are
\begin{equation}
\hat s  =   \frac{Q^2}{z} (1-z), \quad \hat t  =   -\frac{Q^2}{z}(1-y), \quad
\hat u  =   -\frac{Q^2}{z} y.
\end{equation}
The square brackets from (\ref{eq:q_qg}) and $y$ integral from the phase space (\ref{eq:2plePH}) leads to
\begin{eqnarray}
\int_0^1 \left[y(1-y)\right]^{-\epsilon/2} \left[(1-\frac{\epsilon}{2}) \left(\frac{1-z}{1-y} + \frac{1-y}{1-z} \right) + 2 \frac{zy}{(1-z)(1-y)} \right],
\end{eqnarray}
which may be evaluated using a generic identity
\begin{equation}
 \int_0^1 dy y^{\alpha-1}(1-y)^{\beta-1} = \frac{\Gamma(\alpha)\Gamma(\beta)}{\Gamma(\alpha+\beta)},
\end{equation}
resulting in
\begin{equation}
 \frac{\Gamma^2(1-\epsilon/2)}{\Gamma(1-\epsilon)} \left[-\frac{2-\epsilon}{\epsilon} \left( 1-z+\frac{2z}{1-z}\frac{1}{1-\epsilon} \right) + \frac{1-\epsilon/2}{2(1-\epsilon)(1-z)} \right].
\end{equation}
Altogether, 
\begin{eqnarray}
-g^{\mu\nu} \left(M\hat W^{q}_{\mu\nu} \right)_{\big| {\rm Real}} & = & -g^{\mu\nu} \left| \mathcal{M}_{\gamma^* q \rightarrow qg} \right|_{\mu\nu}^2 d \Pi^{(2)} \frac{1}{4\pi} \label{eq:Real1} \\
& = & e_q^2 (1-\epsilon/2) C_F  \frac{\alpha_s}{2\pi} \left( \frac{4\pi\mu^2}{Q^2} \right)^{\epsilon/2} \frac{\Gamma(1-\epsilon/2)}{\Gamma(1-\epsilon)} \nonumber \\
& & \hspace{-4cm} \left( \frac{z}{1-z} \right)^{\epsilon/2} \left[-\frac{2-\epsilon}{\epsilon} \left( 1-z+\frac{2z}{1-z}\frac{1}{1-\epsilon} \right) + \frac{1-\epsilon/2}{2(1-\epsilon)(1-z)} \right], \nonumber
\end{eqnarray}
where the designation ``Real'' reminds that this is a contribution from the tree-level real gluon emission. In this expression the collinear $y \rightarrow 1$ divergences (gluon being emitted in the direction of incoming quark) are manifest as explicit $1/\epsilon$-poles which now remain finite for $\epsilon < 0$. However, the expression is also singular as $z \rightarrow 1$ corresponding to the vanishing energy of the radiated gluon or gluon being collinear with the outgoing quark. These singularities can be made explicit by a distribution identity
\begin{equation}
\frac{1}{(1-z)^{1+\epsilon}} = -\frac{1}{\epsilon} \, \delta(1-z) + \frac{1}{(1-z)_+} - \epsilon \left[ \frac{\log(1-z)}{1-z}\right]_+,
\end{equation}
where the plus-distributions are defined as in Eq.~(\ref{eq:plus1}). Applying this identity to the lowest line of Eq.~(\ref{eq:Real1}) gives
\begin{eqnarray}
& & \frac{8}{\epsilon^2} \, \delta(1-z) - \frac{2}{\epsilon} \left[ \frac{1+z^2}{1-z} -\frac{3}{2} \, \delta(1-z) \right] + (1+z^2)\left[ \frac{\log(1-z)}{1-z}\right]_+ \nonumber \\
& & - \frac{3}{2}\frac{1}{(1-z)_+} - \frac{1+z^2}{1-z}\log z + 3 - z + \frac{7}{2} \, \delta(1-z),
\end{eqnarray}
and we arrive at
\begin{eqnarray}
& & \hspace{-1cm} -g^{\mu\nu} \left(M\hat W^{q}_{\mu\nu} \right)_{\big| {\rm Real}} = e_q^2 (1-\epsilon/2) C_F  \frac{\alpha_s}{2\pi} \left( \frac{4\pi\mu^2}{Q^2} \right)^{\epsilon/2} \frac{\Gamma(1-\epsilon/2)}{\Gamma(1-\epsilon)} \nonumber \\
& & \left\{ \frac{8}{\epsilon^2} \, \delta(1-z) - \frac{2}{\epsilon} \left[ \frac{1+z^2}{1-z} -\frac{3}{2} \, \delta(1-z) \right] + (1+z^2)\left[ \frac{\log(1-z)}{1-z}\right]_+ \right. \nonumber \\
& & \left. - \frac{3}{2}\frac{1}{(1-z)_+} - \frac{1+z^2}{1-z}\log z + 3 - z + \frac{7}{2} \, \delta(1-z) \right\},
\label{eq:Real2}
\end{eqnarray}
The contraction with $P^\mu P^\nu$ is simpler and does not lead to singular behaviour:
\begin{eqnarray}
P^\mu P^\nu \left(M\hat W^{q}_{\mu\nu} \right)_{	\big| {\rm Real}} & = & \frac{1}{4\pi} P^\mu P^\nu \left| \mathcal{M}_{\gamma^* q \rightarrow qg} \right|_{\mu\nu}^2 d \Pi^{(2)} \\
& = & e_q^2 \left( 1-\epsilon/2 \right) C_F \frac{\alpha_s}{2\pi} \frac{Q^2}{4z} \left( \frac{z}{x} \right)^2 \nonumber
\end{eqnarray}

\section{Virtual corrections}

\begin{figure}[h]
\centering
 \includegraphics[scale=0.4]{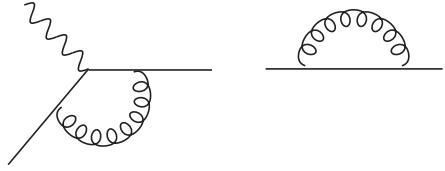}
\caption{Virtual corrections to deeply inelastic scattering.}
\label{Fig:Virt_to_DIS}
\end{figure}
The virtual corrections to the Born cross-section Eq.~(\ref{eq:LO_in_D}) stem from two sources: from loop correction to the photon-quark vertex and from the self-energy corrections to the external legs. These are shown in Fig.~\ref{Fig:Virt_to_DIS}. In the Feynman gauge the 1-loop vertex correction (with massless external quarks) is simply a multiplicative factor $\left[ 1+ \Gamma \right]$ to the tree-level Feynman rule $-ie\gamma_\mu$, with
\begin{equation}
\Gamma = \frac{\alpha_s}{4\pi} C_F \left( \frac{4\pi \mu^2}{Q^2} \right)^{\epsilon/2} \frac{\Gamma(1-\epsilon/2)}{\Gamma(1-\epsilon)} \left[ -\frac{8}{\epsilon^2} - \frac{8}{\epsilon} + \frac{2}{\epsilon_{\rm UV}} - 8 - \frac{\pi^2}{3} \right],
\label{eq:Virtvertex}
\end{equation}
where I have separately indicated the ultraviolet divergence $\epsilon_{\rm UV} > 0$ that occurs in the high end part of the loop momentum. The renormalization constant $Z_F$ for massless quarks is, in turn,
\begin{equation}
Z_F = 1 + \frac{\alpha_s}{4\pi} C_F \left( \frac{2}{\epsilon} - \frac{2}{\epsilon_{\rm UV}} \right).
\label{eq:Virtself}
\end{equation}
The ultraviolet poles cancel, and the total virtual contribution is
\begin{eqnarray}
-g_{\mu\nu} (M \hat{W}^q_{\mu \nu})_{\big| {\rm Virtual}} & = & -g_{\mu\nu} (M \hat{W}^q_{\mu \nu})_{\big| {\rm Born}} \left[ \left|1+ \Gamma \right|^2 Z_F^2 - 1 \right] \label{eq:virtualDIS} \\
& = & e_q^2 \left(1 - {\epsilon}/{2} \right) \delta(1-z) \frac{\alpha_s}{2\pi} C_F \left( \frac{4\pi \mu^2}{Q^2} \right)^{\epsilon/2} \nonumber \\
& & \frac{\Gamma(1-\epsilon/2)}{\Gamma(1-\epsilon)} \left[ -\frac{8}{\epsilon^2} - \frac{6}{\epsilon} - 8 - \frac{\pi^2}{3} \right] \nonumber.
\end{eqnarray}

\section{Total quark contribution}

When the virtual pieces and those from real gluon radiation are combined, the double poles $1/\epsilon^2$ evidently cancel, giving altogether
\begin{eqnarray}
& & -g_{\mu\nu} \left[ (M \hat{W}^q_{\mu \nu})_{\big| {\rm Virtual}} + M \hat{W}^q_{\mu \nu})_{\big| {\rm Real}} \right] = e_q^2 (1-\epsilon/2) C_F  \frac{\alpha_s}{2\pi} \left( \frac{4\pi\mu^2}{Q^2} \right)^{\epsilon/2} \nonumber \\
& & \frac{\Gamma(1-\epsilon/2)}{\Gamma(1-\epsilon)} \left\{  - \frac{2}{\epsilon} \left[ \frac{1+z^2}{1-z} + \frac{3}{2} \, \delta(1-z) \right] + (1+z^2)\left[ \frac{\log(1-z)}{1-z}\right]_+  \nonumber  \right. \\
& & \left. - \frac{3}{2}\frac{1}{(1-z)_+} - \frac{1+z^2}{1-z}\log z + 3 - z - \left( \frac{9}{2} + \frac{\pi^2}{3} \right) \, \delta(1-z) \right\}. \nonumber
\end{eqnarray}

The total quark contribution to the structure functions can now be obtained by folding them with the quark densities
\begin{eqnarray}
M{W}^q_{\mu \nu} & = & \sum_q \int_x^1 \frac{d\xi}{\xi}q_0(\xi) \left[ (M \hat{W}^q_{\mu \nu})_{\big| {\rm Born}} + (M \hat{W}^q_{\mu \nu})_{\big| {\rm Virtual}} + M \hat{W}^q_{\mu \nu})_{\big| {\rm Real}} \right], \nonumber
\end{eqnarray}
where the $q_0(\xi)$ denotes the ``bare'', non-physical density which will eventually go along the redefinition of the quark densities. When the various pieces are put together as instructed in Eq.~(\ref{eq:F12DIM}), 
\begin{eqnarray}
\frac{1}{x} F_2^q & = & \sum_q e_q^2 \int_x^1 \frac{d\xi}{\xi} \, q_0(\xi) 		\\
& & \hspace{-2cm} \left\{ \delta(1-z) - \frac{2}{\epsilon} \, \frac{\alpha_s}{2\pi} P_{qq}(z) \frac{\Gamma(1-\epsilon/2)}{\Gamma(1-\epsilon)} \left( \frac{4\pi\mu^2}{Q^2} \right)^{\epsilon/2} + \frac{\alpha_s}{2\pi} C_q(z) \right\}, \nonumber
\end{eqnarray}
where the coefficient function $C_q(z)$ is defined as
\begin{eqnarray}
C_q(z) & \equiv & C_F \left\{ (1+z^2)\left[ \frac{\log(1-z)}{1-z}\right]_+ - \frac{3}{2}\frac{1}{(1-z)_+} \right. \\
& & \left. - \frac{1+z^2}{1-z}\log z + 3 + 2z - \left( \frac{9}{2} + \frac{\pi^2}{3} \right) \, \delta(1-z) \right\}. \nonumber
\end{eqnarray}

\section{Initial state gluons}

The exact NLO calculation of the initial state gluon contributions proceeds much in a similar fashion as extraction of the quark contributions above. When averaging over the transverse gluon polarization states, one should remember that there are $2-\epsilon$ such states instead of usual $2$. 
\begin{figure}[h]
\centering
 \includegraphics[scale=0.4]{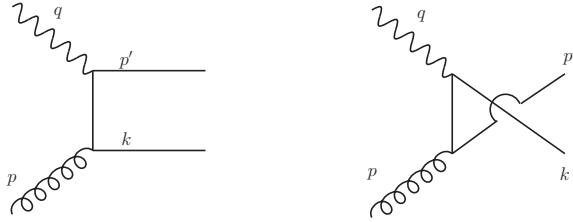}
\caption{The diagrams for $\mathcal{O}(\alpha_s)$ initial state gluon contribution.}
\label{Fig:NLOinitg}
\end{figure}
The squared, matrix element corresponding to the diagrams shown in Fig.~\ref{Fig:NLOinitg}, contracted with $-g_{\mu\nu}$ reads
\begin{eqnarray}
 -g^{\mu\nu} \left| \mathcal{M}_{\gamma^* g \rightarrow q\overline{q}} \right|_{\mu\nu}^2 = \frac{e_q^2}{2} g_s^2 T_R (\mu^2)^{\epsilon/2} 8 \label{eq:g_qq} 
 \left[(1-\epsilon/2) \left(\frac{\hat u}{\hat t} + \frac{\hat t}{\hat u} \right) + 2 \frac{q^2 \hat s}{\hat t \hat u} \right] \nonumber.
\end{eqnarray}
Supplying the phase-space element and doing the angular integrals, we end up with a following gluon contribution
\begin{eqnarray}
& & -g^{\mu\nu} \left(M\hat W^{g}_{\mu\nu} \right) = -\frac{1}{4\pi} g^{\mu\nu} \left| \mathcal{M}_{\gamma^* g \rightarrow q\overline{q}} \right|_{\mu\nu}^2 d \Pi^{(2)} \\
& & \hspace{2.6cm} = e_q^2  \frac{\alpha_s}{2\pi} \left( \frac{4\pi\mu^2}{Q^2} \right)^{\epsilon/2} 2 T_R \nonumber \\
& & \frac{\Gamma(1-\epsilon/2)}{\Gamma(1-\epsilon)} \left\{- \frac{2}{\epsilon} \left[ z^2 + (1-z)^2 \right] + \log \left( \frac{1-z}{z} \right)\left[ z^2 + (1-z)^2 \right]  \right\}. \nonumber
\end{eqnarray}
The analogous results from the contraction with $P^\mu P^\nu$ are
\begin{eqnarray}
P^\mu P^\nu \left| \mathcal{M}_{\gamma^* g \rightarrow q\overline{q}} \right|_{\mu\nu}^2 = \frac{e_q^2}{2-\epsilon} g_s^2 T_R (\mu^2)^{\epsilon/2} 8 \frac{zQ^2}{x^2}(1-z),
\end{eqnarray}
\begin{eqnarray}
P^\mu P^\nu \left(M\hat W^{G}_{\mu\nu} \right) = e_q^2 \frac{\alpha_s}{2\pi} T_R Q^2\frac{z(1-z)}{x^2}.
\end{eqnarray}
From these the we can build the initial state gluon contribution to the structure functions
\begin{eqnarray}
\frac{1}{x} F_2^g(x,Q^2) & = & 2 \sum_q e_q^2 \int_x^1 \frac{d\xi}{\xi} \, g_0(\xi) \\
& & \hspace{-2cm} \left\{ - \frac{2}{\epsilon} \, \frac{\alpha_s}{2\pi} P_{qg}(z) \left( \frac{4\pi\mu^2}{Q^2} \right)^{\epsilon/2} \frac{\Gamma(1-\epsilon/2)}{\Gamma(1-\epsilon)}  + \frac{\alpha_s}{2\pi} C_g(z) \right\}, \nonumber
\end{eqnarray}
where the coefficient function $C_g(z)$ is defined as
\begin{eqnarray}
C_g(z) & \equiv & P_{qg}(z) \log\left(\frac{1-z}{z} \right) +T_R6z(1-z) - P_{qg}(z).
\end{eqnarray}

\section{Total $F_2$ and PDF-schemes}

We are now in a position to add up the quark and gluon contributions to the total $F_2$. The remaining $\epsilon$-dependent terms come with a common factor
\begin{eqnarray}
\frac{2}{\epsilon} \left( \frac{4\pi\mu^2}{Q^2} \right)^{\epsilon/2} \frac{\Gamma(1-\epsilon/2)}{\Gamma(1-\epsilon)} & = & \frac{2}{\epsilon} -\gamma_E + \log(4\pi) + \log \left(\frac{\mu^2}{Q^2} \right) \\
& = & \frac{1}{\hat \epsilon} + \log \left(\frac{\mu^2}{Q^2} \right), \nonumber
\end{eqnarray}
where I introduced a short-hand notation absorbing the Euler-Mascheroni constant $\gamma_E$ and $\log(4\pi)$ to $1/{\hat \epsilon}$. Writing the total $F_2$ explicitly, we have
\begin{eqnarray}
\frac{1}{x} F_2(x,Q^2) & = & \, \, \, \, \sum_{q,\overline{q}} e_q^2 q_0 \otimes \left\{ 1 - \frac{\alpha_s}{2\pi} \left[ \frac{1}{\hat \epsilon} + \log \left(\frac{\mu^2}{Q^2} \right) \,  \right]P_{qq} + \frac{\alpha_s}{2\pi} C_q \right\} \nonumber \\
& + & 2 \sum_q e_q^2 g_0 \otimes \left\{ \, \, \, \, \, - \frac{\alpha_s}{2\pi} \left[ \frac{1}{\hat \epsilon} + \log \left(\frac{\mu^2}{Q^2} \right) \, \right] P_{qg} + \frac{\alpha_s}{2\pi} C_g \right\} \nonumber.
\end{eqnarray}
Based on the discussion of Chapter~\ref{DGLAPevolution}, the collinear divergences, the $1/\epsilon$ pieces, should be absorbed into the redefinition of the quark and gluon densities which, as already noted, are not unique. In the dimensional regularization framework, the general NLO definitions of the scale-dependent quark and gluon distributions at a certain factorization scale $Q_f^2$ can be written as
\begin{eqnarray}
q(x,Q_f^2) & \equiv &  q_0 \otimes \left\{ 1 - \frac{\alpha_s}{2\pi} \left[ \frac{1}{\hat \epsilon} + \log \left(\frac{\mu^2}{Q_f^2} \right) \, \right]P_{qq} + \frac{\alpha_s}{2\pi} f_q^{\rm scheme} \right\} \nonumber \\
& + & g_0 \otimes \left\{ \, \, \, \, \, \, - \frac{\alpha_s}{2\pi} \left[ \frac{1}{\hat \epsilon} + \log \left(\frac{\mu^2}{Q_f^2} \right) \,  \right]P_{qg} + \frac{\alpha_s}{2\pi} f_g^{\rm scheme} \right\} \nonumber \\
\label{eq:NLOdefinitions} \\ 
g(x,Q_f^2) & \equiv &  g_0 \otimes \left\{ 1 - \frac{\alpha_s}{2\pi} \left[ \frac{1}{\hat \epsilon} + \log \left(\frac{\mu^2}{Q_f^2} \right) \, \right]P_{gg} + \frac{\alpha_s}{2\pi} h_g^{\rm scheme} \right\} \nonumber \\
& & \hspace{-1cm} + \sum_{q,\overline{q}} q_0 \otimes \left\{ \, \, \, \, \, \, - \frac{\alpha_s}{2\pi} \left[ \frac{1}{\hat \epsilon} + \log \left(\frac{\mu^2}{Q_f^2} \right) \,  \right]P_{gq} + \frac{\alpha_s}{2\pi} h_q^{\rm scheme} \right\} \nonumber,
\end{eqnarray}
where $f_{q,g}^{\rm scheme}$, $h_{q,g}^{\rm scheme}$ are arbitrary (finite) functions. It should be understood that these terms are just the first ones of a whole series which formally exponentiate and give rise to the DGLAP evolution as discussed in Chapter~\ref{DGLAPevolution}. With these definitions the expression for the NLO structure function $F_2$ becomes
\begin{eqnarray}
& & \hspace{-0.5cm} \frac{1}{x} F_2(x,Q^2)  = \\
& & \hspace{-0.5cm}\sum_{q,\overline{q}} e_q^2 q(Q_f^2) \otimes \left\{ 1 - \frac{\alpha_s}{2\pi} \log \left(\frac{Q_f^2}{Q^2} \right) \,  P_{qq}  + \frac{\alpha_s}{2\pi} \left[ C_q - f_q^{\rm scheme} \right] \right\} \nonumber \\
& & \hspace{-1.05cm} + 2 \sum_q e_q^2 g(Q_f^2) \otimes \left\{ \, \, \, \, \, \, - \frac{\alpha_s}{2\pi} \log \left(\frac{Q_f^2}{Q^2} \right) \,  P_{qg} + \frac{\alpha_s}{2\pi} \left[ C_g - f_g^{\rm scheme} \right] \right\} \nonumber,
\end{eqnarray}
where
\begin{eqnarray}
C_q(z) & \equiv & C_F \left\{ (1+z^2)\left[ \frac{\log(1-z)}{1-z}\right]_+ - \frac{3}{2}\frac{1}{(1-z)_+} \right. \\
& & \left. - \frac{1+z^2}{1-z}\log z + 3 + 2z - \left( \frac{9}{2} + \frac{\pi^2}{3} \right) \, \delta(1-z) \right\} 
\nonumber \\
C_g(z) & \equiv & P_{qg}(z) \log\left(\frac{1-z}{z} \right) +T_R6z(1-z) - P_{qg}(z).
\end{eqnarray}

Obviously, there are arbitrarily may ways to define the scheme --- the two most common ones are
\begin{itemize}
 \item {\bf $\overline{\rm MS}$-scheme:} \\
In this scheme only the collinear divergence and the regularization-framework-originating $-\gamma_E + \log(4\pi)$ are absorbed into the definition, i.e. one chooses $f_q^{\rm scheme}=f_g^{\rm scheme}=h_q^{\rm scheme}=h_g^{\rm scheme}=0$.
 \item {\bf DIS-scheme:} \\
This scheme is defined by choosing $f_q^{\rm scheme}=C_q$ and $f_g^{\rm scheme}=C_g$, which maintains the simple naive parton model form of the cross-section at $Q^2=Q_f^2$. How to define $h_q^{\rm scheme}$ and $h_g^{\rm scheme}$ in DIS-scheme is, however, more or less a matter of convention. Usual choice \cite{Diemoz:1987xu} is dictated by the momentum conservation
$$
\int_0^1 x \left[ \sum_{q,\overline{q}} q^{\overline{\rm MS}} + g^{\overline{\rm MS}} \right] = \int_0^1 x \left[ \sum_{q,\overline{q}} q^{\overline{\rm DIS}} + g^{\overline{\rm DIS}} \right],
$$
which requires $h_g^{\rm scheme}=-2 n_f C_g$ and $h_q^{\rm scheme}=-C_q$.
\end{itemize}



\pagebreak
For completeness, I record here also the expressions for $F_1$:
\begin{eqnarray}
F_1(x,Q^2)  = \frac{1}{2x} F_2(x,Q^2) & - & \,\,\,\, \sum_{q,\overline{q}} e_q^2 q(Q_f^2) \otimes \left[ \frac{\alpha_s}{2\pi} C_F z \right] \label{eq:NLO_F1} \\
& - & 2 \sum_q e_q^2 g(Q_f^2) \otimes \left[\frac{\alpha_s}{2\pi}T_R 2z(1-z)\right] \nonumber.
\end{eqnarray}

\section{Numerical estimate}

\begin{figure}[h!]
\begin{center}
\includegraphics[scale=0.55]{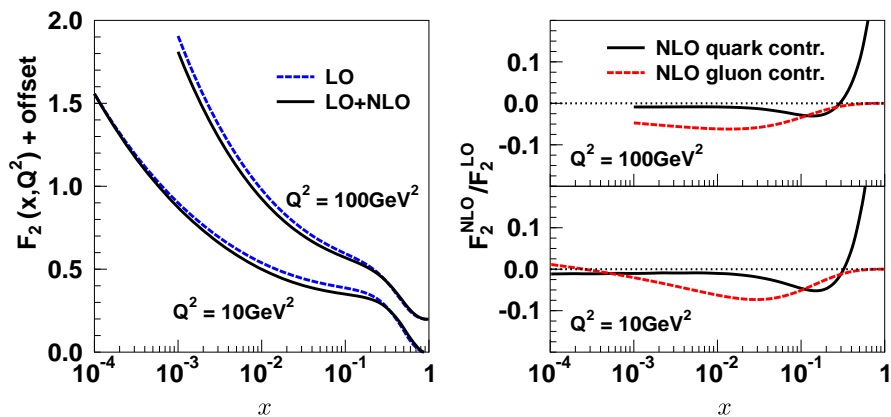}
\end{center}
\caption{The left panel shows the behaviour of absolute proton $F_2$ at $Q^2=10 \, {\rm GeV}^2$ and $Q^2=100 \, {\rm GeV}^2$ (with offset +0.2) calculated with CTEQ6L1 leading order parton densities in the leading-order and in the next-to-leading order. The panels on the right display the NLO contributions from the quark and gluon channels separately normalized by the leading-order $F_2$.}
\label{fig:DISexample}
\end{figure}

The numerical calculation of deeply inelastic structure functions and cross-sections at NLO is relatively simple with only single integrals to be numerically evaluated. In order to get in touch with the expected size of the NLO corrections,
I have computed the proton $F_2$ both in the leading and in the next-to-leading order ($\overline{\rm MS}$-scheme) with the \emph{same} set of parton densities (CTEQ6L1 \cite{Pumplin:2002vw}). The Fig.~\ref{fig:DISexample} presents the results at two typical scales, at $Q^2=10 \, {\rm GeV}^2$ and at $Q^2=100 \, {\rm GeV}^2$.
In most $x$-values the expected NLO corrections are of the order of few percents and only the quark contribution shows a growing behaviour at very large $x$. Although the leading-order term dominates and $F_2$ as such is not very sensitive e.g. to the gluon content of the nucleon\footnote{Gluons, however, are the driving force in the DGLAP evolution of small-$x$ quarks and thus their effect on $F_2$ is significant but indirect.}, such terms cannot, however, be neglected in a precision analysis. An example of a quantity that is directly much more sensitive to the gluons is the longitudinal structure function $F_L \equiv F_2 - 2xF_1$ which vanishes at leading order, as can be seen from Eq.~\ref{eq:NLO_F1}.

\vspace{0.3cm}
Also the NNLO coefficient functions for the deeply inelastic structure functions are nowadays known \cite{Zijlstra:1992qd}, and an analogous estimate as shown here reveals the expected magnitude of the NNLO vs. NLO corrections similar to the NLO vs. LO shown here. In other words, the NNLO terms in the coefficient functions are still important.

%% file: DYNLO.tex
\chapter{Drell-Yan NLO cross-section}
\label{DrellYanNLO}

The Drell-Yan dilepton production in nucleon-nucleon collisions is another process which can be employed in probing the parton distributions. Here, I will derive the double-differential NLO cross-section $d^2\sigma/dM^2dy_R$, where $M^2$ and $y_R$ denote the invariant mass and the rapidity of the lepton pair. A detailed computation of this particular cross-section has proven to be surprisingly difficult to find from the literature: Up to my knowledge, such can only be found in \cite{Kubar:1980zv} which employs a massive gluon scheme to regularize the collinear and infrared singularities. However, in order to rigorously employ the NLO definitions of the scale dependent parton densities discussed in the context of deeply inelastic scattering in Chapter 2, the dimensional regularization methods are to be used. The results obtained in this section coincide with those given in \cite{Anastasiou:2003yy}.

\section{Leading-order calculation}
\begin{figure}[h!]
\begin{center}
\includegraphics[scale=0.5]{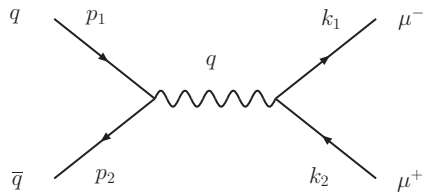}
\end{center}
\caption{Leading-order parton diagram for Drell-Yan dilepton production.}
\label{fig:DYLO}
\end{figure}
In the leading order, the lepton pair production proceeds by annihilation of quark and antiquark as shown in Fig.~\ref{fig:DYLO}. The unpolarized, color-averaged square of the matrix element corresponding to this diagram can be compactly written as
\begin{equation}
 |\mathcal{M}_0|^2 = \frac{e^4e_q^2}{12q^4}L^{\mu\nu}H_{\mu\nu}^0,
\end{equation}
where, in $N$ dimensions,
\begin{eqnarray}
 L^{\mu\nu} & \equiv & {\rm Tr}\left[\slashed{k}_1 \gamma^\mu \slashed{k}_2 \gamma^\nu \right] \label{eq:LOLandH} \\
 H^0_{\mu\nu} & \equiv & {\rm Tr}\left[\slashed{p}_2 \gamma_\mu \slashed{p}_1 \gamma_\nu \right] \nonumber.
\end{eqnarray}
When the leptonic tensor $L^{\mu\nu}$ is integrated over the 2-particle phase space, the dependence on momenta $k_1$ and $k_2$ washes out and the result only depends on the virtual photon momentum $q$. Moreover, since $q_\mu L^{\mu\nu} = q_\nu L^{\mu\nu} = 0$, the Lorentz structure is necessarily
\begin{equation}
 \int d\Pi^{(2)} L^{\mu\nu} = \left( g^{\mu\nu} - \frac{q^\mu q^\nu}{q^2} \right) L(q^2),
\end{equation}
where
$$L(q^2)= \frac{1}{N-1}\int d\Pi^{(2)} \left(g_{\mu\nu}L^{\mu\nu}\right).$$
Similarly, $q^\mu H^0_{\mu\nu} = q^\nu H^0_{\mu\nu} = 0$ and therefore
\begin{equation}
 H^0_{\mu\nu} \int d\Pi^{(2)} L^{\mu\nu} =  \left(g^{\mu\nu}H^0_{\mu\nu}\right)L(q^2).
\end{equation}
Thus, the cross-section splits into independently calculable leptonic and hadronic pieces
\begin{eqnarray}
\hat\sigma_0^q & = & \frac{1}{2\hat s} \frac{e^4e_q^2}{12q^4} \int d\Pi^{(2)}  L^{\mu\nu}H_{\mu\nu}^0 \label{eq:LODYtot} \\
& = & \frac{1}{2\hat s} \frac{e^4e_q^2}{12q^4} \left(g^{\mu\nu}H^0_{\mu\nu}\right) \frac{1}{N-1} \int d\Pi^{(2)} \left(g_{\mu\nu}L^{\mu\nu}\right) \nonumber,
\end{eqnarray}
simplifying the calculation. From (\ref{eq:LOLandH}), we have
\begin{eqnarray}
 g_{\mu\nu} L^{\mu\nu} & = & -2(2-\epsilon) q^2 \\
 g^{\mu\nu} H^0_{\mu\nu} & = & -2(2-\epsilon) \hat s \nonumber.
\end{eqnarray}
Since the leptonic quantity $\frac{1}{N-1} \int d\Pi^{(2)} \left(g_{\mu\nu}L^{\mu\nu}\right)$ will be common also for the higher order diagrams, it is unnecessary to drag its exact $N$-dimensional expression throughout the calculation --- what matters is the hadronic part $g^{\mu\nu} H^0_{\mu\nu}$. In the $N \rightarrow 4$ limit $\int d\Pi^{(2)}=\frac{1}{8\pi}$, and (\ref{eq:LODYtot}) gives
\begin{equation}
 \hat \sigma_0^q = \frac{4 \pi \alpha^2}{9q^2}e_q^2.
\end{equation}
The invariant mass $M^2$, and the rapidity $y_R$ of the produced lepton pair are defined as
\begin{eqnarray}
M^2 & \equiv & (k_1 + k_2)^2 = q^2 \\
y_R & \equiv & \frac{1}{2} \log \frac{(k_{1}^0+k_{2}^0)+(k_{1}^3+k_{2}^3)}{(k_1^0+k_2^0)-(k_{1}^3+k_{2}^3)} = \frac{1}{2} \log \frac{q^0+q^3}{q^0-q^3}.
\end{eqnarray}
At leading order, $k_1+k_2 = p_1+p_2$, and in the center-of-mass frame of the colliding nucleons
\begin{eqnarray}
p_1 = \frac{\sqrt{s}}{2} \xi_1 (1,0,0,1), \qquad
p_2 = \frac{\sqrt{s}}{2} \xi_2 (1,0,0,-1),
\label{eq:CMSinHAD}
\end{eqnarray}
it follows that
\begin{eqnarray}
M^2 = \xi_1 \xi_2 s, \qquad
y_R = \frac{1}{2} \log \frac{\xi_1}{\xi_2}.
\end{eqnarray}
Thus, the double-differential partonic cross-section in these kinematic variables can be written as
\begin{eqnarray}
\frac{d^2\hat \sigma_0^q}{dM^2dy_R} & = & \hat \sigma_0^q  \delta(M^2 - \xi_1 \xi_2 s) \delta \left( y_R - \frac{1}{2} \log \frac{\xi_1}{\xi_2} \right) \\
& = & \frac{4 \pi \alpha_{\rm em}^2}{9\hat s M^2}e_q^2 \,\, \delta(1 - z) \delta \left( y_R - \frac{1}{2} \log \frac{\xi_1}{\xi_2} \right). \nonumber
\end{eqnarray}
where $z \equiv M^2/\hat s$. To obtain the corresponding hadronic cross-section we integrate over the parton densities of the incoming nucleons and sum over all flavors
$$
\frac{d^2 \sigma_0}{dM^2dy_R} = \int_0^1 d\xi_1 d\xi_2 \sum_q \frac{d^2\hat \sigma_0^q}{dM^2dy_R}  \left[ q^{(1)}(\xi_1)\overline{q}^{(2)}(\xi_2) + q^{(2)}(\xi_2)\overline{q}^{(1)}(\xi_1) \right].
$$
Performing the integrals we find
\begin{eqnarray}
\frac{d^2\sigma_0}{dM^2dy_R} = \frac{4 \pi \alpha^2}{9 s M^2} \sum_q e_q^2 \left[ q^{(1)}(x_1)\overline{q}^{(2)}(x_2) + ( 1 \leftrightarrow 2 ) \right], \nonumber
\end{eqnarray}
where
\begin{equation}
x_1 \equiv \sqrt{\tau} e^{y_R}, \quad x_2 \equiv \sqrt{\tau} e^{-y_R}, \quad \tau \equiv \frac{M^2}{s}.
\end{equation}

\section{Gluon radiation}
\begin{figure}[h!]
\begin{center}
\includegraphics[scale=0.4]{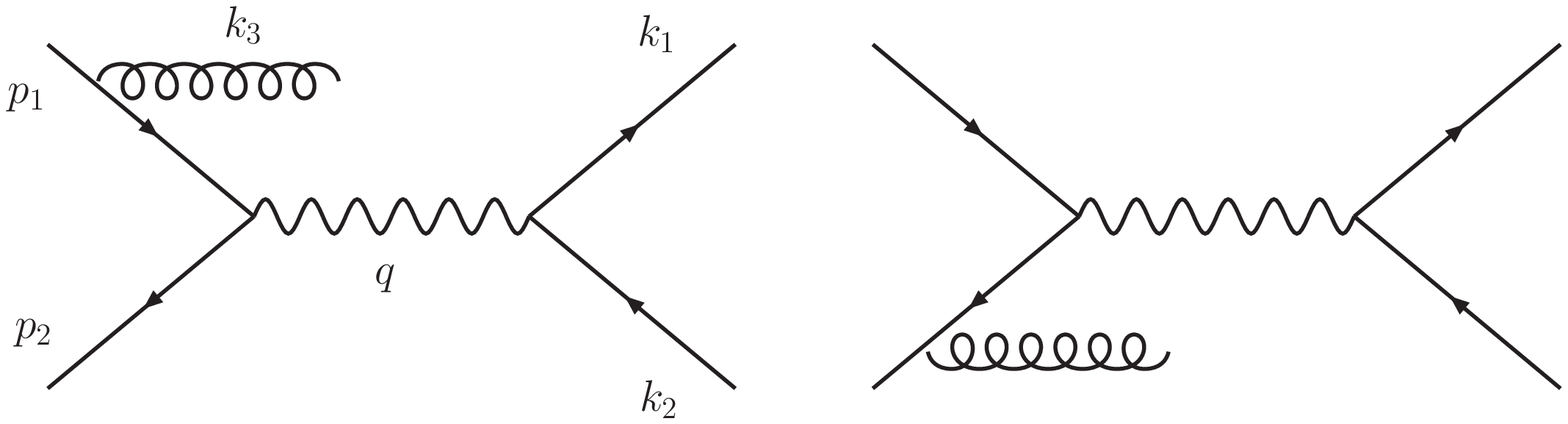}
\end{center}
\caption{Gluon radiation in Drell-Yan dilepton production to NLO pQCD.}
\label{fig:NLODYg}
\end{figure}

The QCD corrections to the Born-level cross-section from emission of one additional unobserved gluon are shown in Fig.~\ref{fig:NLODYg}. The corresponding partonic cross-section for such process can be written as
\begin{equation}
\hat \sigma_R^q = \frac{1}{2\hat s} \frac{1}{q^4} \frac{1}{\underbrace{2 \times 2}_{\rm spin} \times \underbrace{3 \times 3}_{\rm color}} e^4e_q^2 \int d \Pi^{(3)} L^{\mu\nu} H_{\mu\nu}^R,
\end{equation}
where $L^{\mu\nu}$ is the same leptonic tensor as in the previous section, and $H_{\mu\nu}^R$ is obtained from the spin- and color-summed square of diagrams in Fig.~\ref{fig:NLODYg}. In order to separate the leptonic and hadronic pieces as was done in the leading order, we use an identity
\begin{equation}
1 = \int \frac{dM^2}{2\pi} \int \frac{d^{N-1}q}{(2\pi)^{N-1}2E_q}(2\pi)^N \delta^{(N)}(q-k_1-k_2)_{\Big|(k_1+k_2)^2=M^2},
\end{equation}
to rewrite the 3-particle phase space as
\begin{eqnarray}
d \Pi^{(3)} & = & \frac{d^{N-1}k_1}{(2\pi)^{N-1}2E_{k_1}} \frac{d^{N-1}k_2}{(2\pi)^{N-1}2E_{k_2}}  \frac{d^{N-1}k_3}{(2\pi)^{N-1}2E_{k_3}} \\
& & (2\pi)^N \delta^{(N)}(p_1+p_2-k_1-k_2-k_3) \nonumber \\
& = & \frac{dM^2}{2\pi} \left[ \frac{d^{N-1}k_1}{(2\pi)^{N-1}2E_{k_1}} \frac{d^{N-1}k_2}{(2\pi)^{N-1}2E_{k_2}} (2\pi)^N \delta^{(N)}(q-k_1-k_2) \right] \nonumber \\
& & \hspace{0.93cm} \left[ \frac{d^{N-1}k_3}{(2\pi)^{N-1}2E_{k_3}} \frac{d^{N-1}q}{(2\pi)^{N-1}2E_{q}} (2\pi)^N \delta^{(N)}(p_1+p_2-k_3-q) \right] \nonumber \\
& = & \frac{dM^2}{2\pi} d \Pi^{(2)}_L d \Pi^{(2)}_H \nonumber.
\end{eqnarray}
By this trick,
\begin{eqnarray}
\frac{d\hat \sigma_R^q}{dM^2} & = & \frac{e^4e_q^2}{72\hat s M^4} \frac{1}{2\pi} \frac{1}{N-1} \left( \int d \Pi^{(2)}_L g_{\mu\nu} L^{\mu\nu} \right) \left( \int d \Pi^{(2)}_G g^{\mu\nu} H_{\mu\nu}^R \right) \nonumber \\
& = & \frac{4\pi \alpha_{em}^2}{9\hat s M^2} \frac{1}{6\pi} \frac{1}{2(2-N)} \left( \int d \Pi^{(2)}_G g^{\mu\nu} H_{\mu\nu}^R \right),
\end{eqnarray}
where I already took the $\epsilon \rightarrow 0$ limit of the leptonic part $\frac{1}{N-1} \int d \Pi^{(2)}_L g_{\mu\nu} L^{\mu\nu}$. In the present case, we define the Mandelstam variables
\begin{eqnarray}
 \hat s & \equiv & (p_1+p_2)^2 = \frac{M^2}{z} \nonumber \\
 \hat t & \equiv & (p_1-k_3)^2 = -\frac{M^2}{z}(1-z)(1-y) \nonumber \\
 \hat u & \equiv & (p_2-k_3)^2 = -\frac{M^2}{z}(1-z)y \nonumber,
\end{eqnarray}
where I have introduced the angular variable $y \equiv \frac{1}{2}(1+\cos \theta) \in [0,1]$, with $\theta$ referring to the angle between incoming quark and emitted gluon in the center-of-mass frame of the incoming quark and antiquark
\begin{eqnarray}
p_1  & = & (\sqrt{\hat s}/2, 0, 0, \ldots,  \sqrt{\hat s}/2) \nonumber \\
p_2  & = & (\sqrt{\hat s}/2, 0, 0, \ldots, -\sqrt{\hat s}/2) \nonumber \\
k_3  & = & (|{\bf k_3}|, \ldots, |{\bf k_3}|\cos \theta); \quad |{\bf k_3}|=({\hat s - M^2})/({2\sqrt{\hat s}}) \nonumber \\
q    & = & (q^0, \ldots, -|{\bf k_3}|\cos \theta); \quad \,\,\,\, q^0=({\hat s + M^2})/({2\sqrt{\hat s}}) \nonumber
\end{eqnarray}
and $z \equiv M^2/ \hat s \in [\tau,1] $. The rapidity $y_R^*$ of the produced lepton pair in this frame is
\begin{equation}
y_R^* = \frac{1}{2} \log \frac{1-y(1-z)}{z+y(1-z)},
\end{equation}
and owing to the additivity under Lorentz-boosts, the rapidity $y_R$ in the nucleon-nucleon center-of-mass frame is
\begin{equation}
y_R = \frac{1}{2} \log \frac{\xi_1}{\xi_2} + y_R^* = \frac{1}{2} \log \left[\frac{\xi_1}{\xi_2} \frac{1-y(1-z)}{z+y(1-z)}\right].
\end{equation}
Thus, the wanted doubly differential cross-section is obtained by
\begin{equation}
\frac{d^2\hat \sigma_R^q}{dy_RdM^2} = \frac{d\hat \sigma_R^q}{dM^2} \delta \left(y_R - \frac{1}{2} \log \left[\frac{\xi_1}{\xi_2} \frac{1-y(1-z)}{z+y(1-z)}\right] \right).
\end{equation}
Having now fixed the kinematics, we can write down the squared matrix element, contracted with the metric tensor $g^{\mu\nu}$,
\begin{equation}
g^{\mu\nu} H_{\mu\nu}^R = -g^2 12 C_F \left(1-\epsilon/2 \right) \left[(2-\epsilon) \left( \frac{\hat u}{\hat t} + \frac{\hat t}{\hat u} \right) + 4\frac{\hat s M^2}{\hat t \hat u} \right].
\end{equation}
In terms of the variables $z$ and $y$ this becomes
$$
g^{\mu\nu}H_{\mu\nu}^R = -g^2 12 C_F \left(2-\epsilon \right) \left[(1-\epsilon/2) \left( \frac{1-y}{y} + \frac{y}{1-y} \right) + \frac{2z}{(1-z)^2y(1-y)} \right].
$$
A similar calculation that led to Eq.~(\ref{eq:2plePH}) gives the 2-particle phase space
\begin{equation}
\int d \Pi_H^{(2)} = \frac{1}{8\pi} \frac{1}{\Gamma(1-\frac{\epsilon}{2})} \left(\frac{4\pi}{M^2}\right)^{\epsilon/2} (1-z)^{1-\epsilon}z^{\epsilon/2} \int_0^1 dy \left[y(1-y)\right]^{-\epsilon/2}.
\end{equation}
At this moment, the various poles that occur in the kinematic corners $z \rightarrow 1$, $y \rightarrow 1$, $y \rightarrow 0$ should be made explicit by using the following distribution identities
\begin{eqnarray}
\frac{1}{(1-z)^{1+\epsilon}} & = & -\frac{1}{\epsilon}\delta(1-z) + \frac{1}{(1-z)_+} - \epsilon \left[ \frac{\log(1-z)}{1-z} \right]_+ \nonumber \\
\frac{1}{z^{1+\epsilon }} & = & -\frac{1}{\epsilon}\delta(z) + \frac{1}{z_+} - \epsilon \left[ \frac{\log z}{z} \right]_+ . \nonumber
\end{eqnarray}
After some algebra, we find the following stack of terms
\begin{eqnarray}
& & \frac{(1-z)^{1-\epsilon}z^{\epsilon/2}}{\left[y(1-y)\right]^{\epsilon/2}} \left[(1-\epsilon/2) \left( \frac{1-y}{y} + \frac{y}{1-y} \right) + \frac{2z}{(1-z)^2y(1-y)} \right] \nonumber \\
& = & \delta(1-z) \left[\delta(1-y) + \delta(y) \right] \left( \frac{4}{\epsilon^2} + \frac{3}{\epsilon} \right) \nonumber \\
& + & \delta(1-z) \left\{ -\frac{2}{\epsilon} \left[ \frac{1}{(1-y)_+} + \frac{1}{y_+} \right] + \left[ \frac{\log(1-y)}{1-y} \right]_+ + \left[\frac{\log y}{y} \right]_+ + \right. \nonumber \\
& & \hspace{1.7cm} \left. \frac{\log(1-y)}{y} + \frac{\log y}{1-y} \right\} \nonumber \\
& + & \left[\delta(1-y) + \delta(y) \right] \left\{ -\frac{2}{\epsilon}\left[ \frac{1+z^2}{(1-z)_+} +  \frac{3}{2}\delta(1-z) \right] \right. + \nonumber \\ 
& & \hspace{3cm} \left. 2(1+z^2) \left[ \frac{\log(1-z)}{1-z} \right]_+ - \right. \nonumber \\
& & \hspace{3cm} \left.  \frac{(1+z^2)\log z}{1-z} + (1-z) \right\} \nonumber \\
& + & \frac{1+z^2}{(1-z)_+} \left[ \frac{1}{(1-y)_+} + \frac{1}{y_+} \right] - 2(1-z). \nonumber
\end{eqnarray}
All terms in the third line above are irrelevant as they vanish under integration: for example
\begin{eqnarray}
& & \int_0^1 dy \, \delta \left(y_R - \frac{1}{2} \log \left[\frac{\xi_1}{\xi_2} \frac{1-y(1-z)}{z+y(1-z)}\right] \right) \delta(1-z) \left(\frac{\log y}{y}\right)_+ \nonumber \\
& = & \delta \left(y_R - \frac{1}{2} \log \frac{\xi_1}{\xi_2} \right) \delta(1-z) \underbrace{\int_0^1 dy \left(\frac{\log y}{y}\right)_+}_{=0} = 0. \nonumber
\end{eqnarray}
We also notice the following analytic result
\begin{eqnarray}
& & \int_0^1 dy \, \delta \left(y_R - \frac{1}{2} \log \left[\frac{\xi_1}{\xi_2} \frac{1-y(1-z)}{z+y(1-z)}\right] \right) \delta(1-z) \left[\frac{\log (1-y)}{y}+\frac{\log y}{1-y}\right] \nonumber \\
& = & \delta \left(y_R - \frac{1}{2} \log \frac{\xi_1}{\xi_2} \right) \delta(1-z) 2 \underbrace{\int_0^1 dy \frac{\log y}{1-y}}_{-\zeta(2)=-\frac{\pi^2}{6}} \nonumber \\
& = & -\frac{\pi^2}{3} \, \delta \left(y_R - \frac{1}{2} \log \frac{\xi_1}{\xi_2} \right) \delta(1-z), \nonumber
\end{eqnarray}
where $\zeta(2)$ refers to the Riemann zeta-function. 
Performing the remaining $\delta$-function-restricted $y$-integrals, we reach the final form of the partonic cross-section:
\begin{eqnarray}
& & \frac{d^2\hat \sigma_R^q}{dy_RdM^2} = \frac{4\pi \alpha_{em}^2}{9\hat s M^2} \frac{\alpha_s}{2\pi}C_F \left(\frac{4\pi \mu^2}{M^2}\right)^{\epsilon/2} \frac{1}{\Gamma(1-\epsilon/2)} \\
& & \left\{ \delta \left(y_R - \frac{1}{2} \log \frac{\xi_1}{\xi_2} \right) \delta(1-z) \left[\frac{8}{\epsilon^2} + \frac{6}{\epsilon} - \frac{\pi^2}{3} \right] \right. + \nonumber \\
& &  \left[\delta \left(y_R - \frac{1}{2} \log \frac{z\xi_1}{\xi_2}\right) + \delta \left(y_R - \frac{1}{2} \log \frac{\xi_1}{z\xi_2}\right) \right] \times \nonumber \\
& & \left\{ -\frac{2}{\epsilon}\left[ \frac{1+z^2}{(1-z)_+} +  \frac{3}{2}\delta(1-z) \right] \right. + \nonumber \\ 
& & \hspace{0.3cm} \left. 2(1+z^2) \left[ \frac{\log(1-z)}{1-z} \right]_+ - \frac{(1+z^2)\log z}{1-z} + (1-z) \right\} + \nonumber \\
& & \left. J(z,\xi_1,\xi_2) \left[\frac{1+z^2}{(1-z)_+} \left[ \left(\frac{1}{1-y_0}\right)_+ + \left(\frac{1}{y_0}\right)_+ \right] - 2(1-z)\right]\right\}, \nonumber
\end{eqnarray}
where I have defined
\begin{eqnarray}
y_0 & \equiv & \frac{1}{2} \left[1-\frac{1+z}{1-z} \tanh \left(y_R - \frac{1}{2}\log \frac{\xi_1}{\xi_2} \right) \right] \\
J(z,\xi_1,\xi_2) & \equiv & \frac{dy}{dy_R} = \frac{1}{2} \frac{1+z}{1-z} {\rm sech}^2 \left(y_R - \frac{1}{2}\log \frac{\xi_1}{\xi_2} \right).
\end{eqnarray}

\section{Virtual corrections}
The virtual 1-loop diagrams are essentially same as for deeply inelastic scattering, Eqs.~(\ref{eq:Virtvertex}) and (\ref{eq:Virtself}), but the vertex contribution must be analytically continued to the time-like region $Q^2=-q^2 \rightarrow -M^2<0$. The virtual contributions may be written as 
\begin{eqnarray}
& & \hspace{-0.9cm} \frac{d^2\hat \sigma^q}{dy_RdM^2}_{\big| {\rm Virtual}} =
\frac{4\pi \alpha_{em}^2}{9\hat s M^2} \frac{\alpha_s}{2\pi}C_F \frac{\left(\frac{4\pi \mu^2}{M^2}\right)^{\epsilon/2}}{\Gamma(1-\epsilon/2)} \left[ -\frac{8}{\epsilon^2} - \frac{6}{\epsilon} - 8 +\pi^2 \right] \nonumber \\
& & \hspace{2.3cm} \delta \left(y_R - \frac{1}{2} \log \frac{\xi_1}{\xi_2} \right) \delta(1-z).
\end{eqnarray}
Adding the virtual and real gluon emission contributions, the infrared $1/\epsilon^2$-poles cancel, giving
\begin{eqnarray}
& & \frac{d^2\hat \sigma_R^q}{dy_RdM^2} + \frac{d^2\hat \sigma^q}{dy_RdM^2}_{\big| {\rm Virtual}} = \frac{4\pi \alpha_{em}^2}{9\hat s M^2} \label{eq:realplusvirt} \\
& & \left\{ \delta \left(y_R - \frac{1}{2} \log \frac{\xi_1}{\xi_2} \right) \delta(1-z) \frac{\alpha_s}{2\pi}C_F \left[\frac{2\pi^2}{3} - 8 \right] \right. + \nonumber \\
& &  \left[\delta \left(y_R - \frac{1}{2} \log \frac{z\xi_1}{\xi_2}\right) + \delta \left(y_R - \frac{1}{2} \log \frac{\xi_1}{z\xi_2}\right) \right] \alpha_s f_\epsilon(z) + \nonumber \\
& & \left. J(z,\xi_1,\xi_2) \frac{\alpha_s}{2\pi}C_F  \left[\frac{1+z^2}{(1-z)_+} \left[ \left(\frac{1}{1-y_0}\right)_+ + \left(\frac{1}{y_0}\right)_+ \right] - 2(1-z)\right]\right\}, \nonumber
\end{eqnarray}
where
\begin{eqnarray}
 \alpha_s f_\epsilon(z) & \equiv & \frac{\alpha_s}{2\pi} P_{qq}(z) \left[ -\frac{1}{\hat \epsilon} + \log \frac{M^2}{\mu^2} \right] \\
& + & \frac{\alpha_s}{2\pi} C_F \left\{ 2(1+z^2) \left[ \frac{\log(1-z)}{1-z} \right]_+ - \frac{(1+z^2)\log z}{1-z} + (1-z) \right\} \nonumber.
\end{eqnarray}
To turn the partonic cross-section above to a hadronic one, we integrate over the parton densities
$$
\frac{d^2 \sigma^q}{dy_RdM^2} = \int_0^1 d\xi_1 d\xi_2 \left[ \frac{d^2\hat \sigma^q}{dy_RdM^2}_{\big| {\rm Born}} + \frac{d^2\hat \sigma_R^q}{dy_RdM^2} + \frac{d^2\hat \sigma^q}{dy_RdM^2}_{\big| {\rm Virtual}}\right] H_0(\xi_1,\xi_2),
$$
where
$$
H_0(\xi_1,\xi_2) \equiv \sum_q e_q^2 \left[ q^{(1)}_0(\xi_1)\overline{q}^{(2)}_0(\xi_2) + q^{(2)}_0(\xi_2)\overline{q}^{(1)}_0(\xi_1) \right].
$$
The $0$-subscript above is to remind us that these are still ``bare'' parton densities, anticipating their replacement by the scale-dependent ones. Performing the integrals constrained by the delta-functions, we obtain
\begin{eqnarray}
 \frac{d^2 \sigma^q}{dy_RdM^2} & = & \frac{4\pi \alpha_{em}^2}{9 s M^2} \left\{ H_0(x_1,x_2) \left[ 1 + \frac{\alpha_s}{2\pi}C_F \left(\frac{2\pi^2}{3} - 8 \right) \right] \right. \label{eq:Bareqq} \\
& + & \alpha_s \int_{x_1}^1 \frac{d\xi_1}{\xi_1} H_0(\xi_1,x_2) f_\epsilon(\frac{x_1}{\xi_1}) + \alpha_s \int_{x_2}^1 \frac{d\xi_2}{\xi_2} H_0(x_1,\xi_2) f_\epsilon(\frac{x_2}{\xi_2}) \nonumber \\
& + & \frac{\alpha_s}{2\pi}C_F \int_{x_1}^1 \frac{d\xi_1}{\xi_1} \int_{x_2}^1 \frac{d\xi_2}{\xi_2} H_0(\xi_1,\xi_2) J(z,\xi_1,\xi_2) \nonumber \\
& & \left. \left[\frac{1+z^2}{(1-z)_+} \left[ \left(\frac{1}{1-y_0}\right)_+ + \left(\frac{1}{y_0}\right)_+ \right] - 2(1-z)\right]\right\}, \nonumber
\end{eqnarray}

\section{Initial state gluons}
\begin{figure}[h!]
\begin{center}
\includegraphics[scale=0.45]{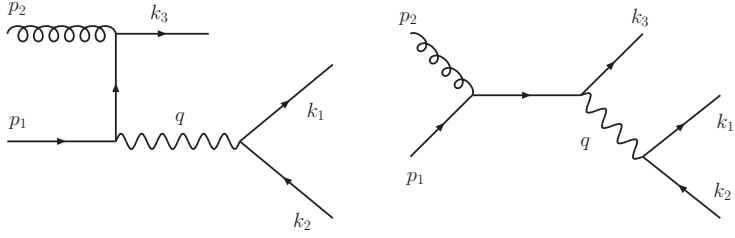}
\end{center}
\caption{Gluon-quark subprocess in Drell-Yan dilepton production.}
\label{fig:NLODYgq}
\end{figure}
The second type of $\mathcal{O}(\alpha_s)$ process that contributes to the Drell-Yan cross-section is the one with a gluon in the initial state. The two diagrams for such process are shown in Fig.~\ref{fig:NLODYgq}. The corresponding partonic cross-section can be written as
\begin{equation}
\hat \sigma^{qG} = \frac{1}{2\hat s} \frac{1}{q^4} \frac{1}{\underbrace{2 \times (2-\epsilon)}_{\rm spin} \times \underbrace{3 \times 8}_{\rm color}} e^4e_q^2 \int d \Pi^{(3)} L^{\mu\nu} H_{\mu\nu}^{qG},
\end{equation}
and it follows that
\begin{eqnarray}
\frac{d^2\hat \sigma^{qG}}{dy_RdM^2} & = & \frac{4\pi \alpha_{em}^2e_q^2}{9\hat s M^2} \frac{1}{2\pi} \frac{-1}{8(2-\epsilon)^2} \left( \int d \Pi^{(2)} g^{\mu\nu} H_{\mu\nu}^{qG} \right) \\ 
& & \delta \left(y_R - \frac{1}{2} \log \left[\frac{\xi_1}{\xi_2} \frac{1-y(1-z)}{z+y(1-z)}\right] \right). \nonumber
\end{eqnarray}
The hadronic part $H_{\mu\nu}^{qG}$ is the color- and spin-summed square of the diagrams shown in Fig.~\ref{fig:NLODYgq}
$$
g^{\mu\nu} H_{\mu\nu}^{qG} = -g^2_s (\mu^2)^{-\epsilon/2} 64 T_R \left(1-\epsilon/2 \right) \left[(1-\epsilon/2) \left( \frac{-\hat u}{\hat s} + \frac{-\hat s}{\hat u} \right) - 2\frac{\hat t M^2}{\hat s \hat u} \right].
$$
In terms of variables $y$ and $z$, defined as in the previous section, this reads
\begin{eqnarray}
g^{\mu\nu}H_{\mu\nu}^{qG} & = & -g^2_s (\mu^2)^{-\epsilon/2} 64 T_R \left(1-\epsilon/2 \right) \\
& & \left\{(1-\epsilon/2) \left[y(1-z) + \frac{1}{y(1-z)} \right] - \frac{2(1-y)z}{y} \right\}. \nonumber
\end{eqnarray}
The phase-space is
$$
\int d \Pi^{(2)} = \frac{1}{8\pi} \frac{1}{\Gamma(1-\frac{\epsilon}{2})} \left(\frac{4\pi}{M^2}\right)^{\epsilon/2} (1-z)^{1-\epsilon}z^{\epsilon/2} \int_0^1 dy \left[y(1-y)\right]^{-\epsilon/2}. \nonumber
$$
In this case only collinear $y \rightarrow 0$ singularities are present, which can be turned into explicit $1/ \epsilon$-poles by distributions:
\begin{eqnarray}
& & \hspace{-0.6cm} \left[y(1-y)\right]^{-\epsilon/2}(1-z)^{1-\epsilon}z^{\epsilon/2}\left\{(1-\epsilon/2) \left[y(1-z) + \frac{1}{y(1-z)} \right] - \frac{2(1-y)z}{y} \right\} \nonumber \\
& & = \delta(y) \left\{ -\frac{2}{\epsilon}\left[z^2+(1-z)^2 \right] + \left[z^2+(1-z)^2 \right] \log \frac{(1-z)^2}{z} + 1  \right\} \nonumber \\
& & \hspace{1.5cm}  + \, 2z(1-z) + (1-z)^2y + \frac{z^2+(1-z)^2}{y_+}. \nonumber
\end{eqnarray}
Putting all factors together, we have
\begin{eqnarray}
& & \hspace{-0.5cm}\frac{d^2\hat \sigma_R^{qG}}{dy_RdM^2} = \frac{4\pi \alpha_{em}^2e_q^2}{9\hat s M^2} \frac{\alpha_s}{2\pi}T_R \left(\frac{4\pi \mu^2}{M^2}\right)^{\epsilon/2} \frac{1}{\Gamma(2-\epsilon/2)} \\
& & \hspace{-0.5cm} \left\{ \delta \left(y_R - \frac{1}{2} \log \frac{\xi_1}{z\xi_2}\right) \left[ -\frac{2}{\epsilon}\left[z^2+(1-z)^2 \right] + \left[z^2+(1-z)^2 \right] \log \frac{(1-z)^2}{z} + 1  \right] \right. \nonumber \\
& & \hspace{-0.5cm} \left. + J(z,\xi_1,\xi_2) \left[ 2z(1-z) + (1-z)^2 y_0 + \frac{z^2+(1-z)^2}{y_{0+}}  \right]\right\}. \nonumber
\end{eqnarray}
Defining
\begin{equation}
K_0^{qG}(\xi_1,\xi_2) \equiv \sum_q e_q^2 \left[ \left( q^{(1)}_0(\xi_1) + \overline{q}^{(1)}_0(\xi_1)\right) g_0^{(2)}(\xi_2) \right],
\end{equation}
the hadronic cross-section becomes
\begin{eqnarray}
 \frac{d^2 \sigma_R^{qG}}{dy_RdM^2} & = & \frac{4\pi \alpha_{em}^2}{9 s M^2} \left\{ \alpha_s \int_{x_2}^1 \frac{d\xi_2}{\xi_2} K_0^{qG}(x_1,\xi_2) g_{\epsilon}(\frac{x_2}{\xi_2})\right. \label{eq:BareqG} \\
& + & \frac{\alpha_s}{2\pi}T_R \int_{x_1}^1 \frac{d\xi_1}{\xi_1} \int_{x_2}^1 \frac{d\xi_2}{\xi_2} K_0^{qG}(\xi_1,\xi_2) J(z,\xi_1,\xi_2) \nonumber \\
& & \left. \left[2z(1-z) + (1-z)^2y_0 + \frac{z^2+(1-z)^2}{y_{0+}} \right]\right\}, \nonumber
\end{eqnarray}
where
\begin{eqnarray}
 \alpha_s g_\epsilon(z) & \equiv & \frac{\alpha_s}{2\pi} P_{qG}(z) \left[ -\frac{1}{\hat \epsilon} + \log \frac{M^2}{\mu^2} \right] \nonumber \\
& + & \frac{\alpha_s}{2\pi} \left\{  P_{qG}(z) \log \left[\frac{(1-z)^2}{z} - 1 \right] + T_R \right\} \nonumber.
\end{eqnarray}
The contribution from the mirror process is obtained similarly. Defining
\begin{equation}
K_0^{Gq}(\xi_1,\xi_2) \equiv \sum_q e_q^2 \left[ g_0^{(1)}(\xi_1) \left( q^{(2)}_0(\xi_2) + \overline{q}^{(2)}_0(\xi_2)\right) \right],
\end{equation}
the result is
\begin{eqnarray}
 \frac{d^2 \sigma_R^{Gq}}{dy_RdM^2} & = & \frac{4\pi \alpha_{em}^2}{9 s M^2} \left\{ \alpha_s \int_{x_1}^1 \frac{d\xi_1}{\xi_1} K_0^{Gq}(\xi_1,x_2) g_{\epsilon}(\frac{x_1}{\xi_1})\right. \label{eq:BareGq} \\
& + & \frac{\alpha_s}{2\pi}T_R \int_{x_1}^1 \frac{d\xi_1}{\xi_1} \int_{x_2}^1 \frac{d\xi_2}{\xi_2} K_0^{Gq}(\xi_1,\xi_2) 
J(z,\xi_1,\xi_2) \nonumber \\
& & \left. \left[2z(1-z) + (1-z)^2(1-y_0) + \frac{z^2+(1-z)^2}{(1-y_0)_+} \right]\right\}. \nonumber
\end{eqnarray}

\section{Finite result}

The results derived above still contain all collinear $1/\epsilon$-poles. The universality of the parton distributions requires that it must be possible to remove these singularities by the same definition as has been applied in deeply inelastic scattering. To first order in strong coupling we can invert the definitions (\ref{eq:NLOdefinitions}) to write
\begin{eqnarray}
& & H_0(x_1,x_2)  =  H(x_1,x_2,Q_f^2)  \\
& + & \frac{\alpha_s}{2\pi} \int_{x_1}^1 \frac{d\xi_1}{\xi_1} H(\xi_1,x_2,Q_f^2) \left\{ P_{qq} (z) \left[ \frac{1}{\hat \epsilon} + \log \left(\frac{\mu^2}{Q_f^2} \right) \, \right] - f_q^{\rm scheme}(z) \right\}_{\Big{|}z=\frac{x_1}{\xi_1}} \nonumber \\
& + & \frac{\alpha_s}{2\pi} \int_{x_2}^1 \frac{d\xi_2}{\xi_2} H(x_1,\xi_2,Q_f^2) \left\{ P_{qq} (z) \left[ \frac{1}{\hat \epsilon} + \log \left(\frac{\mu^2}{Q_f^2} \right) \, \right] - f_q^{\rm scheme}(z) \right\}_{\Big{|}z=\frac{x_2}{\xi_2}} \nonumber \\
& + & \frac{\alpha_s}{2\pi} \int_{x_1}^1 \frac{d\xi_1}{\xi_1} K^{Gq}(\xi_1,x_2,Q_f^2) \left\{ P_{qg} (z) \left[ \frac{1}{\hat \epsilon} + \log \left(\frac{\mu^2}{Q_f^2} \right) \, \right] - f_g^{\rm scheme}(z) \right\}_{\Big{|}z=\frac{x_1}{\xi_1}} \nonumber \\
& + & \frac{\alpha_s}{2\pi} \int_{x_2}^1 \frac{d\xi_2}{\xi_2} K^{qG}(x_1,\xi_2,Q_f^2) \left\{ P_{qg} (z) \left[ \frac{1}{\hat \epsilon} + \log \left(\frac{\mu^2}{Q_f^2} \right) \, \right] - f_g^{\rm scheme}(z) \right\}_{\Big{|}z=\frac{x_2}{\xi_2}} \nonumber
\end{eqnarray}
and
\begin{equation}
 K_0^{qG,Gq}(x_1,x_2)  =  K^{qG,Gq}(x_1,x_2,Q_f^2) + \mathcal{O}(\alpha_s).
\end{equation}
Inserting these expressions to (\ref{eq:Bareqq}), all $1/\epsilon$-poles in (\ref{eq:Bareqq}), (\ref{eq:BareqG}) and (\ref{eq:BareGq}) cancel giving the final, finite results:
\begin{eqnarray}
 \frac{d^2 \sigma^{q\overline{q}}}{dy_RdM^2} & = & \frac{4\pi \alpha_{em}^2}{9 s M^2} \left\{ H(x_1,x_2,Q_f^2) \left[ 1 + \frac{\alpha_s}{2\pi}C_F \left(\frac{2\pi^2}{3} - 8 \right) \right] \right.  \\
& + & \alpha_s \int_{x_1}^1 \frac{d\xi_1}{\xi_1} H(\xi_1,x_2,Q_f^2) f(\frac{x_1}{\xi_1}) + \alpha_s \int_{x_2}^1 \frac{d\xi_2}{\xi_2} H(x_1,\xi_2,Q_f^2) f(\frac{x_2}{\xi_2}) \nonumber \\
& + & \frac{\alpha_s}{2\pi}C_F \int_{x_1}^1 \frac{d\xi_1}{\xi_1} \int_{x_2}^1 \frac{d\xi_2}{\xi_2} H(\xi_1,\xi_2,Q_f^2) \frac{1}{2} \frac{1+z}{1-z} {\rm sech}^2 \left(y_R - \frac{1}{2}\log \frac{\xi_1}{\xi_2} \right) \nonumber \\
& & \left. \left[\frac{1+z^2}{(1-z)_+} \left[ \left(\frac{1}{1-y_0}\right)_+ + \left(\frac{1}{y_0}\right)_+ \right] - 2(1-z)\right]\right\}, \nonumber
\end{eqnarray}
\begin{eqnarray}
 \frac{d^2 \sigma_R^{qG+Gq}}{dy_RdM^2} & = & \frac{4\pi \alpha_{em}^2}{9 s M^2} \left\{ \alpha_s \int_{x_2}^1 \frac{d\xi_2}{\xi_2} K^{qG}(x_1,\xi_2,Q_f^2) g(\frac{x_2}{\xi_2})\right.\\
 & + & \hspace{1.5cm} \alpha_s \int_{x_1}^1 \frac{d\xi_1}{\xi_1} K^{Gq}(\xi_1,x_2,Q_f^2) g(\frac{x_1}{\xi_1}) \nonumber \\
& + & \frac{\alpha_s}{2\pi}T_R \int_{x_1}^1 \frac{d\xi_1}{\xi_1} \int_{x_2}^1 \frac{d\xi_2}{\xi_2} \frac{1}{2} \frac{1+z}{1-z} {\rm sech}^2 \left(y_R - \frac{1}{2}\log \frac{\xi_1}{\xi_2} \right) \nonumber \\
& & \left[ K^{qG}(\xi_1,\xi_2,Q_f^2) \left[2z(1-z) + (1-z)^2y_0 + \frac{z^2+(1-z)^2}{y_{0+}} \right] \right. \nonumber \\
& + & \left. \left. K^{Gq}(\xi_1,\xi_2,Q_f^2) \left[2z(1-z) + (1-z)^2(1-y_0) + \frac{z^2+(1-z)^2}{(1-y_0)_+} \right] \right] \right\}, \nonumber
\end{eqnarray}
where
\begin{eqnarray}
 \alpha_s f(z) & \equiv & \frac{\alpha_s}{2\pi} P_{qq}(z) \log \frac{M^2}{Q_f^2} - \frac{\alpha_s}{2\pi}f_q^{\rm scheme}(z) \nonumber \\
& + & \frac{\alpha_s}{2\pi} C_F \left\{ 2(1+z^2) \left[ \frac{\log(1-z)}{1-z} \right]_+ - \frac{(1+z^2)\log z}{1-z} + (1-z) \right\} \nonumber
\\
 \alpha_s g(z) & \equiv & \frac{\alpha_s}{2\pi} P_{qg}(z) \log \frac{M^2}{Q_f^2} - \frac{\alpha_s}{2\pi}f_g^{\rm scheme}(z) \nonumber \\
& + & \frac{\alpha_s}{2\pi} \left\{  P_{qg}(z) \log \left[\frac{(1-z)^2}{z} - 1 \right] + T_R \right\} \nonumber.
\end{eqnarray}

It is not very difficult to integrate these expressions over $y_R$ to recover the differential $d\sigma/dM^2$ cross-section given e.g in \cite{Furmanski:1981cw}.

\section{Numerical implementation}

The presence of double integrals makes the numerical calculation of Drell-Yan cross-section somewhat more challenging than the deeply inelastic scattering. Especially, one should pay attention how to evaluate integrals involving a product of plus-distributions.
\begin{figure}[h!]
\begin{center}
\includegraphics[scale=0.70]{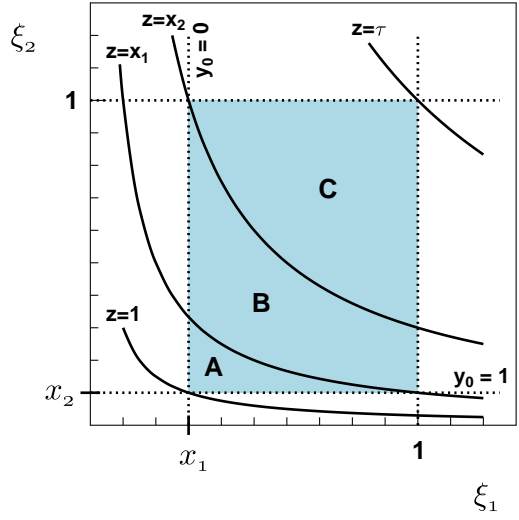}
\end{center}
\caption{The integration regions for $y_R>0$. The constant $y_0$-lines and $z$-hyperbola are indicated.}
\label{fig:DYintegfig}
\end{figure}
First, by change of variables
\begin{equation}
\int_{x_1}^1 \frac{d\xi_1}{\xi_1} \int_{x_2}^1 \frac{d\xi_2}{\xi_2} \frac{1}{2} \frac{1+z}{1-z} {\rm sech}^2 \left(y_R - \frac{1}{2}\log \frac{\xi_1}{\xi_2} \right) = \int \hspace{-0.2cm} \int_{A+B+C} \frac{dz}{z} dy_0,
\end{equation}
where the division of the rectangular integration region in $(\xi_1,\xi_2)$-plane to $A$, $B$ and $C$ parts are indicated in Fig.~\ref{fig:DYintegfig} for $y_R>0$. Among other terms, the NLO contribution to quark-antiquark process involves a term
$$
\frac{1+z^2}{(1-z)_+} \frac{H(\xi_1,\xi_2)}{(1-y_0)_+}.
$$
According to the definition of the plus-distributions, 
\begin{eqnarray}
 \int \hspace{-0.2cm} \int_{A} \frac{dz}{z} dy_0 \frac{1+z^2}{(1-z)_+} \frac{H(\xi_1,\xi_2)}{(1-y_0)_+} = \int_{x_1}^1 \frac{dz}{z} \frac{1+z^2}{(1-z)_+} \int_0^1 dy_0 \frac{H(\xi_1,\xi_2)-H(\frac{x_1}{z},x_2)}{1-y_0}, \nonumber
\end{eqnarray}
where
\begin{equation}
 \xi_1 = x_1 \sqrt{\frac{1}{z}\frac{z+y(1-z)}{1-y(1-z)}}, \qquad \xi_2 = \frac{\tau}{z\xi_1}. \nonumber
\end{equation}
Applying the definition of plus-distributions again to the remaining $z$-integral, we have
\begin{eqnarray}
& & \hspace{-0.5cm} \int_{0}^1 \frac{dz}{(1-z)} \int_0^1 dy_0 \left[\frac{1+z^2}{z} \frac{H(\xi_1,\xi_2)-H(\frac{x_1}{z},x_2)}{1-y_0} - 2 \frac{H(x_1,x_2)-H(x_1,x_2)}{1-y_0}\right] \nonumber \\
& & \hspace{-1.0cm} - \int_{0}^{x_1} \frac{dz}{(1-z)} \int_0^1 dy_0 \left[\frac{1+z^2}{z} \frac{H(\xi_1,\xi_2)-H(\frac{x_1}{z},x_2)}{1-y_0} \right], \nonumber
\end{eqnarray}
where the subtraction term evidently vanishes, leaving
\begin{eqnarray}
 \int \hspace{-0.2cm} \int_{A} \frac{dz}{z} dy_0 \frac{1+z^2}{(1-z)_+} \frac{H(\xi_1,\xi_2)}{(1-y_0)_+} = \int_{x_1}^1 \frac{dz}{z} \frac{1+z^2}{1-z} \int_0^1 dy_0 \frac{H(\xi_1,\xi_2)-H(\frac{x_1}{z},x_2)}{1-y_0}. \nonumber
\end{eqnarray}
The regions $B$ and $C$ avoid both the $z=1$ and the $y=1$ singularities,
\begin{eqnarray}
 \int \hspace{-0.2cm} \int_{B} \frac{dz}{z} dy_0 \frac{1+z^2}{(1-z)_+} \frac{H(\xi_1,\xi_2)}{(1-y_0)_+} = \int_{x_2}^{x_1} \frac{dz}{z} \frac{1+z^2}{1-z} \int_0^{y_0(\xi_1=1)} dy_0 \frac{H(\xi_1,\xi_2)}{1-y_0} \nonumber \\
 \int \hspace{-0.2cm} \int_{C} \frac{dz}{z} dy_0 \frac{1+z^2}{(1-z)_+} \frac{H(\xi_1,\xi_2)}{(1-y_0)_+} = \int_{\tau}^{x_2} \frac{dz}{z} \frac{1+z^2}{1-z} \int_{y_0(\xi_2=1)}^{y_0(\xi_1=1)} dy_0 \frac{H(\xi_1,\xi_2)}{1-y_0}. \nonumber
\end{eqnarray}
Since $H(\xi_1,\xi_2) = 0$ for $\xi_1,\xi_2 \ge 0$, the $y_0$-integrals for regions $B$ and $C$ may be extended to range from $0$ to $1$, and all integrals can be grouped neatly together:
\begin{eqnarray}
 \int \hspace{-0.2cm} \int_{A+B+C} \frac{dz}{z} dy_0 \frac{1+z^2}{(1-z)_+} \frac{H(\xi_1,\xi_2)}{(1-y_0)_+} = \int_{\tau}^1 \frac{dz}{z} \frac{1+z^2}{1-z} \int_0^1 dy_0 \frac{H(\xi_1,\xi_2)-H(\frac{x_1}{z},x_2)}{1-y_0}. \nonumber
\end{eqnarray}
Although we considered here the special case $y_R>0$, the result above is valid also for $y_R \le 0$. The treatment of other integrals involving plus-distributions is a straightforward extension to what was presented above and for completeness, I record here the $\overline{MS}$-scheme cross-sections in a form which no longer explicitly involves plus-distributions and can thus easily be turned into a computer code:

\begin{eqnarray}
 \frac{d^2 \sigma^{q\overline{q}}}{dy_RdM^2} & = & \frac{4\pi \alpha_{em}^2}{9 s M^2} \left\{ H(x_1,x_2,Q_f^2) \left[ 1 + \frac{\alpha_s}{2\pi}C_F \left(\frac{2\pi^2}{3} - 8 \right.\right. \right. \nonumber \\
& & \hspace{4.35cm} + \left. \left. 2\log^2 (1-x_1) + 2\log^2 (1-x_2) \right) \right] \nonumber \\
& + & \frac{\alpha_s}{2\pi}C_F \int_{x_1}^1 dz \frac{2\log(1-z)}{1-z} \left[ \frac{1+z^2}{z} H(\frac{x_1}{z},x_2,Q_f^2) - 2H(x_1,x_2,Q_f^2)\right] \nonumber \\
& + & \frac{\alpha_s}{2\pi}C_F \int_{x_2}^1 dz \frac{2\log(1-z)}{1-z} \left[ \frac{1+z^2}{z} H(x_1,\frac{x_2}{z},Q_f^2) - 2H(x_1,x_2,Q_f^2)\right] \nonumber \\
& + & \alpha_s \int_{x_1}^1 \frac{d\xi_1}{\xi_1} H(\xi_1,x_2,Q_f^2) \tilde f(\frac{x_1}{\xi_1}) + \alpha_s \int_{x_2}^1 \frac{d\xi_2}{\xi_2} H(x_1,\xi_2,Q_f^2) \tilde  f(\frac{x_2}{\xi_2}) \nonumber \\
& + & \frac{\alpha_s}{2\pi}C_F \int_{\tau}^1 \frac{dz}{z} \int_0^1 dy_0 \left[ -2(1-z)H(\xi_1,\xi_2) \right. \nonumber \\
& & \left.\left. +\frac{1+z^2}{1-z} \frac{H(\xi_1,\xi_2)-H(\frac{x_1}{z},x_2)}{1-y_0} + \frac{1+z^2}{1-z} \frac{H(\xi_1,\xi_2)-H(x_1,\frac{x_2}{z})}{y_0} \right] \right\} \nonumber 
\end{eqnarray}
\begin{eqnarray}
 \frac{d^2 \sigma_R^{qG+Gq}}{dy_RdM^2} & = & \frac{4\pi \alpha_{em}^2}{9 s M^2} \left\{ \alpha_s \int_{x_2}^1 \frac{d\xi_2}{\xi_2} K^{qG}(x_1,\xi_2,Q_f^2) \tilde g(\frac{x_2}{\xi_2})\right.\\
 & + & \hspace{1.5cm} \alpha_s \int_{x_1}^1 \frac{d\xi_1}{\xi_1} K^{Gq}(\xi_1,x_2,Q_f^2) \tilde g(\frac{x_1}{\xi_1}) \nonumber \\
& + & \frac{\alpha_s}{2\pi}T_R \int_{\tau}^1 \frac{dz}{z} \int_0^1 dy_0 \left[ K^{qG}(\xi_1,\xi_2,Q_f^2) \left[2z(1-z) + (1-z)^2y_0 \right] \right. \nonumber \\
& + & \left. \left(z^2+(1-z)^2\right) \frac{K^{qG}(\xi_1,\xi_2)-K^{qG}(x_1,\frac{x_2}{z})}{y_0}  \right] \nonumber \\
& + & \frac{\alpha_s}{2\pi}T_R \int_{\tau}^1 \frac{dz}{z} \int_0^1 dy_0 \left[ K^{Gq}(\xi_1,\xi_2,Q_f^2) \left[2z(1-z) + (1-z)^2(1-y_0) \right] \right. \nonumber \\
& + & \left.\left. \left(z^2+(1-z)^2\right) \frac{K^{Gq}(\xi_1,\xi_2)-K^{Gq}(\frac{x_1}{z},x_2)}{1-y_0}  \right] \right\} \nonumber
\end{eqnarray}
where
\begin{eqnarray}
 \alpha_s \tilde f(z) & \equiv & \frac{\alpha_s}{2\pi} \left\{ P_{qq}(z) \log \frac{M^2}{Q_f^2}  +  C_F \left[ - \frac{(1+z^2)\log z}{1-z} + (1-z)\right] \right\} \nonumber
\\
 \alpha_s \tilde g(z) & \equiv & \frac{\alpha_s}{2\pi} \left\{ P_{qg}(z) \log \frac{M^2}{Q_f^2} + P_{qg}(z) \log \left[\frac{(1-z)^2}{z} - 1 \right] + T_R  \right\} \nonumber.
\end{eqnarray}

\section{Numerical estimate}
\begin{figure}[h!]
\begin{center}
\includegraphics[scale=0.55]{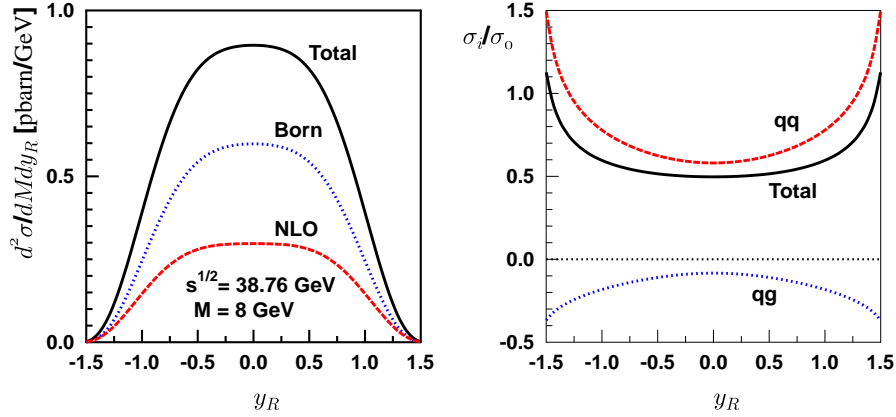}
\end{center}
\caption{An example of the rapidity distributions in Drell-Yan process computed in the $\overline{\rm MS}$-scheme. The left panel shows the behaviour of the absolute cross-section for $\sqrt{s}=38.76 \, {\rm GeV}$ and $M=8 \, {\rm GeV}$ using CTEQ6L1 leading-order parton densities and indicating the leading order and next-to-leading order contributions separately. In the right panel, the relative contributions normalized to the leading order cross-section $\sigma_0$ of quark-antiquark and (anti)quark-gluon processes are shown.}
\label{fig:DYexample}
\end{figure}

In order to get a feeling about the size of the NLO corrections to the leading order cross-section, Fig.~\ref{fig:DYexample} shows a numerical result in the $\overline{\rm MS}$-scheme with the factorization scale choice $Q_f^2=M^2$, for a kinematical configuration $\sqrt{s}=38.76 \, {\rm GeV}$ (corresponding to a fixed target experiment with $800 \, {\rm GeV}$ proton beam) and $M=8 \, {\rm GeV}$ using the CTEQ6L1 leading-order parton distribution functions. Unlike in the deeply inelastic scattering, the contribution of the NLO terms is rather large --- always at least 50\% and increasing when going away from the midrapidity. It is interesting to notice that the quark-gluon contribution remains always negative, partly cancelling the large positive contribution of quark-quark subprocesses.

\vspace{0.3cm}
The Drell-Yan rapidity distribution, computed here to NLO, is nowadays known still one power higher in $\alpha_s$ (NNLO) \cite{Anastasiou:2003yy}. The size of the NNLO corrections relative to the NLO are not, however, as large as the NLO relative to LO and the perturbative expansion seems to stabilize.

%% file: hadronprod.tex
\chapter{Inclusive hadron production}

For a deeply inelastic scattering event the experimental signature typically consists of the scattered lepton and a narrow shower, a \emph{jet} of hadrons, originating from the struck parton that balances the transverse momentum of the scattered lepton. The \emph{hadronization} --- how the partons transform themselves into a cascade of hadrons --- is a non-perturbative process and beyond the reach of pQCD-tools. As long as we are not concerned about the structure of the jet, we can simply ignore such process. This kind of final state is said to be \emph{fully inclusive}. However, as the jets consist of hadrons, with a suitable detector they can be identified and their momentum measured. In this Chapter I shortly describe how such indentified hadron production cross-sections are calculated in pQCD, especially in hadron-hadron collisions. 

\section{Fragmentation functions}

\begin{figure}[h]
\centering
 \includegraphics[scale=0.45]{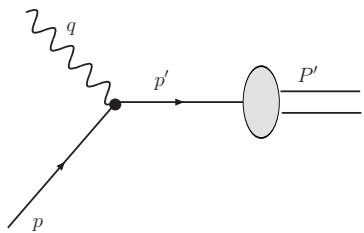}
\caption{Hadron production in deeply inelastic scattering.}
\label{Fig:hadr_DIS}
\end{figure}

Let us return to the leading-order deeply inelastic scattering, where a high-$Q^2$ photon knocks a quark $q$ to an escape-course from the nucleon triggering off a jet as shown schematically in Fig.~\ref{Fig:hadr_DIS}. 

\begin{figure}[h]
\centering
 \includegraphics[scale=0.40]{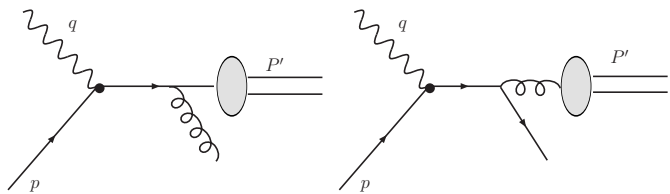}
\caption{Dominant gluon radiation graphs for deeply inelastic hadron production.}
\label{Fig:hadr2_DIS}
\end{figure}

\pagebreak
The number density of hadrons $h$ carrying a fraction $z$ of the jet energy is described by a \emph{fragmentation function} $D^h_q(z)$, where the $q$ indicates the parton that initiated the fragmentation. 
Consequently, the leading-order cross-section for single particle (plus anything else) production in the deeply inelastic scattering (SIDIS) is 
\begin{equation}
\frac{d^2\sigma_0^h}{dxdzdQ^2} = \frac{d^2\hat{\sigma_{0}}}{dxdQ^2} \sum_q e_q^2 f_q(x) D^h_q(z),
\label{eq:hadrDIS_partonmodel}
\end{equation}
where the energy-fraction $z$ may be expressed in an invariant form as
\begin{equation}
z \equiv \frac{P \cdot P'}{P \cdot q} = \frac{E_{\rm hadron}}{\nu},
\end{equation}
where the latter equality refers to the target rest frame.
Beyond leading order, however, collinear divergences due to e.g collinear gluon radiation from the outgoing quark emerge. Here, I would emphasize that these divergences remain only because the final state is not inclusive enough: If it was not for the fragmentation functions --- if we would not care what the quark will eventually become of --- the divergences above would exactly cancel against the loop diagrams. In the axial gauge, the dominant $\mathcal{O}(\alpha_s)$ logarithm originates from the quark-splitting diagram shown in Fig.~\ref{Fig:hadr2_DIS}, 
and the contribution to the cross-section can be expressed as
\begin{eqnarray}
\frac{d^2\hat{\sigma_{0}}}{dxdQ^2} \sum_q e_q^2 f_q(x) \left\{ \left[1 + \frac{\alpha_s}{2\pi} \log \left( \frac{Q^2}{m^2} \right) P_{qq} \right] \otimes D^h_q + \frac{\alpha_s}{2\pi} \log \left( \frac{Q^2}{m^2} \right) P_{qg} \otimes D^h_g \right\} \nonumber,
\end{eqnarray}
where $\otimes$ again denotes the convolution integral and the splitting functions $P_{qq}$ and $P_{gq}$ are the same as earlier. These leading logarithmic terms can be resummed essentially in the same way as was done in Chapter \ref{DGLAPevolution} for the initial state radiation. Then, by absorbing these logarithms into the definition of scale-dependent fragmentation functions $D^h(z,Q^2)$, they are seen to follow the same DGLAP equations 
\begin{eqnarray}
Q^2 \frac{\partial D_{q_i}^h(x,Q^2)}{\partial Q^2} & = & \frac{\alpha_s(Q^2)}{2\pi} \left[P_{q_iq_j} \otimes D_{q_j}^h(Q^2) + P_{q_ig} \otimes D_g^h (Q^2) \right] \\
Q^2 \frac{\partial D_{g}^h(x,Q^2)}{\partial Q^2} & = & \frac{\alpha_s(Q^2)}{2\pi} \left[P_{gg} \otimes D_{g}^h(Q^2) + P_{gq_j} \otimes D^h_{q_j} (Q^2) \right], \nonumber 
\label{eq:completeDGLAP_for_frag}
\end{eqnarray}
as the parton distribution functions do. At the leading logarithmic approximation, the splitting kernels are exactly the same as derived in the initial state parton branching, but at higher orders they become different being still related by a proper analytic continuation \cite{Stratmann:1996hn}. As was argued in Chapter \ref{DGLAPevolution}, the collinear divergences between the incoming and outgoing partons do not interfere. That is, they independently factorize, and the first pQCD-improved version of Eq.~(\ref{eq:hadrDIS_partonmodel}) would be the one in which the parton distributions and fragmentation functions are scale-dependent,	
\begin{equation}
\frac{d^3\sigma^h}{dxdzdQ^2} = \frac{d^2\hat{\sigma_{0}}}{dxdQ^2} \sum_q e_q^2 f_q(x,Q^2) D^h_q(z,Q^2).
\end{equation}
For an exact NLO calculation, one should consider the lepton-parton processes
$$
\ell(k) + p^i_1 \rightarrow p^j_2 + {\ell}(k') + X \nonumber,
$$
and adopt the dimensional regularization methods to calculate the differential cross-sections
$$
\frac{d^3 \hat \sigma_\epsilon^{\ell + p^i_1 \rightarrow p^j_2 + {\ell} + X}}{d\hat{x}d\hat{z}dQ^2}, \qquad \hat{x} = \frac{Q^2}{2p_1^i \cdot q}, \quad \hat{z} = \frac{P \cdot p_2^j}{P \cdot q}.
$$
The diagrams for such calculation are the same as those displayed in Chapter \ref{Deeply_inelastic_scattering_at_NLO}, but the phase-space integrals should be constrained by an additional delta function $\delta(\hat{z}-({P \cdot p_2^j})/({P \cdot q}))$. I will not get into details here \cite{Altarelli:1979kv}, but at the end, the partonic cross- sections will retain several $1/\epsilon$-poles from the collinear divergences. Multiplying these expressions by the ``bare'' parton densities and fragmentation functions one can construct the hadronic cross-section
\begin{eqnarray}
\hspace{-0.5cm}\frac{d^3\sigma^h}{dxdzdQ^2} & = & \sum_{i,j} \int_x^1 \frac{d\xi_1}{\xi_1} \int_z^1 \frac{d\xi_2}{\xi_2} f_{i0}(\xi_1) \frac{d^3 \hat \sigma_\epsilon^{\ell + p^i_1 \rightarrow p^j_2 + {\ell} + X}}{d\hat{x}d\hat{z}dQ^2} D^{j \rightarrow h}_0(\xi_2),
\end{eqnarray}
where $p_1^i = \xi_1 P$ and $p_2^j = P'/\xi_2$. Defining the scale-dependent fragmentation functions similarly to the parton distributions in Eq.~(\ref{eq:NLOdefinitions}), and writing the bare quantities $f_{i0}(x)$ and $D_0(x)$ in terms of the scale-dependent ones, one can take the $\epsilon \rightarrow 0$ limit
\begin{eqnarray}
\hspace{-0.2cm}\frac{d^3\sigma^h}{dxdzdQ^2} & = & \sum_{i,j} \int_x^1 \frac{d\xi_1}{\xi_1} \int_z^1 \frac{d\xi_2}{\xi_2} f_{i}(\xi_1,Q_{\rm fac}^2) \frac{d^3 \hat \sigma^{\ell + p^i_1 \rightarrow p^j_2 + {\ell} + X}}{d\hat{x}d\hat{z}dQ^2} D^{j \rightarrow h}(\xi_2,Q_{\rm frag}^2), \nonumber
\end{eqnarray}
where $d \hat \sigma$ depends on $Q^2$, $Q_{\rm frag}^2$, and $Q_{\rm fac}^2$. This is a cross-section guaranteed to be free from collinear divergences.

\vspace{0.3cm}
The fragmentation functions can be determined by analyzing experimental data by a similar procedure which is used to extract the parton densities (to be described later) \cite{Kniehl:2000fe,de Florian:2007hc,deFlorian:2007aj,Albino:2008fy}. The cleanest enviroment to extract the fragmentation functions are the $e^+e^-$-induced processes due to less crowded final state (compared to the collisions involving hadrons), and as they are not complicated by the parton distribution functions. However, the $e^+e^-$-data alone cannot constrain all components of the fragmentation functions well, and therefore some latest analyses like \cite{deFlorian:2007aj} are complemented by SIDIS measurements just introduced, and also by data from hadron-hadron collisions \cite{de Florian:2007hc,deFlorian:2007aj,Albino:2008fy} to be described shortly. The resulting fragmentation functions obviously somewhat depend on the particular set of parton densities used in the analysis, and the most complete analysis would combine both to a single analysis.

\section{Single-inclusive hadron production in hadronic collisions}

The fragmentation function formalism can also be applied to inclusive production of large transverse momentum hadrons in hadron-hadron collisions:
\begin{equation}
{\rm H_1}(K_1) + {\rm H_2}(K_2) \rightarrow {H_3}(K_3) + X \nonumber,
\end{equation}
where $K_1$ and $K_2$ denote the incoming hadron momenta, $K_3$ is the momentum of the observed hadron, and $X$ is, as usual, anything. The calculation of this cross-section begins by computing the invariant partonic cross-sections
\begin{equation}
p_3^0 \frac{d\hat{\sigma}_\epsilon(p_1^i + p_2^j \rightarrow p_3^\ell + X)}{d^3p_3}.
\end{equation}
At NLO \cite{Aversa:1988vb}, these consist of three pieces: purely tree-level $2 \rightarrow 2$ and $2 \rightarrow 3$ diagrams, and $2 \rightarrow 2$ diagrams decorated with loops. As earlier, there will be various divergences: collinear, infrared, and ultraviolet. The singularities appearing as ${1}/{\epsilon^2}$-poles cancel between the real and virtual contributions, and when the ultraviolet divergences from the loop integrals are subtracted according to the adopted renormalization prescription, only the collinear ${1}/{\epsilon}$-poles remain. Multiplying these quantities by the (bare) parton densities and fragmentation functions and integrating over the available phase-space, the invariant hadronic cross-section becomes


\begin{eqnarray}
\label{eq:1sthadroncrosssection}
E_3 \frac{d\sigma(H_1 + H_2 \rightarrow H_3 + X)}{d^3K_3}  & = & \\
& & \hspace{-4cm} \sum _{ijl} \int dx_1 \int dx_2 \int \frac{dx_3}{x_3^2} f_{i0}^{H_1}(x_1) f_{j0}^{H_2}(x_2) D^{l \rightarrow H_3}_0(x_3) \nonumber\\
 & & \hspace{-4cm} \, p_3^0 \frac{d\hat{\sigma}_\epsilon(p_1^i + p_2^j \rightarrow p_3^l, \mu^2_{\rm ren})}{d^3p_3}
_{
\begin{array}{l}
\vline \; p_1 = x_1K_1 \\
\vline \; p_2 = x_2K_2 \\
\vline \; p_3 = K_3/x_3 \\
\end{array}
}  \nonumber,
\end{eqnarray}
where the additional factor $1/x_3^2$ originates from $d^3 K_3/E_3 =x_3^2 d^3 p_3/p_3$. Writing the bare quantities $f_{i0}(x)$ and $D_0(x)$ in terms of the scale-dependent ones, the remaining collinear divergences again cancel and a finite result is obtained in the $\epsilon \rightarrow 0$ limit,
\begin{eqnarray}
E_3 \frac{d\sigma(H_1 + H_2 \rightarrow H_3 + X)}{d^3K_3}  & = & \\
& & \hspace{-4cm} \sum _{ijl} \int_{x_1^{\rm min}}^1 dx_1 \int_{x_2^{\rm min}}^1 dx_2 \int_{x_3^{\rm min}}^1 \frac{dx_3}{x_3^2} f_i^{H_1}(x_1,\mu^2_{\rm fact}) f_j^{H_2}(x_2,\mu^2_{\rm fact}) \nonumber\\
 & & \hspace{-4cm} D_{l \rightarrow H_3}(x_3,\mu^2_{\rm frag}) \, p_3^0 \frac{d\hat{\sigma}(p_1^i + p_2^j \rightarrow p_3^l, \mu^2_{\rm ren})}{d^3p_3}
_{
\begin{array}{l}
\vline \; p_1 = x_1K_1 \\
\vline \; p_2 = x_2K_2 \\
\vline \; p_3 = K_3/x_3 \\
\end{array}
}  \nonumber.
\end{eqnarray}

The cross-sections are often reported specifying the center-of-mass energy $\sqrt{s}$ of the hadron-hadron collision, the transverse momentum $p_T$, and the rapidity $y_R$ of the observed hadron. In terms of these variables, the integration limits in the expression above are
\begin{equation}
x_1^{\rm min} = \frac{p_T e^{y_R}}{\sqrt{s}-p_T e^{-y_R}}, \quad x_2^{\rm min} = \frac{x_1 p_T e^{-y_R}}{x_1 \sqrt{s}-p_T e^{y_R}}, \quad x_3^{\rm min} = \frac{2 p_T \cosh y_R}{\sqrt{x_1x_2s}}. \nonumber
\end{equation}

Similarly to the Drell-Yan dilepton production, the leading-order approximation to the inclusive hadron production turns out to undershoot the experimental data \cite{Eskola:2002kv} by a typical factor of $\sim 2$, depending on the kinematical variables and scale choices. The larger cross-sections at NLO improve such situation nicely \cite{Aurenche:1999nz}, although at low $\sqrt{s} \lesssim 60 \, {\rm GeV}$ the theory still seems to undershoot the data. For more recent work discussing RHIC data for constraining the fragmentation functions, see e.g. \cite{de Florian:2007hc,deFlorian:2007aj}.

%% file: solveDGLAP.tex
\chapter{Solving the DGLAP equations}
\label{SolvingDGLAP}

Global QCD-analyses, to be described later in Chapter \ref{AboutGlobal}, require an efficient way of solving the DGLAP evolution equations. Being integro-differential equations, there is not much that can be done purely analytically but numerical methods are to be used. Several methods to accomplish this has been developed --- for an elementary account for couple of treatments at leading order, see \cite{Kumano:2004dw}. At leading order, the DGLAP equations are still fairly easy to solve but the technical difficulties significantly increase when going to higher orders (NLO \& NNLO). The method I describe in this section is based on \cite{Santorelli:1998yt} and it has been employed in the publications \cite{EPS08-paperi,EPS09-paperi} of this thesis. For description of further methods and available codes see e.g. \cite{Salam:2008qg,Vogt:2004ns}.

\section{Decomposition of the DGLAP equations}

The full set of evolution equations to be solved can be written as
\begin{eqnarray}
Q^2 \frac{\partial {q_i}}{\partial Q^2} & = & \frac{\alpha_s}{2\pi} \left[\sum_k P_{q_i q_k} \otimes {q_k} + \sum_k P_{q_i \overline{q}_k} \otimes \overline{q}_k + P_{qg} \otimes g  \right] \nonumber \\
Q^2 \frac{\partial {\overline{q}_i}}{\partial Q^2} & = & \frac{\alpha_s}{2\pi} \left[\sum_k P_{\overline{q}_i q_k} \otimes {q_k} + \sum_k P_{\overline{q}_i \overline{q}_k} \otimes \overline{q}_k + P_{\overline{q}g} \otimes g \right] \label{eq:allDGLAP} \\
Q^2 \frac{\partial {g}}{\partial Q^2} & = & \frac{\alpha_s}{2\pi} \left[ \sum_k P_{g q_k} \otimes {q_k} \,\, + \sum_k P_{g \overline{q}_k} \otimes \overline{q}_k \,\, + P_{gg} \otimes g \right], \nonumber
\end{eqnarray}
where the arguments of parton densities and the strong coupling are not displayed. A useful decomposition \cite{Curci:1980uw,Furmanski:1981cw} of the splitting functions $P_{qq}$ and $P_{\overline{q}q}$ is to separate the flavor-preserving ``valence'' and possibly-flavor-changing ``sea'' parts as
\begin{eqnarray}
P_{q_i q_k} & = & \delta_{ik} P_{q q}^V + P_{q q}^S \label{eq:splitting_of_splitting} \\
P_{q_i \overline{q}_k} & = & \delta_{ik} P_{q \overline{q}}^V + P_{q\overline{q}}^S  \nonumber.
\end{eqnarray}
At leading order only $P_{q q}^V$ in (\ref{eq:splitting_of_splitting}) is non-zero, but at NLO they all are non-trivial, but respect the following relations:
\begin{eqnarray}
P_{qq}^V & = & P_{\overline{q}\overline{q}}^V, \qquad \qquad P_{q\overline{q}}^V = P_{\overline{q}{q}}^V, \qquad
P_{qq}^S = P_{q\overline{q}}^S = P_{\overline{q}q}^S = P_{\overline{q}\overline{q}}^S \nonumber \\
P_{q_ig} & = & P_{\overline{q}_ig} \equiv P_{qg}, \hspace{0.25cm} P_{gq_i} = P_{g\overline{q}_i} \equiv P_{gq},
\end{eqnarray}
which are reflections from the charge-conjugation invariance and the SU(3) flavor symmetry, but can also be easily understood on the basis of the Feynman diagrams\footnote{At NNLO, however, $P_{qq}^S \neq P_{q\overline{q}}^S$}. By defining
\begin{eqnarray}
P_{\pm} & \equiv & P_{qq}^V \pm P_{q\overline{q}}^V \nonumber \\
P_{\rm FF} & \equiv & P_{+} + 2n_f P_{qq}^S \\
P_{\rm FG} & \equiv & 2n_fP_{qg} \nonumber \\
P_{\rm GF} & \equiv & P_{gq} \nonumber,
\end{eqnarray}
where $n_f$ is the number of active flavors, and
\begin{eqnarray}
q_i^{\pm} \equiv q_i \pm \overline{q}_i, \qquad q^{\pm} \equiv \sum_{i}^{n_f} q_i^{\pm},
\label{eq:plus_and_minus_distr}
\end{eqnarray}
the set of equations (\ref{eq:allDGLAP}) can written as
\begin{eqnarray}
\frac{d}{d\log Q^2} 
\left(
\begin{array}{l}
q^{+} \\
g
\end{array}
\right)
& = &
\frac{\alpha_s}{2\pi}
\left(
\begin{array}{cc}
P_{\rm FF} & P_{\rm FG} \\
P_{\rm GF} & P_{\rm GG}
\end{array}
\right)
\left(
\begin{array}{l}
q^{+} \\
g
\end{array}
\right) \label{eq:evo1}
\\
\frac{dq_i^{-}}{d\log Q^2} & = & \frac{\alpha_s}{2\pi} \, P_{-} \otimes q_i^{-} \label{eq:evo2} \\
\frac{d}{d\log Q^2} \left[ q_i^{+}-\frac{1}{n_f}q^{+} \right] & = & \frac{\alpha_s}{2\pi} P_{+} \otimes \left[ q_i^{+}-\frac{1}{n_f}q^{+} \right] \label{eq:evo3}. \\
\nonumber
\end{eqnarray}
The densities $q_i^{-}$ and $q_i^{+}-({1}/{n_f})q^{+}$ evolve independently, whereas $q^{+}$ and $g$ are coupled. The strategy to solve the evolution of individual flavors $q_i$, is to substitute $q^{+}$ derived from (\ref{eq:evo1}) to result of (\ref{eq:evo3}) and use (\ref{eq:plus_and_minus_distr}). A good reference containing the expressions for all splitting functions needed to solve (\ref{eq:evo1})-(\ref{eq:evo3}) is \cite{Ellis:1991qj}.

\section{The Taylor expansion}

To keep the subsequent discussion as transparent as possible, let us consider the simplest evolution equation, namely that for the valence quarks $q_v(x,Q^2) \equiv q^-(x,Q^2)$ (with $P=P_-$),
\begin{eqnarray}
 Q^2 \frac{\partial }{\partial Q^2}q_v(x,Q^2) & = & \frac{\alpha_s(Q^2)}{2\pi} \int_x^1 \frac{d\xi}{\xi} P(\frac{x}{\xi}) q_v(\xi,Q^2) \label{eq:valevo} \\
 & = & \frac{\alpha_s(Q^2)}{2\pi} P \otimes q_v, \nonumber
\end{eqnarray}
with a given initial condition $q_v(x,Q_0^2)$. To make the $Q^2$-evolution appear as linear as possible, it is useful to define a new evolution variable
\begin{equation}
 t \equiv \frac{2}{\beta_0} \log \frac{\alpha_s(Q_0^2)}{\alpha_s(Q^2)},
\end{equation}
where $\beta_0 = \frac{11}{3}C_G - \frac{4}{3} T_R n_f$ appears in the QCD renormalization group equation
\begin{equation}
Q^2 \frac{d\alpha_s(Q^2)}{dQ^2} = - \alpha_s(Q^2) \left[ \beta_0 \frac{\alpha_s(Q^2)}{4\pi} + \beta_1 \left(\frac{\alpha_s(Q^2)}{4\pi}\right)^2 + \ldots \right].
\end{equation}
Trading the $Q^2$-derivative with $t$-derivative, we have
\begin{equation}
 Q^2 \frac{d}{dQ^2} = \frac{\alpha_s(Q^2)}{2\pi} \left[1 + \frac{\beta_1}{2\beta_0} \frac{\alpha_s(Q^2)}{2\pi} \right] \frac{d}{dt} + \mathcal{O}(\alpha_s^3).
\end{equation}
With this change of evolution variable, the Eq.~(\ref{eq:valevo}) reads
\begin{equation}
\left[1 + \frac{\beta_1}{2\beta_0} \frac{\alpha_s(Q^2)}{2\pi} \right] \frac{d}{dt}q_v(x,t) = P \otimes q_v(t).
\label{eq:evolution_with_t}
\end{equation}
To the NLO accuracy, the splitting function $P$ is of the form
\begin{equation}
P(z) = P^{(0)}(z) + \frac{\alpha_s}{2\pi} P^{(1)}(z),
\end{equation}
and we may write Eq.~(\ref{eq:evolution_with_t}) as
\begin{eqnarray}
\frac{\partial }{\partial t}q_v(x,t) & = & {\Omega} \otimes q_v(t), \label{eq:evoint}
\end{eqnarray}
where 
\begin{equation}
{\Omega} \equiv P^{(0)} + \frac{\alpha_s(Q^2)}{2\pi} \left(P^{(1)} - \frac{\beta_1}{2\beta_0}P^{(0)} \right).
\end{equation}
The very crux of the matter here is to expand $q_v(t)$ as a Taylor series around the initial scale $t_0 = t(Q_0^2)=0$
\begin{equation}
q_v(x,t) = \sum_{k=0}^{\infty} \frac{t^k}{k!} \frac{\partial ^k q_v(x,t=0)}{\partial t^k},
\label{eq:original_Taylor}
\end{equation}
where ${\partial ^k q_v(x,0)}/{\partial t^k}$ are multiple derivatives
\begin{eqnarray}
\frac{\partial ^0q_v(t)}{\partial t^0} & = & q_v(t) \nonumber \\
\frac{\partial q_v(t)}{\partial t} & = & {\Omega} \otimes q_v(t) \nonumber \\
\frac{\partial ^2q_v(t)}{\partial t^2} & = & \frac{\partial {\Omega}}{\partial t} \otimes q_v(t) +  {\Omega} \otimes \frac{\partial q_v(t)}{\partial t} \nonumber \\
\frac{\partial ^3q_v(t)}{\partial t^3} & = & \frac{\partial ^2{\Omega}}{\partial t^2} \otimes q_v(t) +  2 \frac{\partial {\Omega}}{\partial t} \otimes \frac{\partial q_v(t)}{\partial t} + {\Omega} \otimes \frac{\partial ^2q_v(t)}{\partial t^2} \nonumber \\
\frac{\partial ^4q_v(t)}{\partial t^4} & = & \frac{\partial ^3{\Omega}}{\partial t^3} \otimes q_v(t) +  3 \frac{\partial ^2{\Omega}}{\partial t^2} \otimes \frac{\partial q_v(t)}{\partial t} + 3 \frac{\partial {\Omega}}{\partial t} \otimes \frac{\partial ^2q_v(t)}{\partial t^2} + {\Omega} \otimes \frac{\partial ^3q_v(t)}{\partial t^3} \nonumber \\
& \vdots & \nonumber
\end{eqnarray}
By using the lower-order derivatives in the expression for the higher derivatives, the $n$th one we can be written as
\begin{equation}
\frac{\partial ^nq_v(t)}{\partial t^n} = {M}^{(n)} \otimes q_v(t),
\end{equation}
where each ${M}^{(k)}$ can be iteratively computed from previous ones
\begin{eqnarray}
M^{(0)} & = & 1 \nonumber \\
M^{(1)} & = & {\Omega}^{(0)} \nonumber \\
M^{(2)} & = & {\Omega}^{(1)} + {\Omega}^{(0)} \otimes M^{(1)} \nonumber \\
M^{(3)} & = & {\Omega}^{(2)} + 2 {\Omega}^{(1)} \otimes M^{(1)} + {\Omega}^{(0)} \otimes M^{(2)} \nonumber \\
M^{(4)} & = & {\Omega}^{(3)} + 3 {\Omega}^{(2)} \otimes M^{(1)} + 3 {\Omega}^{(1)} \otimes M^{(2)} + {\Omega}^{(0)} \otimes M^{(3)}  \nonumber \\
& \vdots & \nonumber
\end{eqnarray}
where $\Omega^{(0)} \equiv \Omega $, and $ \Omega^{(k)} \equiv d^k \Omega(t=0)/{dt^k} $ for $k \ge 1$. In general,
\begin{eqnarray}
{M}^{(k)} & = & \sum_{n=0}^{k-1}
{{k-1} \choose {n}}
{\Omega}^{(n)} \otimes M^{(k-1-n)} \\
\Omega^{(0)} & = & P^{(0)} + \,\,\,\,\,\, \frac{\alpha_s(Q_0^2)}{2\pi} \left(P^{(1)} - \frac{\beta_1}{2\beta_0}P^{(0)} \right) \\
\Omega^{(n)} & = & \left(-\frac{\beta_0}{2} \right)^{n} \frac{\alpha_s(Q_0^2)}{2\pi} \left(P^{(1)} - \frac{\beta_1}{2\beta_0}P^{(0)} \right)
\end{eqnarray}
where ${a \choose b} = \frac{a!}{b!(a-b)!}$ is the usual binomial coefficient. Thus, the Taylor series (\ref{eq:original_Taylor}) becomes
\begin{equation}
q_v(x,t) = \left[ \sum_{k=0}^{\infty} \frac{t^k}{k!} M^{(k)} \right] \otimes q_v(x,0). \label{eq:finalTaylor}
\end{equation}
The crucial feature to be noticed is that the for each $x$, the evolution functions $M^{(k)}$ are \emph{independent of the parton density} $q_v$. Also, the magnitude of $t$ in the physically conceivable domain is $\sim 0.1$, and one can expect the series to converge with a reasonable number of terms in the expansion.

\vspace{0.5cm}
Since the parton densities tend to generally diverge as $x^{-\delta}, \, (\delta > 0)$ towards $x \rightarrow 0$, it is numerically more stable to damp such bad behaviour by writing the evolution not for the absolute parton density $q(x,Q^2)$ itself, but rather for the momentum distribution $\hat q(x,Q^2) \equiv x q(x,Q^2)$. Multiplying the expansion (\ref{eq:finalTaylor}) by $x$,
\begin{equation}
\hat q_v(x,t) = \left[ \sum_{k=0}^{\infty} \frac{t^k}{k!} M^{(k)} \right] \hat \otimes \, \hat q_v(x,0),
\end{equation}
where the ``hatted'' convolution $\hat \otimes$ should be understood as
$$
f_1 \, \hat\otimes \, f_2 \, \hat\otimes \ldots \hat\otimes \, \hat q = \int_x^1 d\xi_1 f_1(\xi_1) \int_{\frac{x}{\xi_1}}^1 f_2(\xi_2) \ldots \int_{\frac{x}{\xi_1 \xi_2 \cdots \xi_{N-1}}}^1 f_N(\xi_N) \hat q(\frac{x}{\xi_1 \xi_2 \cdots \xi_N}).
$$

\section{Integration}

\begin{figure}[!h]
\centering
 \includegraphics[scale=0.40]{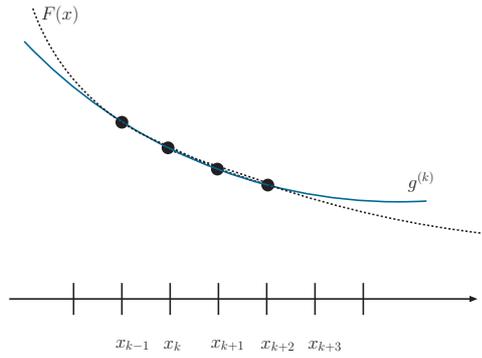}
\caption{Illustration of a polynomial fit to the function $F(x)$ used in the interval $x \in [x_k,x_{k+1}]$. The black dots denote the matching points.}
\label{Fig:polynomialfit}
\end{figure}

In order to actually calculate the very formal Taylor expansion written down in the above section, one needs to evaluate a series of ``nested'' integrals, each one of the general form 
\begin{equation}
I \equiv \int_x^1 d\xi P(\xi) F(\frac{x}{\xi}), \label{eq:genconv}
\end{equation}
where $F$ is a result from a similar integral. To accomplish this task it is useful to break the $x$-interval $[x,1]$ into $N$ smaller sub-intervals by a discrete grid $(x_0=x, x_1, x_2, \ldots, x_{N-1},x_N=1)$, and approximate the function $F(x)$ in each interval by a simpler one, like a polynomial, for which the integrals against the splitting functions can be \emph{analytically} evaluated. In other words, we write
\begin{equation}
 F(x) \approx \sum_{\ell=1}^m a_\ell^{(k)} g_\ell(x), \qquad \forall x \in [x_k, x_{k+1}],
\label{eq:Fapprox}
\end{equation}
where $a_\ell^{(k)}$ are coefficients which naturally depend on $F(x)$. For example, if we employ a 3rd order polynomial, $g_\ell(x)=x^{\ell-1}$, $m=4$, the coefficients $a_\ell^{(k)}$ can be taken to satisfy
\begin{equation}
\left(
\begin{array}{c}
 F(x_{k-1}) \\
 F(x_{k}) \\
 F(x_{k+1}) \\
 F(x_{k+2})
\end{array}
\right)
=
\left(
\begin{array}{cccc}
 1 & x_{k-1} & x_{k-1}^2 & x_{k-1}^3 \\
 1 & x_{k} & x_{k}^2 & x_{k}^3 \\
 1 & x_{k+1} & x_{k+1}^2 & x_{k+1}^3 \\
 1 & x_{k+2} & x_{k+2}^2 & x_{k+2}^3
\end{array}
\right)
\left(
\begin{array}{c}
 a_1^{(k)} \\
 a_2^{(k)}\\
 a_3^{(k)}\\
 a_4^{(k)}
\end{array}
\right),
\label{eq:F_a_matching}
\end{equation}
that is, the coefficients of the polynomial are chosen to match the $F(x)$ at four points around $x_k$ as illustrated in Fig.~(\ref{Fig:polynomialfit}).
By inverting (\ref{eq:F_a_matching}),
\begin{equation}
a_\ell^{(k)} = \sum_{r=1}^4 G_{\ell r}^{(k)} F(x_{k+r-2}),
\label{eq:acoefficients}
\end{equation}
where
$$
G_{\ell r}^{(k)} =
\left(
\begin{array}{cccc}
 1 & x_{k-1} & x_{k-1}^2 & x_{k-1}^3 \\
 1 & x_{k} & x_{k}^2 & x_{k}^3 \\
 1 & x_{k+1} & x_{k+1}^2 & x_{k+1}^3 \\
 1 & x_{k+2} & x_{k+2}^2 & x_{k+2}^3
\end{array}
\right)^{-1}_{\ell r}.
$$ 

The various terms in the splitting functions $P$ can be grouped to the following categories
\begin{equation}
 P(z) = \frac{A(z)}{(1-z)_+} + B(z) + C \delta(1-z),
\end{equation}
and inserting such expression to (\ref{eq:genconv}), we find
\begin{eqnarray}
I_{\big|x=x_i} & = & x_i \int_{x_i}^1 \frac{dz}{z} \left[ \frac{A(x_i/z)F(z)-A(1)F(x_i)}{z-x_i} + \frac{1}{z^2}B(x_i/z)F(z) \right] \nonumber \\
& & + \Big[ C + A(1) \log (1-x_i) \Big] F(x_i). \nonumber
\end{eqnarray}
Decomposing the integrals above as
$
\int_{x_i}^1 = \sum_{k=i}^{N-1} \int_{x_k}^{x_{k+1}},
$
and using the approximation (\ref{eq:Fapprox}), we find
\begin{eqnarray}
I_{\big|x=x_i} & \approx & \sum_{\ell=1}^m a_\ell^{(i)} \left( \beta_\ell^{(i)} + \rho_{i\ell}^{(i)} \right) + \sum_{k=i+1}^{N-1} \sum_{\ell=1}^m a_\ell^{(k)} \left( \gamma_{k\ell}^{(i)} + \rho_{k\ell}^{(i)} \right) \\
& + & \Big[ C + A(1) \log (1-x_i) -A(1) \sigma^i \Big] F(x_i) \nonumber,
\end{eqnarray}
where
\begin{eqnarray}
\beta_\ell^{(i)} & \equiv & x_i \int_{x_i}^{x_{i+1}} \frac{dz}{z} \frac{A(x_i/z)g_\ell(z)-A(1) g_\ell(x_i)}{z-x_i} \label{eq:Pcoeff} \\
\rho_{k\ell}^{(i)} & \equiv & x_i \int_{x_k}^{x_{k+1}} \frac{dz}{z^2} B\left(\frac{x_i}{z}\right) g_\ell(z) \nonumber \\
\gamma_{k\ell}^{(i)} & \equiv & x_i \int_{x_k}^{x_{k+1}} \frac{dz}{z} \frac{A(x_i/z)g_\ell(z)}{z-x_i} \nonumber \\
\sigma^{(i)} & \equiv & x_i \int_{x_{i+1}}^{1} \frac{dz}{z} \frac{1}{z-x_i} \nonumber.
\end{eqnarray}
Substituting here the coefficients $a_\ell^{(k)}$ from Eq.~(\ref{eq:acoefficients}), the integral $I_{\big|x=x_i}$ can be written as
\begin{equation}
I_{\big|x=x_i} = \int_{x_i}^1 d\xi P(\xi) F(\frac{x_i}{\xi}) = \sum_{k=0}^N P_{ik} F(x_k),
\end{equation}
where the entries of the matrix $P_{ik}$ can be computed from
\begin{eqnarray}
P_{ik} & \equiv & \sum_{r=1}^4 \sum_{\ell=1}^4 \left[ G_{\ell r}^{(i)} \left( \beta_\ell^{(i)} + \rho_{i\ell}^{(i)} \right) \right]_{\Big| k=i+r-2} \\
& + & \sum_{r=1}^4 \sum_{\ell=1}^4 \sum_{n=i+1}^{N-1} \left[ G_{\ell r}^{(n)} \left( \gamma_{n\ell}^{(i)} + \rho_{n\ell}^{(i)} \right) \right]_{\Big| k=n+r-2} \nonumber \\
& + & \Big[ C + A(1) \log (1-x_i) - A(1) \sigma^i \Big]_{\Big| k=i}. \nonumber
\end{eqnarray}
In this way, the splitting functions $P$ and consequently also the functions $\Omega^{(n)}$ and $M^{(n)}$ derived from those, become matrices and the multi-dimensional integrals reduces to mundane matrix multiplication
\begin{equation}
\hat q_v(x_i,t) = \left[ \sum_{k=0}^{\infty} \sum_{n=0}^{N} \frac{t^k}{k!} M^{(k)}_{in} \right] \hat \otimes \, \hat q_v(x_n,0), \label{eq:evomatrix}
\end{equation}
where the matrices $M$ do not depend on the form of the parton density, but only on the $x$-grid, and the initial scale $Q_0^2$. That is, they can be computed once and for all. The hard manual work is to analytically evaluate the integrals in Eq.~(\ref{eq:Pcoeff}) for all splitting functions. At leading order the expressions are still reasonable to compute even by hand, but already at NLO-level the expressions become long --- involving special functions like polylogarithms and Riemann zeta functions --- and use of a symbolic computer program like {\ttfamily Mathematica} is, in practice, mandatory.

\section{Numerical test of the method}

\begin{figure}[!h]
 \includegraphics[scale=0.38]{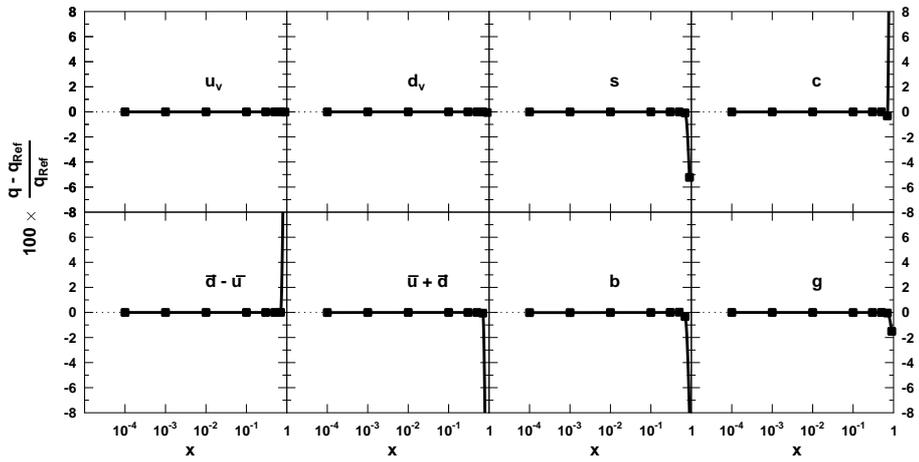}
\caption{Comparison to the benchmark NLO partons \cite{Giele:2002hx} for valence quarks, sea quarks, and gluons.}
\label{Fig:PDFcomp}
\end{figure}

Using the NLO splitting functions given in \cite{Ellis:1991qj} (and also the leading-order ones discussed in Chapter \ref{DGLAPevolution}), I have constructed a Fortran code for calculating the evolution with the method described above. In order to verify the accuracy of the method, I have tested it against the benchmark values of parton distribution functions given in Ref.~\cite{Giele:2002hx}. This reference contains carefully cross-checked results for evolution of a given initial parametrization of partons from $2 \, {\rm GeV}^2$ up to $10000 \, {\rm GeV}^2$ with an unambiguously defined $\alpha_s$. For doing the comparison, I have constructed an $x$-grid from $x=10^{-4}$ to $x=1$ with 100 logarithmic intervals in a range $10^{-4}\ldots 10^{-1}$ and 100 linear intervals in a range $10^{-1}\ldots 1$. I have truncated the Taylor expansion to include the first 9 terms. The results for specific combinations of partons are displayed in Fig.~\ref{Fig:PDFcomp} as a relative difference to the benchmark partons in percents. 
The evidently excellent agreement with the benchmark partons proves the applied integration method accurate as well as that already the 9th order Taylor expansion seems to convergence nicely. Only at very large $x$, where the parton densities are numerically very small --- typically falling off like $(1-x)^\alpha$, $\alpha \gtrsim 3$ --- a third-order polynomial does not optimally fit the input densities and a higher order expansion would be needed. However, for the purposes of our global PDF analyses the very large-$x$ region is rather irrelevant for all other parton types except maybe the valence quarks which are numerically larger.

%% file: global.tex
\chapter{About global QCD analyses}
\label{AboutGlobal}

Within the pQCD-improved parton model, the hadronic cross-sections for hard scattering processes can be calculated through the factorization theorem folding the universal, scale-dependent PDFs $f_i(x,Q^2_f)$ with the perturbatively calculable pieces $d\hat{\sigma}$, formally
\begin{eqnarray}
d\sigma^{A + B \rightarrow c + X} & = & \sum_{i,j=q,\overline{q},g} f_i^A(Q^2_f) \otimes f_j^B(Q^2_f) \otimes d\hat{\sigma}^{ij\rightarrow c + X}(\mu^2,Q^2_f). \label{eq:fac}
\end{eqnarray}
The physical content of this expression has been discussed in the preceding chapters. Thus, experimental measurements provide information about the PDFs $f_i(Q^2_f)$ as well as about the underlying parton dynamics $d\hat{\sigma}$. This is, in short, the central idea behind the global QCD-analyses of PDFs. Here the word ``global'' is related to the universality-hypothesis of the PDFs i.e. their process-independence: As much experimental data as possible should be considered simultaneously to find whether this is really true --- can they all be described by same set of PDFs. If not, it may be a sign of factorization breakdown or perhaps discovering new physics beyond the Standard Model. However, due to the enormous complexity of the present-day accelerator-based experiments, one should also be cautious of not being misled by e.g. fluctuations in the data that might not respect any textbook statistics. It should be emphasized that the global analyses do not only constrain the PDFs, but also various fundamental parameters like the strong coupling $\alpha_s$, the heavy quark masses, and even the elements of the CKM-matrix.

\vspace{0.5cm}
The modern global analyses employing data from several free proton experiments was ushered in by works of Morfin and Tung \cite{Morfin:1990ck}\footnote{Sadly, Wu-Ki Tung passed away during the writing of this thesis in March 2009.}, triggering an enormous effort which today demonstrate a huge success with continuously increasing amount of data accommodated in the analysis. The leading groups in this domain are nowadays the CTEQ \cite{Nadolsky:2008zw}, the MSTW \cite{Martin:2009iq}, and the NNPDF \cite{Ball:2008by} collaborations, but various other parties like the HERAPDF consortium (see e.g. \cite{Dittmar:2009ii}) which often focus only on a more restricted data input, exist.

\vspace{0.5cm}
It is well known that when the cross-sections measured with \emph{nuclear} targets are compared to the free proton results, the two are not identical, but various \emph{nuclear effects} are observed \cite{Arneodo:1992wf,Geesaman:1995yd,Norton:2003cb}. Although, the QCD factorization is not as well-grounded theorem in the case of bound nucleons (see e.g. Ref. \cite{Armesto:2006ph}), the pioneering work \cite{Eskola:1998iy,Eskola:1998df} and subsequent analyses like \cite{Hirai:2007sx,deFlorian:2003qf}, and especially the article \cite{EPS09-paperi} of this thesis, have nevertheless revealed that such conjecture holds to a very good precision in describing the world data from deep inelastic scattering and Drell-Yan dilepton measurements. In effect, only the shapes of the PDFs are modified by the presence of the nuclear environment. In other words, although the strong, non-perturbative nuclear binding has an effect on the quark-gluon structure of bound nucleons, the partons at high $Q^2$ appear to largely obey the same QCD dynamics as do their free counterparts.

\vspace{0.5cm}
In what follows, I will describe some general features of global QCD-analyses paying special attention to the nuclear PDFs in the light of the publications of this thesis. I will keep the discussion here quite condensed, yet logical. For much more pedantic description about the free proton global fits with comprehensive reference list, consult e.g. the very profound MSTW paper \cite{Martin:2009iq}. Also, the lecture notes from the series of Summer Schools ran by the CTEQ collaboration are an inexhaustible source of pedagogic up-to-date material. Much more technicalities about the nuclear PDF fits can be found in the publications of this thesis \cite{EKPS-paperi,EPS08-paperi,EPS09-paperi}.

\section{Choice of experimental data}

As mentioned above, the guideline in a global QCD-analysis is to keep it really global, i.e. include as many different types of data as possible --- ideally all. In practice, however, it is necessary to somewhat restrict what data is accepted and what is not.

\vspace{0.3cm}
One obvious restriction is that the factorization framework is only applicable when the process is ``hard'' i.e. the invariant scale $Q^2$ inherent for the whatever process is large $Q^2 \gg \Lambda^2_{\rm QCD}$. For example, in deeply inelastic scattering typical kinematical cuts are
$$
Q^2 \ge 4 \, {\rm GeV}^2, \qquad M_X^2 = (P+q)^2 = M^2+Q^2\frac{1-x}{x} \ge 12 \, {\rm GeV}^2,
$$ 
where the latter condition is to keep away from the resonances. Beyond such limits it may be necessary to account also for the higher-twist $Q^{-2n}$ contributions (e.g. by parametrizing them as they are usually poorly known), as well as target-mass corrections \cite{Schienbein:2007gr,Accardi:2008ne}.

\vspace{0.5cm}
It sometimes happens that independent data sets are not compatible with each other. In a case where there are several measurements for the same observable and only one data set disagrees with the rest, it may be possible to rule out the one measurement as being ``wrong'' or ``not fully understood'', and concentrate solely on the others. However, when there are not too many independent measurements, it may be necessary to completely abandon that type of data for safety i.e. for not being too biased by subjective choices. An example of this kind of issue has been the direct photon production \cite{Huston:1995vb,Aurenche:1998gv,Aurenche:2006vj} which is nowadays not included in the latest free-proton PDF analyses despite its potential ability to constrain gluons. However, due to complexity of the modern collider experiments, small mutual inconsistencies between independent data sets are more a rule than an exception. This is eventually reflected in the PDF uncertainty analysis, necessitating various extensions to the strict rules of ideal statistics.

\vspace{0.5cm}
Typical processes employed in the free-proton PDF analyses include
\begin{itemize}
\item Deeply inelastic scattering related measurements
\item Drell-Yan dilepton production
\item Rapidity distributions in heavy boson ($W^\pm$ and $Z$) production
\item Jet measurements
\end{itemize}
The sensitivity of these data types for different PDF-components is extensively documented e.g. in \cite{Martin:2009iq}, and I will not go to details here.

\subsubsection{Bound protons}

\begin{figure}[h!]
\begin{center}
\hspace{-1.5cm}\includegraphics[scale=0.40]{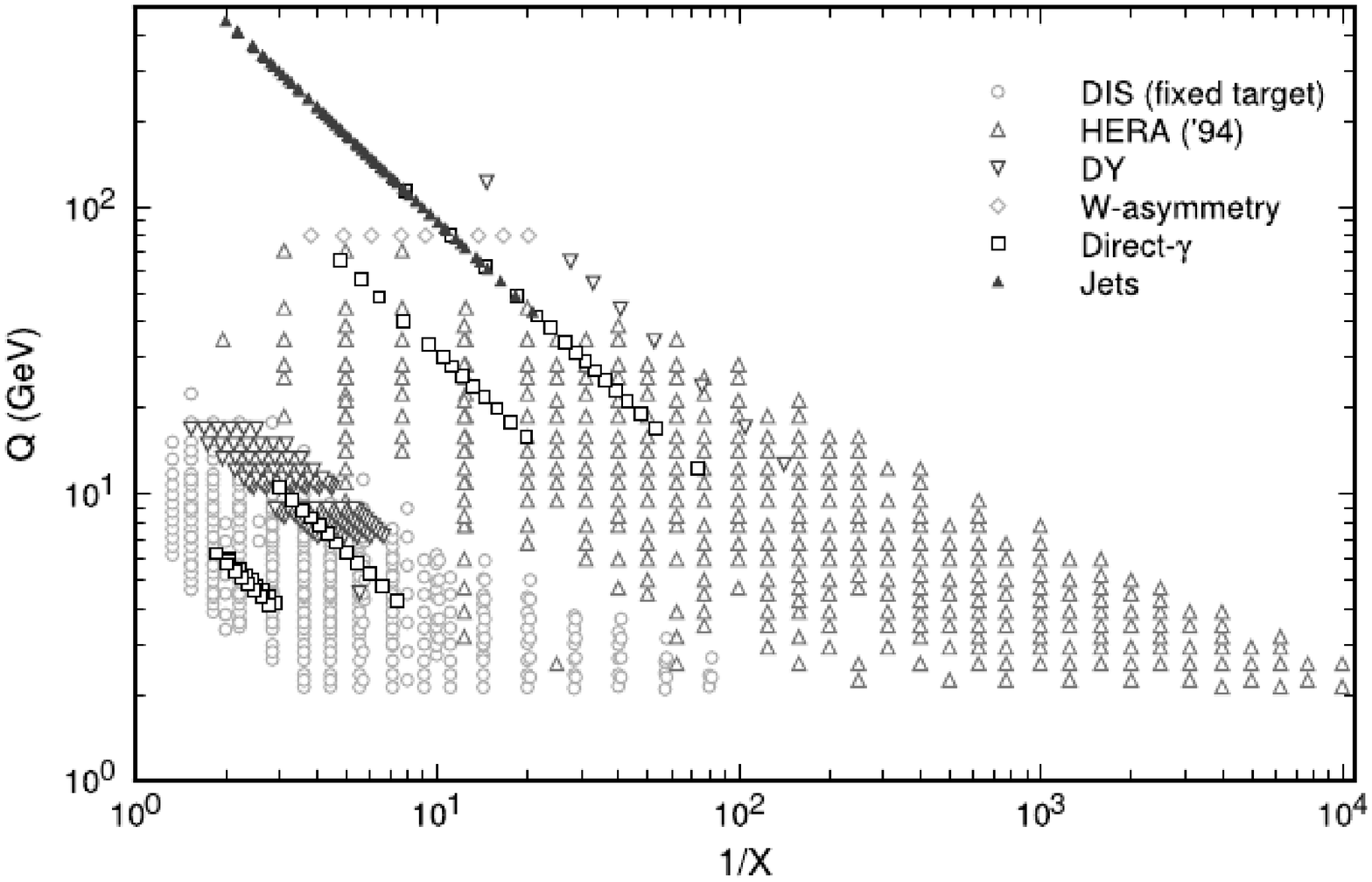}
\includegraphics[scale=0.70]{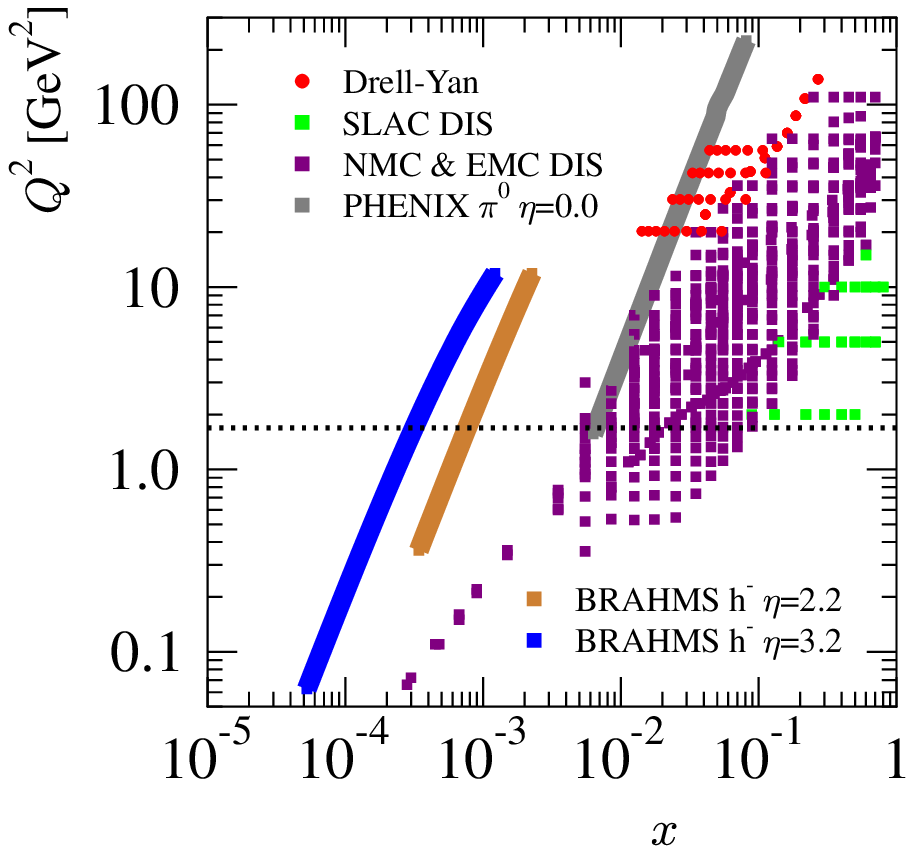}
\end{center}
\caption{Kinematical reach of data employed in typical global analyses of the free proton (upper panel) \cite{Pumplin:2003ek} and nuclear (lower panel) PDFs. On the upper panel, the kinematical cuts have been already applied, whereas on the lower panel the dotted line denote the cut imposed in \cite{EKPS-paperi,EPS08-paperi,EPS09-paperi} for the DIS data.}
\label{fig:xQdata}
\end{figure}

In the case of the nuclear PDF studies the variety, amount and kinematical reach of the available data is much smaller, see Fig.~\ref{fig:xQdata}. 
For this reason, typical nuclear PDF studies have so far adopted the standpoint that the free-proton PDFs are taken as fixed, fully known, and only the nuclear modifications suggested by the data are inferred. The data conventionally utilized in the nuclear PDF studies come not as absolute cross-sections, but as ratios of nuclear cross-sections to the free-nucleon ones. In such quantities, things like electroweak corrections and higher-twist contributions can be expected to largely cancel, and consequently it is reasonable to push the kinematical cuts little further down than is usually done if absolute cross-sections are in question. For several years, the only available precision data for nuclear PDF analyses were from deeply inelastic scattering and Drell-Yan dilepton production, as shown by the dots in Fig.~\ref{fig:xQdata}, but those data alone could not constrain the nuclear gluon PDFs very convincingly. Indeed, the usual procedure has been to fix a large part of the nuclear modifications for gluons by hand invoking reasoning based on physical intuition. To relax the need for such assumptions, our latest global analysis in the nuclear sector \cite{EPS09-paperi} includes data from inclusive $\pi^0$-production (the gray line in Fig.~\ref{fig:xQdata}) measured recently at BNL-RHIC \cite{Adler:2006wg}. These data are not yet very precise, but provide a handle to some extent improve the determination of the nuclear gluons.

\section{The traditional approach}

The conventional procedure of global analyses can be summarized as follows:

\subsubsection{A.}

\begin{figure}[h!]
\begin{center}
\includegraphics[scale=0.7]{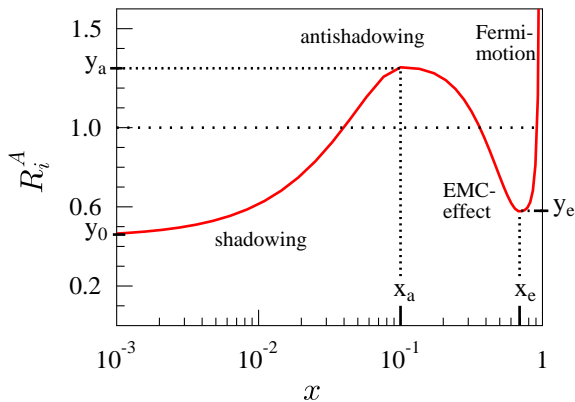}
\end{center}
\caption{Schematic picture of the nuclear modifications to PDFs.}
\label{fig:Rexample}
\end{figure}

{\bf The PDFs are first parametrized at a chosen initial scale $Q_0^2$ imposing the sum rules.} For absolute free-proton PDFs the functional form of the parametrization is typically
$$
f_{i}(x,Q_0^2) = x^{-\alpha} (1-x)^\beta F(x),
$$
where the function $F(x)$ varies from one analysis to another. Ideally, this function should be as flexible as possible, but too much freedom may induce unphysical, completely artificial features to the PDFs. At the end, how much complexity one should build in $F(x)$, depends on the diversity and accuracy of the input data. In the nuclear PDF analyses, usually the \emph{nuclear modification factors} $R_i^A(x,Q_0^2)$ encoding the relative difference to the free proton PDFs are parametrized. A typical parametrization indicating which $x$-regions are meant by the commonly used terms: shadowing, antishadowing, EMC-effect, and Fermi-motion, is shown in Fig.~\ref{fig:Rexample}.
For example, in \cite{EKPS-paperi,EPS08-paperi,EPS09-paperi} we define the bound proton PDFs by
\begin{equation}
f_{i}^A(x,Q^2) \equiv R_{i}^A(x,Q^2) f_{i}^{p}(x,Q^2),
\label{eq:partondefinition}
\end{equation}
for each parton flavor $i$. Above $f_{i}^{p}(x,Q^2)$ refers to a fixed set of the free-proton PDFs, and we consider three different modification factors: $R_V^A(x,Q_0^2)$ for both $u$ and $d$ valence quarks, $R_S^A(x,Q_0^2)$ for all sea quarks, and $R_G^A(x,Q_0^2)$ for gluons. Also their $A$-dependence is parametrized but the nuclear corrections for the Deuteron ($A=2$) which are expected to be of order 1..2\% \cite{Hirai:2007sx,Badelek:1994qg}, are neglected.

In principle there is no real physics reason for assuming flavor-independent nuclear modification factors for valence and sea quarks at $Q_0^2$.  Unfortunately, the presently available experimental data does not allow to determine such flavor dependence. However, as was shown in \cite{NuTeV-paperi}, a difference between $R_{d_V}^A(x,Q_0^2)$ and $R_{u_V}^A(x,Q_0^2)$ would affect the extraction of the Weinberg weak-mixing angle $\sin^2 \theta_{\rm W}$ from neutrino scattering off an Iron target. Concretely, such undertaking was done by the Fermilab NuTeV collaboration which measured the Paschos-Wolfenstein ratio \cite{Paschos:1972kj}
\begin{equation}
R^-_{\rm Fe}(x,Q^2) \equiv \frac{ d\sigma_{\rm Fe}^{\nu, {\rm NC}}/dxdQ^2 - d\sigma_{\rm Fe}^{\overline{\nu}, {\rm NC}}/dxdQ^2}
{ d\sigma_{\rm Fe}^{\nu, {\rm CC}}/dxdQ^2 - d\sigma_{\rm Fe}^{\overline{\nu}, {\rm CC}}/dxdQ^2} \approx \frac{1}{2} - \sin^2 \theta_{\rm W},
\label{PWint}
\end{equation}
where NC and CC refer to the charged ($W^\pm$ mediated) and neutral ($Z$ mediated) current processes. Surprisingly, the $\sin^2 \theta_{\rm W}$ found, lied three standard deviations away from the world average --- an observation known as the \emph{NuTeV anomaly} \cite{Zeller:2001hh}. Among others sources possible, the simple relation above receives a correction from non-isoscalarity of Iron which is of the form
\begin{equation}
R^-_{\rm Fe}(x,Q^2) \approx \left( \frac{1}{2} - \sin^2 \theta_{\rm W} \right) \left[1 + f\left(\frac{A-2Z}{A}\frac{u_V^A-d_V^A}{u_V^A+d_V^A}\right) \right],
\end{equation}
where $f$ is a certain function, and $u_V^A,d_V^A$ are the average up and down valence-quark distributions in a bound nucleon. Thus, a difference between $R_{d_V}^{\rm Fe}$ and $R_{u_V}^{\rm Fe}$ would affect the NuTeV analysis. As found in \cite{NuTeV-paperi}, such difference can be quite large without really showing up in the global analyses and could very naturally reduce the anomaly by one standard deviation or so. There is, however, also correction originating from an asymmetric strange sea, $\int dxx[s(x)-\overline{s}(x)] \neq 0$, which is not very well constrained either and which could explain even the whole NuTeV anomaly alone \cite{RojoJuanDIS2009}.

\subsubsection{B.}

{\bf The absolute PDFs are evolved from the parametrization scale $Q_0^2$ to other perturbative scales $Q^2 > Q_0^2$ by the DGLAP equations.} On the practical level, an efficient numerical solver for the parton evolution is critical for a successful global analysis with a reasonable computing time. One, working, solution was discussed in Chapter 4. Another solution would be to make a shortcut and use ready-made package like QCDNUM \cite{QCDNUM}. Under the evolution, also the nuclear modification factors $R_i^A(x,Q^2)$ become scale-dependent, and the initial flavour-independence, if it was assumed, usually disappears. Significant differences between e.g. $R_{d_V}^{A}$ and $R_{u_V}^{A}$ do not, however, seem to build up if they start from the same initial condition. 

\subsubsection{C.}

{\bf The cross-sections are computed using the factorization theorem.} As became explicitly demonstrated for deeply inelastic scattering (Chapter \ref{Deeply_inelastic_scattering_at_NLO}) and Drell-Yan dilepton production (Chapter \ref{DrellYanNLO}), the numerical evaluation of the NLO cross-sections is sometimes quite demanding, involving multiple integrations. Consequently increased computing time is a potential ``killer'' for a global analysis beyond leading order. For example, Monte-Carlo integrations for inclusive jet production NLO cross-sections, with various experimental cuts implemented, would simply be too slow to be always recalculated at every round of parameter iteration. The simplest solution is to calculate so-called ``K-factors'', $\sigma_{\rm NLO}/\sigma_{\rm LO}$, for all data points with an educated guess for the NLO PDFs. Then, one can resort to computing only leading order cross-section, which, when multiplied by such pre-computed factor, serves as an approximation to the NLO one. However, tricks similar as used in Chapter~\ref{SolvingDGLAP} to reduce an integration to a matrix multiplication, are also viable \cite{Kluge:2006xs,Carli:2005ji}. A special feature of the nuclear PDF studies which take the free-proton PDFs fixed, is that sometimes it is possible to perform some integrals also beforehand, see Appendix of Ref. \cite{EPS09-paperi}.

\subsubsection{D.}

{\bf The computed cross-sections are compared to the experimental ones, and the parametrization is varied until the best agreement with the data is reached.} What is meant by ``best agreement'' is, to some extent, a matter of convention. An elementary solution is to tune the parameters by trial and error to establish a parametrization that simply ``looks good'' \cite{Eskola:1998iy} when contrasted with the data. However, nowadays such procedure is replaced by more algorithmic methods based on minimization of a $\chi^2$-function, defined e.g. as in \cite{EPS08-paperi,EPS09-paperi}
\begin{eqnarray}
\chi^2(\{a\})   & \equiv & \sum _N w_N \, \chi^2_N(\{a\}) 
\label{eq:chi2mod_1}
\\	
\chi^2_N(\{a\}) & \equiv & \left( \frac{1-f_N}{\delta_N^{\rm norm}} \right)^2 + \sum_{i \in N}
\left[\frac{ f_N D_i - T_i(\{a\})}{\delta_i}\right]^2.
\label{eq:chi2}
\end{eqnarray}
Within each data set labeled by $N$, $D_i$ denotes the experimental data value with $\delta_i$ point-to-point uncertainty, and $T_i$ is the theory prediction corresponding to a parameter set $\{a\}$.  The weight factors $w_N$ are used to amplify the importance of those data sets whose content is physically relevant, but contribution to $\chi^2$ would otherwize be too small to be noticed by an automated minimization.

In certain cases, an overall relative normalization uncertainty $\delta_N^{\rm norm}$ is specified by the experiment. The normalization factor $f_N \in [1-\delta_N^{\rm norm},1+\delta_N^{\rm norm}]$ is introduced to account for such cases. For each set of fit parameters $\{a\}$, its value is solved by minimizing $\chi^2_N$ making thus the final $f_N$ not something that is put in by hand, but really an output of the analysis. This procedure gives a possibility to reproduce the \emph{shape} (relative magnitude) of the data thus capturing the relevant PDF-physics, and not be mislead by problems with, sometimes model-dependent, normalization. Apart from the experimental normalization uncertainty, there may also be unknown contributions --- like electroweak corrections --- to the absolute computed cross-sections having nothing to do with the PDFs. Thus, there is also a theoretical call for such tunable normalization factor.

Although it may sound straightforward, finding the minimum $\chi^2$ in practice is a non-trivial task as the $\chi^2$ is a highly non-linear functional of the fit parameters. An efficient convergence of a minimization algorithm often relies on the knowledge of the gradient terms and the second-derivative matrix at the given location in the space of fit parameters. Due to the finite accuracy of the DGLAP-solver and numerical multi-dimensional integrations, the $\chi^2$ becomes actually non-continuous at very small parameter intervals, which makes the reliable derivative calculations complicated. Consequently, general-purpose packages like \texttt{MINUIT} \cite{James:1975dr} become easily insufficient for harsh requirements of global analyses and tailor-made add-ons or independent minimizing routines are often needed.

\section{Treatment of heavy flavors}

The calculations of the partonic cross-sections $\hat \sigma$ are enormously simplified by treating all quarks as massless particles. This is perfectly fine if the invariant scales in the cross-sections are much larger than the heavy quark masses (only charm and bottom are relevant to present global analyses). However, rather than completely forgetting the heavy flavor masses a following prescription to feed them in is often adopted:
\begin{itemize}
\item 
The parton distribution for the heavy quark $q_H(x,Q^2)$ remains zero if $Q^2 \le m_H^2$, but follows the DGLAP evolution when $Q^2 > m_H^2$. In other words, the number of flavors ($n_f$) is incremented by one every time a heavy-quark mass-threshold is crossed.

\item 
When the factorization scale $Q_f^2$ in a cross-section is below the heavy-quark mass-threshold, the calculation is performed as this quark flavor did not exist --- even if there is enough center-of-mass energy to physically produce it.
\end{itemize}

This treatment of heavy flavors is known as \emph{Zero-mass variable flavor number scheme} (ZM-VFNS). It is extremely simple but its shortcomings are also rather obvious. In any case, it has been the standard choice in the global analyses until only very recently, and it is also the one used in the publications of this thesis.

\vspace{0.5cm}
Another extreme is to work harder with the partonic cross-sections retaining the full heavy quark mass-dependence and never consider the heavy quarks as partons i.e. keep the number of active flavors fixed in the DGLAP equations. Such scheme is known as \emph{Fixed flavor number scheme} (FFNS). The problem in this scheme is its limited domain of applicability: the partonic cross-section contain terms $\sim \log(Q^2/m_H^2)$ which become unstable at large $Q^2$, where only resummation of the large logarithms would bring the perturbative expansion under control.

\vspace{0.5cm}
The class of schemes that combine the advantages and avoid the shortcomings of both ZM-VFNS and FFNS schemes are called \emph{General-mass variable flavor number schemes} (GM-VFNS). In short, these are chains of FFNS-type of schemes with specific matching conditions. For a pedagogical review, see \cite{Thorne:2008xf}. These have now become the standard in the most comprehensive QCD analyses and providing a clearly improved fit to the data \cite{Nadolsky:2009ge}. It is also possible to construct something called \emph{Intermediate-mass variable flavor number schemes} \cite{Nadolsky:2009ge}, which preserve the simplicity of the ZM scheme, but mimic the full GM scheme by implementing its most relevant kinematical effects. 

\section{Uncertainty analysis}

Finding only the single set of parameters $\{a_0\}$ that optimally fits the experimental data in the sense of giving the minimum $\chi^2$, does not, however, alone represent everything that can be learned about the PDFs. Due to the experimental uncertainties and fluctuations in the data, the partons obtained by steering the fit parameters slightly off from the $\chi^2$-minimum, cannot be right away ruled out as being ``completely wrong''. Quantifying the uncertainties of such origin has become an increasingly important topic in the global PDF analyses. For example, the effects of such uncertainties in the predictions for the LHC ``standard candles'', ${\rm W}^\pm$ and Z production in pp-collisions, are of the order of few percents \cite{Nadolsky:2008zw,Martin:2009iq}. Thus, measuring eventually something which is like 10\% apart from these predictions would be very interesting as such large discrepancy would be difficult to fix by simply re-fitting the PDFs. Several methods for performing the PDF uncertainty analysis exist. Some common ones and the ideas behind them are:

\subsubsection{Lagrange multiplier method \cite{Stump:2001gu}}


Defining
$$
\Psi(\lambda,\{a\}) \equiv \chi^2(\{a\}) + \lambda X(\{a\}),
$$
and minimizing $\Psi(\lambda,\{a\})$ for several fixed values of $\lambda$, gives a sequence of $(\chi^2_\lambda,X_\lambda)$-pairs, corresponding to a best $\chi^2$ that can be achieved with $X$ taking a specific value $X_\lambda$. These pairs comprise a $\chi^2$-profile as a function of $X$ with minimum at $X=X_0=X(\{a_0\})$. By restricting the allowed growth of $\chi^2$ above the minimum $\chi^2_0$, the achievable range for $X$ becomes mapped out. This is a robust procedure, but has a limited applicability as finding the uncertainty for one single PDF-related quantity $X$ requires an ability to perform several global fits.
 
\subsubsection{Monte Carlo technique \cite{Dittmar:2009ii}}
 

The principle is to prepare multitude replica of the original cross-sections $\sigma_i$ by transformation
$$
\sigma_i \rightarrow \sigma_i \left( 1 + \delta_i \, R_i \right),
$$
where $\delta_i$ denotes the experimental uncertainty and $R_i$ is a random number drawn from a Gaussian distribution centered at 0. Performing a fit to each of the prepared replica gives a corresponding set of PDFs. The uncertainty in any quantity $X$ is then estimated from the spread of the individual predictions computed by these sets. Although simple and easy to implement, the problem is that in order to get enough statistics, hundreds of separate PDF fits are required, but still nothing guarantees that the all relevant possibilities are covered \cite{Pumplin:2009nk}.

\subsubsection{The Hessian method \cite{Pumplin:2001ct}}


This method is superior in its usefulness in various applications. As explained in detail in \cite{EPS09-paperi}, it rests on expanding the $\chi^2$ around its minimum $\chi^2_0$ as
\begin{equation}
\chi^2 \approx \chi^2_0 + \sum_{ij} \frac{1}{2} \frac{\partial^2 \chi^2}{\partial a_i \partial a_j} (a_i-a_i^0)(a_j-a_j^0) \equiv \chi^2_0 +  \sum_{ij} H_{ij}(a_i-a_i^0)(a_j-a_j^0),
\label{eq:chi_2approx}
\end{equation}
which defines the Hessian matrix $H$. The non-zero off-diagonal elements in the Hessian matrix are a sign of correlations between the original fit parameters, invalidating the usual error propagation
\begin{equation}
(\Delta X)^2 = \left( \frac{\partial X}{\partial a_1} \cdot \delta a_1 \right)^2 + \left( \frac{\partial X}{\partial a_2} \cdot \delta a_2 \right)^2 + \cdots
\end{equation}
for a PDF-dependent quantity $X$. Therefore, it is useful to diagonalize the Hessian matrix, such that
\begin{equation}
\chi^2 \approx  \chi^2_0 +  \sum_{i} z_i^2,
\label{eq:chi_2diag}
\end{equation}
where each $z_i$ is a certain linear combination of the original parameters around $\{a^0\}$. In these variables, the usual form of the error propagation
\begin{equation}
(\Delta X)^2 = \left( \frac{\partial X}{\partial z_1} \cdot \delta z_1 \right)^2 + \left( \frac{\partial X}{\partial z_2} \cdot \delta z_2 \right)^2 + \cdots
\label{eq:error_better}
\end{equation}
is justified. The practicality of the Hessian method resides in constructing so-called PDF error sets $S_k^\pm$ which are obtained by displacing the fit parameters to the negative/positive direction along each $z_k$ separately such that $\chi^2$ grows by a certain amount $\Delta \chi^2$. Approximating the derivatives in Eq.~(\ref{eq:error_better}) by finite differences, the error formula can be re-written e.g. as
\begin{equation}
(\Delta X)^2 = \frac{1}{4} \sum_k \left[ X(S^+_k)-X(S^-_k) \right]^2,
\label{eq:error_best}
\end{equation}
where $X(S^\pm_k)$ denotes the value of the quantity $X$ computed with the PDF error set $S_k^\pm$. 

The methods above require to specify how much the $\chi^2$ is allowed to grow\footnote{In the Monte Carlo procedure similar role is played by the width of the normal distribution.} above its minimum value $\chi^2_0$, i.e. what is $\Delta \chi^2$. To determine such range is, however, far from being a straightforward exercise. On the contrary, it is a problematic and much debated issue with no universally agreed procedure. As the spirit of the PDF uncertainty analysis is not so much to find statistically ideal answers, but more to map out the \emph{physically relevant} range for the partons, choosing the statistically ideal one-sigma 68\% confidence criterion $\Delta \chi^2 = 1$ \cite{Amsler:2008zzb} would not make much sense. The basic reason being, as mentioned earlier, that in a truly global analysis the data sets seldomly demostrate a perfect mutual harmony, but small inconsistensies tend to exist. From the practical point of view, if the absolute minimum $\chi^2$ is $\sim 1000$ (as in the NLO analysis of this thesis \cite{EPS09-paperi}), it would be very naive to rule out partons which increase $\chi^2$ by only one unit. Instead, adopting a prescription that requires each data set to remain roughly within its 90\%-confidence range \cite{Stump:2001gu,Pumplin:2001ct,EPS09-paperi}, we find $\Delta \chi^2 = 50$ much better motivated. 


\vspace{0.5cm}
Apart from the experimental uncertainties and their treatment described shortly above, there are also other sources of uncertainties. One issue is the adopted form of the parametrization which may introduce a bias and make the uncertainty bands too narrow in the region which is not covered by the data. Such issue can be addressed in a neural network approach, which tries to get rid of the fit-function bias \cite{Ball:2008by}. Although providing a very flexible way to parametrize the PDFs, it has been criticized as being \emph{too} flexible: There are often physical reasons for picking a fit function of a particular form, and too much freedom may lead to unphysical behaviour of PDFs. This is an example that falls to the category of ``Theoretical uncertainties'' \cite{Martin:2003sk} among other things like unknown higher-order pQCD corrections, resummations near the phase space boundaries, higher-twist corrections, and absorptive effects in parton evolution. The size of such uncertainties is often impossible to quantify in practice.

\chapter{Status of global nuclear PDF analyses}

Finally, I review the main results from the nuclear PDF analysis of this thesis, EPS09 \cite{EPS09-paperi}, which is an NLO successor to the pioneering leading-order work EKS98 \cite{Eskola:1998iy,Eskola:1998df}, and articles \cite{EKPS-paperi} and \cite{EPS08-paperi} of this thesis.

\section{Nuclear modifications at LO and NLO}

\begin{figure}[!h]
\center
\includegraphics[scale=0.58]{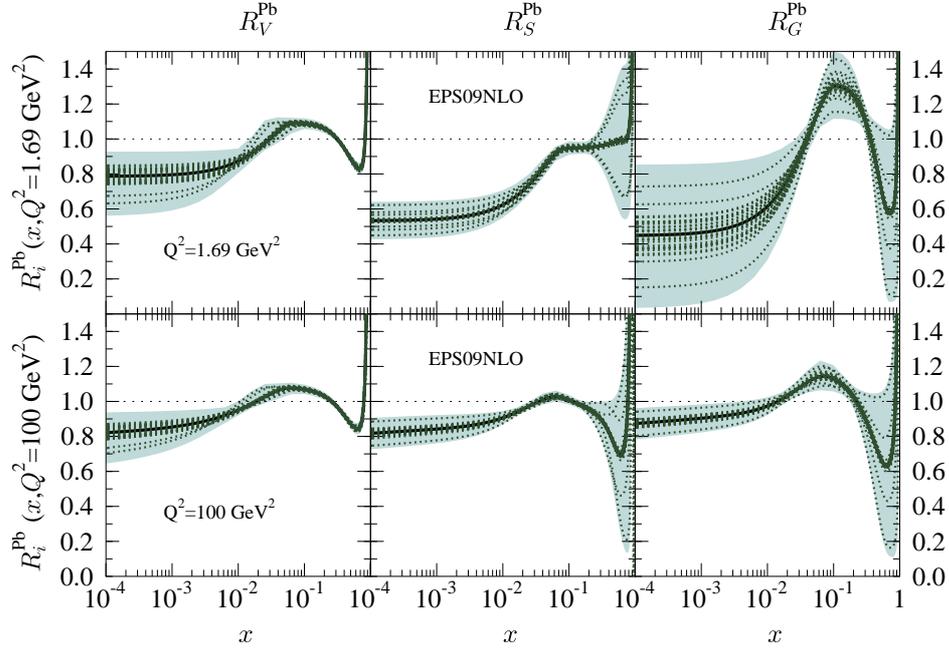}
\caption[]{The nuclear modifications $R_V$, $R_S$, $R_G$ for Lead at an initial scale $Q^2_0=1.69 \, {\rm GeV}^2$ and at $Q^2=100 \, {\rm GeV}^2$. The thick black lines indicate the best-fit results, whereas the dotted green curves denote the individual error sets. The shaded bands are the total uncertainty, computed according to equation close to Eq.~\ref{eq:error_best}.}
\label{Fig:PbPDFs}
\end{figure}

As mentioned, the goal of these studies is to test the QCD factorization and find the process-independent nuclear modifications to the free proton PDFs. The results from EPS09NLO are shown in Fig.~\ref{Fig:PbPDFs} for Lead.
\begin{figure}[!h]
\center
\includegraphics[scale=0.55]{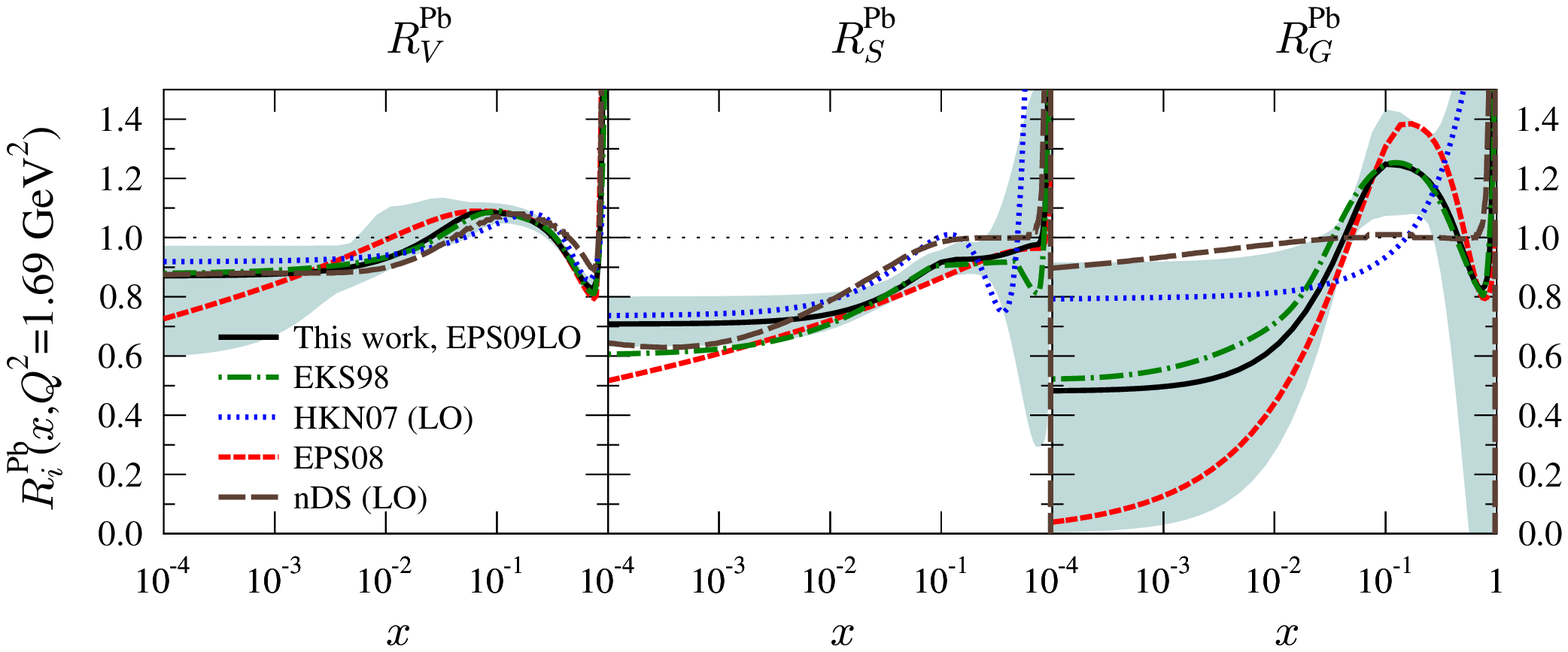}
\caption[]{Comparison of the valence and sea quark, and gluon modifications at $Q^2 = 1.69 \, {\rm GeV}^2$ for Pb nucleus from the leading-order global analyses EKS98 \cite{Eskola:1998iy,Eskola:1998df}, EKPS \cite{EKPS-paperi}, nDS~\cite{deFlorian:2003qf}, HKN07~\cite{Hirai:2007sx}, EPS09LO \cite{EPS08-paperi}, and EPS09LO \cite{EPS09-paperi}.}
\label{Fig:LOnPDFs}
\end{figure}
I have plotted the obtained modifications at two scales, at $Q^2_0=1.69 \, {\rm GeV}^2$ and at $Q^2=100 \, {\rm GeV}^2$, in order to emphasize their scale-dependence. One prominent feature that becomes clearly conveyed by this figure is that even rather large uncertainty band for the initial small-$x$ gluon modification $R_G^A$, shrinks along the scale evolution quite a bit. This is a clear prediction of the factorization approach. The uncertainty band for the valence quarks at large $x$ becomes very small, only $\sim 2 \%$, thanks to large amount of deep inelastic data. However, as the uncertainty becomes so small, neglecting the nuclear effects in Deuteron might not be justified if the PDFs are defined as in (\ref{eq:partondefinition}). This is an example of the experimental uncertainties being small enough that the theoretical uncertainty may actually dominate.

\vspace{0.5cm}
The large uncertainty of the small-$x$ gluons is not very surprising. It was also investigated in our preceding leading order analysis EPS08 \cite{EPS08-paperi}, where we searched for the strongest possible gluon shadowing which was still in agreement with the DIS data, but only barely so. Satisfyingly, the result found in that analysis is very close to the lower uncertainty limit which we now find by the Hessian method in the leading-order version of EPS09 \cite{EPS09-paperi}. This is demonstrated in Fig.~\ref{Fig:LOnPDFs} which displays the low-$Q^2$ nuclear modifications for Lead from several leading-order analyses.

\section{Data vs. theory}

\subsection{Deeply inelastic scattering}

\begin{figure}[!h]
\center
\includegraphics[scale=0.40]{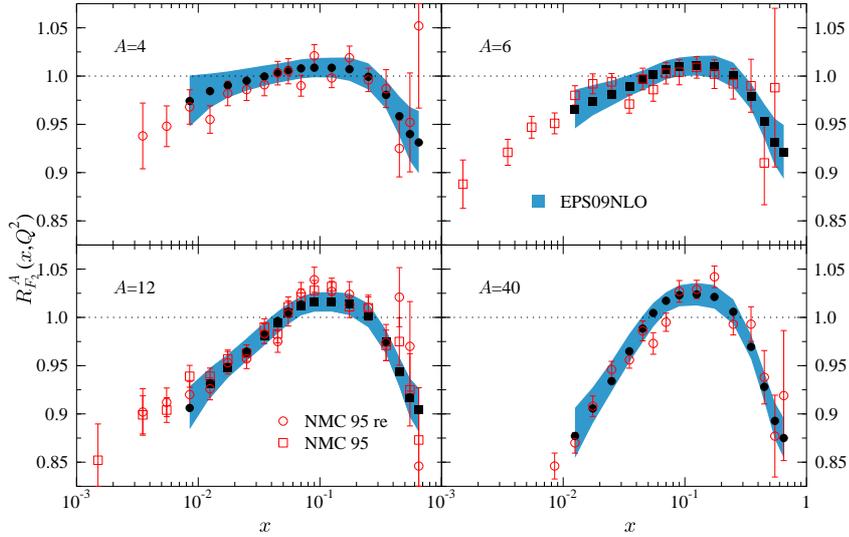}
\caption[]{The calculated NLO $R_{F_2}^A(x,Q^2)$ (filled symbols and error bands) compared with the NMC 95 \cite{Arneodo:1995cs} and the reanalysed NMC 95 \cite{Amaudruz:1995tq} data (open symbols).}
\label{Fig:RF2A1}
\end{figure}
The DIS data are the ``bread and butter'' of all global PDF analyses. In Fig.~\ref{Fig:RF2A1} I show a compilation of measured nuclear modifications for deep inelastic structure functions with respect to Deuterium
\begin{equation}
R_{F_2}^{\rm A}(x,Q^2) \equiv  \frac{F_2^A(x,Q^2)}{F_2^d(x,Q^2)}
\end{equation}
for various nuclei and compare them with the EPS09NLO parametrization. The shaded bands denote the uncertainty derived from the EPS09NLO and --- as I would like to emphasize --- their size is very similar to the error bars in the experimental data. This \emph{a posteriori} validates the quite large $\Delta \chi^2$ which we found as explained in detail in \cite{EPS09-paperi}.

\subsection{Drell-Yan dilepton production}

\begin{figure}[!h]
\center
\hspace{-1.0cm}
\includegraphics[scale=0.35]{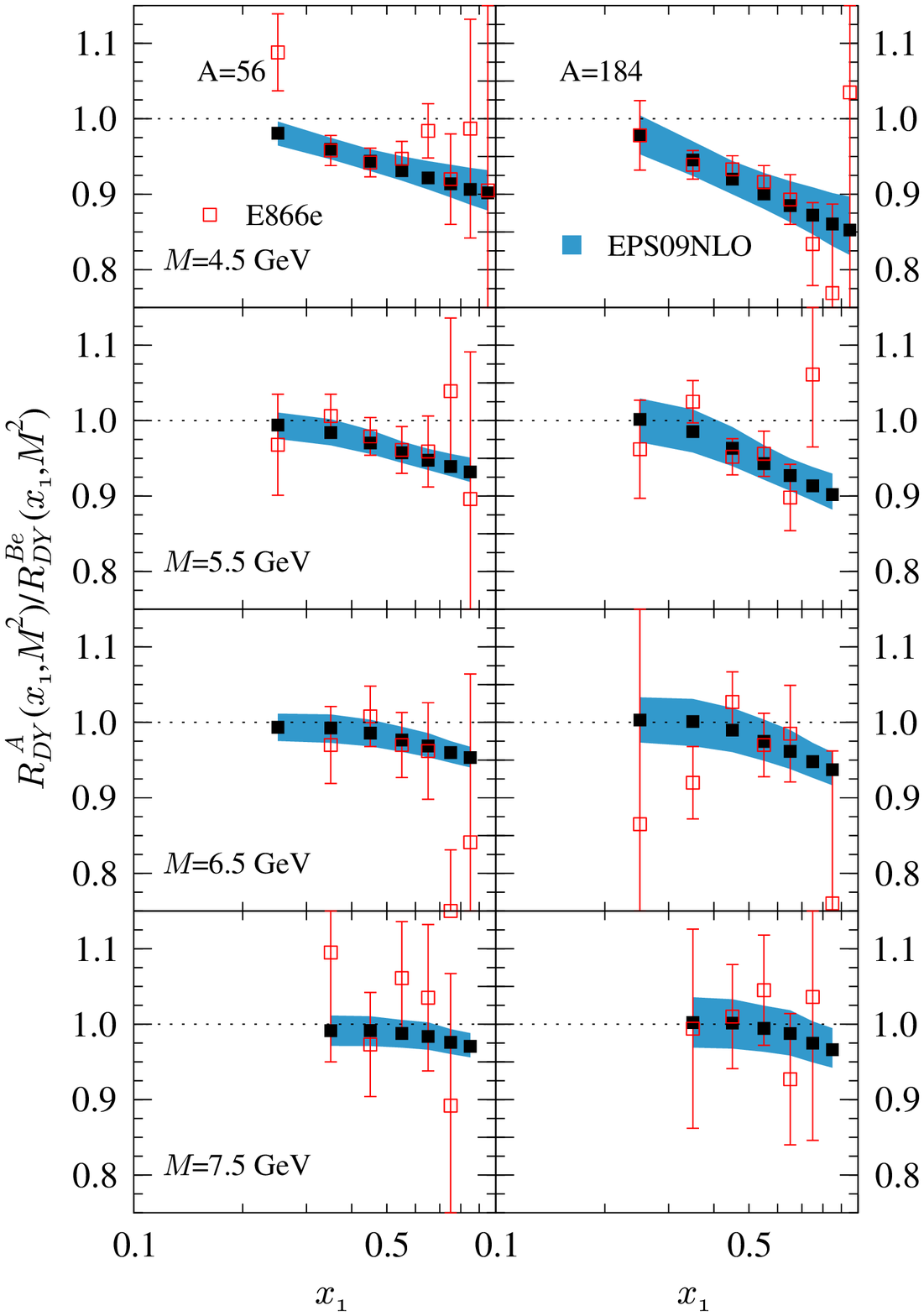}
\hspace{-1.2cm}
\includegraphics[scale=0.37]{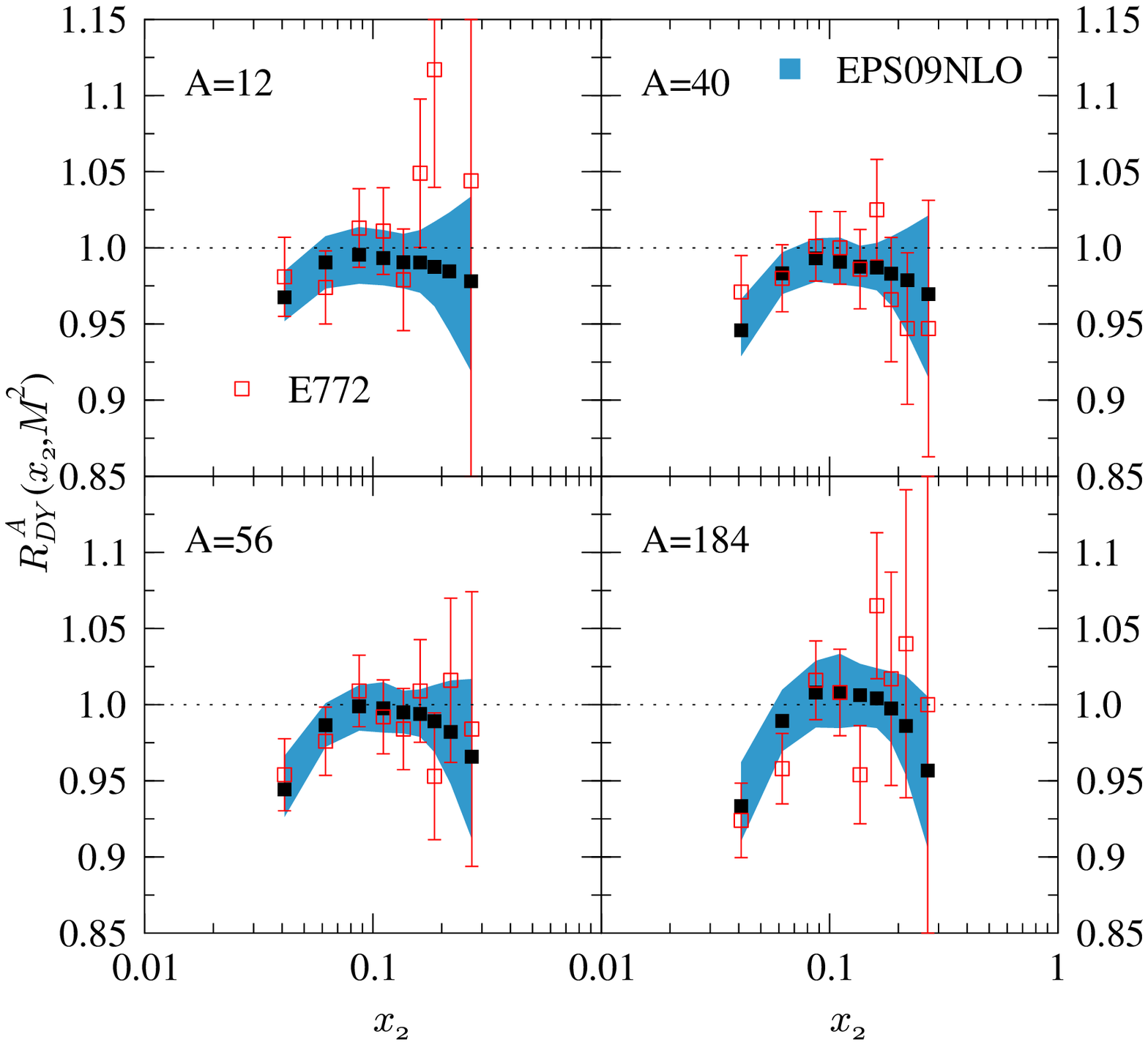}
\caption[]{The computed NLO $R_{\rm DY}^{\rm A}(x,M^2)$ (filled squares and error bands) as a function of $x_1$ (left) and $x_2$ (right) compared with the E866 \cite{Vasilev:1999fa} and the E772 \cite{Alde:1990im} data (open squares).}
\label{Fig:RDY}
\end{figure}

The situation is rather similar with the nuclear effects in the Drell-Yan dilepton data
\begin{equation}
R_{\rm DY}^{\rm A}(x_{1,2},M^2) \equiv \frac{\frac{1}{A}d\sigma^{\rm pA}_{\rm 
DY}/dM^2dx_{1,2}}{\frac{1}{2}d\sigma^{\rm pd}_{\rm DY}/dM^2dx_{1,2}},
\end{equation}
where $M^2$ is the invariant mass of the lepton pair and $x_{1,2} \equiv \sqrt{M^2/s}\,e^{\pm y_R}$. I display the comparison to the E772 and E866 data in Fig.~\ref{Fig:RDY}.

\subsection{Inclusive hadron production}

\begin{figure}[!h]
\centering
\includegraphics[width=20pc]{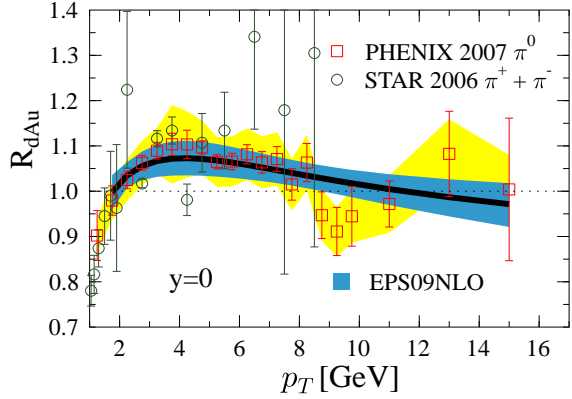}
\caption[]{The computed $R^\pi_{\rm dAu}$ (thick black line and blue error band) at $y=0$ for inclusive pion production compared with the PHENIX \cite{Adler:2006wg} (open squares) and STAR \cite{Adams:2006nd} (open circles) data. The error bars are the statistical uncertainties, and the yellow band around PHENIX data indicate the point-to-point systematic errors. The data have been multiplied by the optimized normalization factor $f_N = 1.03$ for PHENIX and $f_N = 0.90$ for STAR.}
\label{Fig:PHENIX}
\end{figure}

Whereas the free-proton PDF studies can exploit the inclusive jet production data from Fermilab experiments Z0 and CDF to access the large-$x$ gluons, there are not yet\footnote{There is, however, hope that jet measurements performed in nuclear collisions at RHIC could be exploited for the nPDF analyses.}  corresponding nuclear data to use for this purpose. Consequently, the nuclear modifications for the gluons have been largely unknown. To do better job in this respect, almost anything which is sensitive to the gluons is welcome if it reduces the need for theoretical assumptions and fits in with the other data. As we have learned \cite{EPS08-paperi,EPS09-paperi}, potential candidates are the RHIC measurements for inclusive high-$p_T$ pion production, especially for the corresponding nuclear modification
\begin{equation}
R_{\rm dAu}^{\pi}  \equiv  \frac{1}{\langle N_{\rm coll}\rangle} \frac{d^2 N_{\pi}^{\rm dAu}/dp_T dy}{d^2 N_{\pi}^{\rm pp}/dp_T dy} \stackrel{\rm min. bias}{=} \frac{\frac{1}{2A} d^2\sigma_{\pi}^{\rm dAu}/dp_T dy}{d^2\sigma_{\pi}^{\rm pp}/dp_T dy}, \nonumber 
\end{equation}
where the pion transverse momentum is denoted by $p_T$, rapidity by $y_R$, and where $\langle N_{\rm coll}\rangle$ denotes the number of binary nucleon-nucleon collisions. In the factorization framework the cross-sections are computed, schematically, by
$$
\sigma^{A+B \rightarrow \pi + X}  =  \sum_{i,j,k=q,\overline{q},g} f_i^A(\mu^2) \otimes f_j^B \otimes \hat{\sigma}^{ij\rightarrow k + X} \otimes D_{k \rightarrow \pi},
$$
where the additional factor $D_{k \rightarrow \pi}$ is the fragmentation function for parton $i$ to make a pion. Due to the presence of this piece, pion data is not used in the free-proton PDF fits as there are uncertainties in disentangling between the PDF-originated effects and those inherent to the fragmentation functions. However, in the nuclear ratio $R^\pi_{\rm dAu}$, many details of the fragmentation functions seem to cancel: It is reassuring that the shape of $R^\pi_{\rm dAu}$ is practically independent of the particular fragmentation functions used in the calculation --- modern sets like \cite{Kniehl:2000fe,deFlorian:2007aj,Albino:2008fy} all give about equal results in the required kinematical domain ($y_R = 0$). Having made this observation, it should be quite safe to utilize this type of data in the nPDF analysis. The comparison with the PHENIX and STAR data is shown in Fig.~\ref{Fig:PHENIX}. In the computed black curve the downward trend at the large-$p_T$ end is due to the presence of an EMC-effect in the large-$x$ gluons, while the turnover toward small $p_T$ originates from the gluon shadowing. No other effects are needed to reproduce the observed spectra, and the fit does not seem to show a significant tension between this and the other data.

\vspace{0.5cm}
As already mentioned, in the EPS08-article \cite{EPS08-paperi} we studied the chances of a very strong gluon shadowing. This study was inspired by the suppression observed in the nuclear modification $R_{\rm dAu}$ for the negatively-charged hadron yield in the forward rapidities ($\eta=2.2,3.2$) by the BRAHMS collaboration \cite{Arsene:2004ux} in d+Au collisions at RHIC. Although a quite deep shadowing at small-$x$ and at low-$Q^2$ was found out to be well within the experimental uncertainties, we found a clear tension between the BRAHMS and small-$x$ deeply inelastic NMC data, as demonstrated in Fig.~\ref{Fig:weightcomp} below.

\begin{figure}[!h]
\center
\includegraphics[scale=0.50]{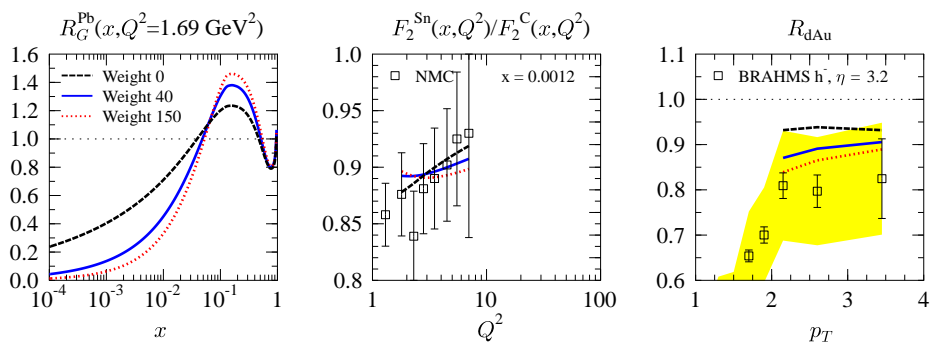}
\caption[]{These three panels stress the tension between the deeply inelastic NMC data and the suppression in the forward rapidity hadron production $R_{\rm dAu}$ measured by the BRAHMS collaboration. The three curves demonstrate the effect of assigning weights $w_N=0,40,150$ for the BRAHMS data in the EPS08 global leading-order analysis: the more weight is given to the BRAHMS data, the worse the $Q^2$-slope in the NMC data becomes reproduced.}
\label{Fig:weightcomp}
\end{figure}

\begin{figure}[!h]
\center
\includegraphics[scale=0.51]{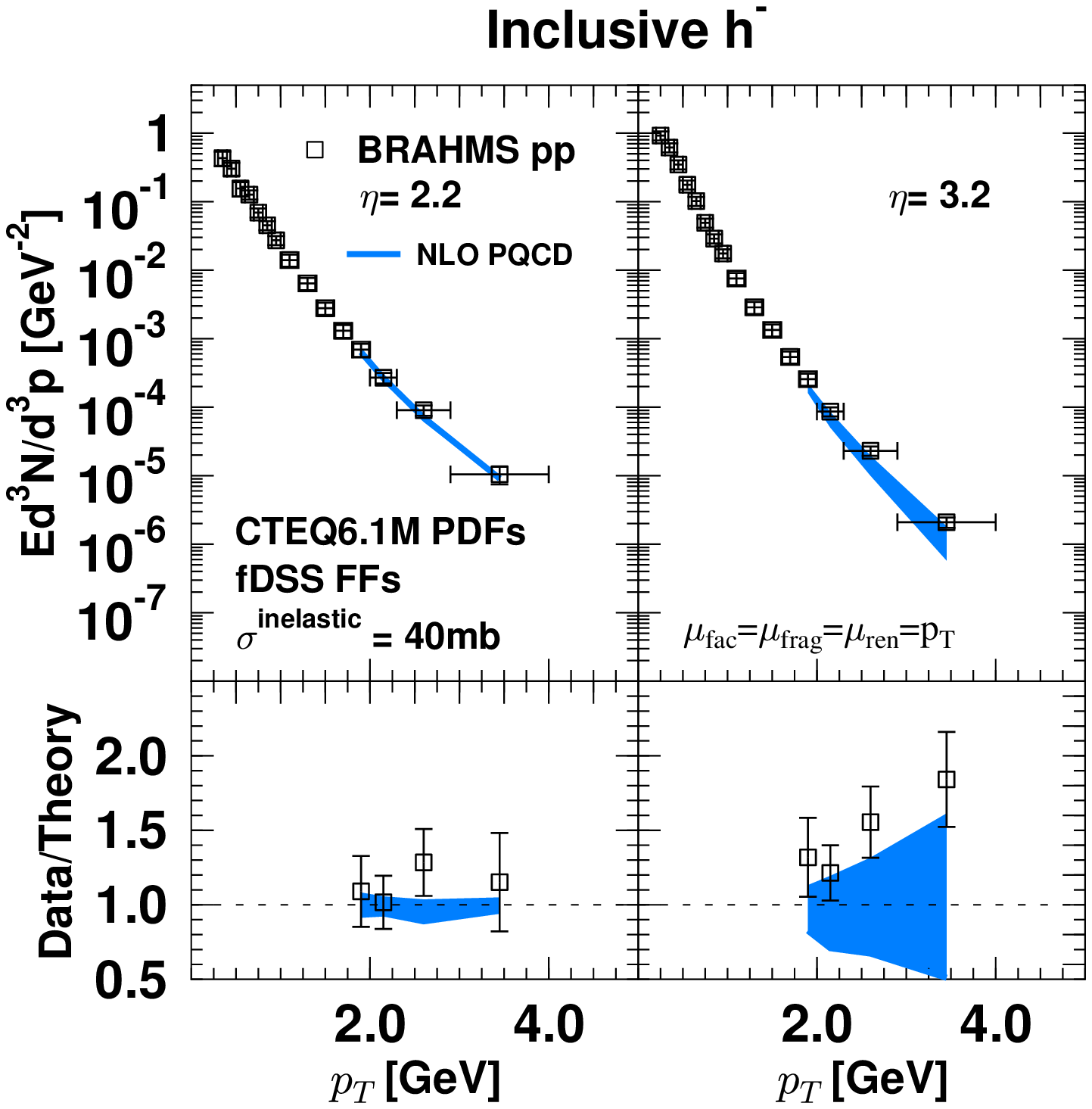}
\includegraphics[scale=0.51]{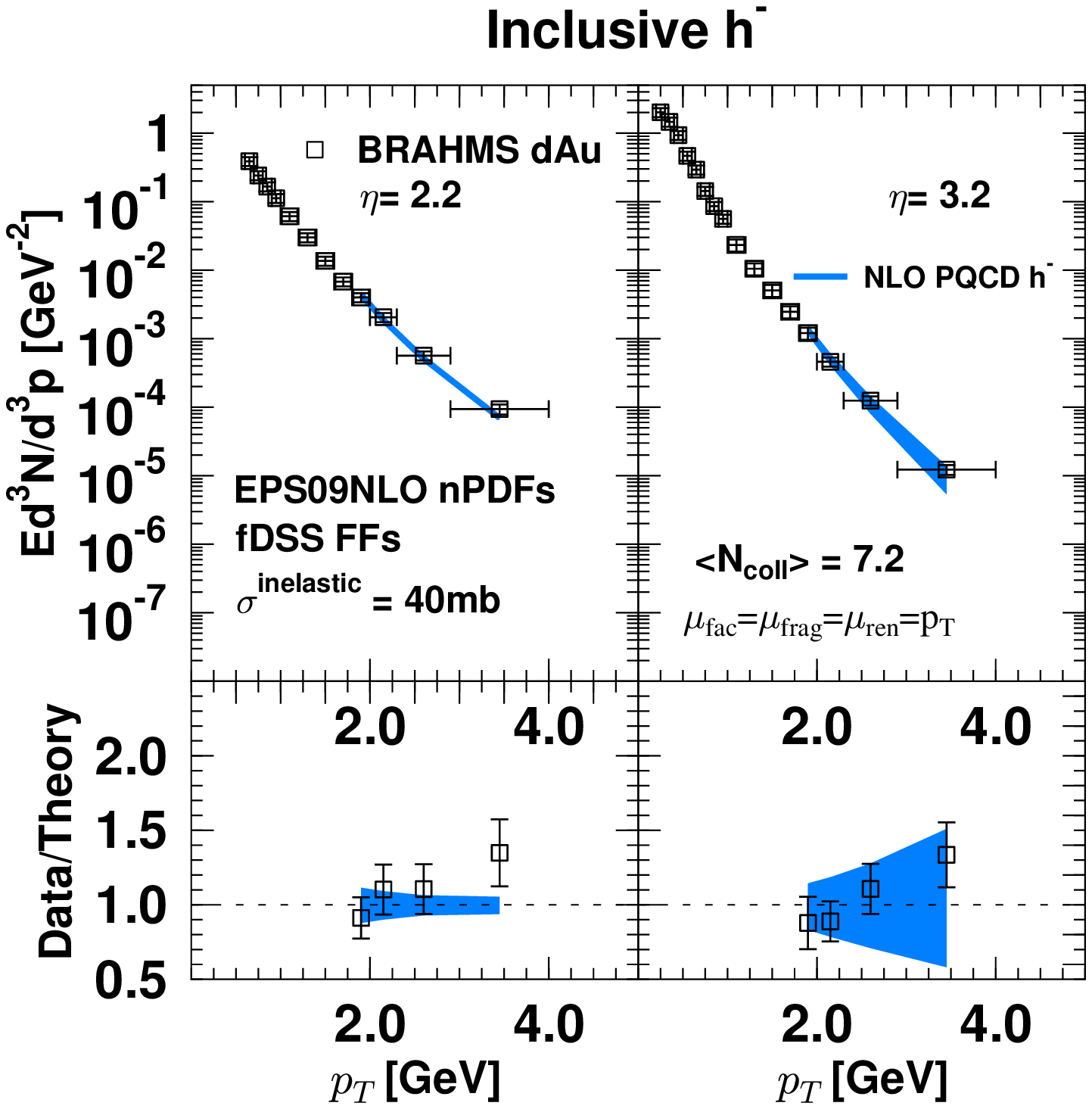}
\caption[]{Inclusive $h^-$ yields in p+p and d+Au collisions. The experimental $p_T$-binned data from BRAHMS \cite{Arsene:2004ux} are shown by open squares with statistical and systematic errors added in quadrature. The blue band indicates the 90\% confidence range derived from the CTEQ6.1M and EPS09 uncertainties. The calculated cross-sections have been averaged over the $p_T$-bin width as in the data.}
\label{Fig:BRAHMS1}
\end{figure}

\newpage
Being aware that such issue existed, we decided not to use the BRAHMS data in the subsequent EPS09 \cite{EPS09-paperi} analysis. There is, however, a deeper reason for not to include the BRAHMS data in EPS09. The Fig.~\ref{Fig:BRAHMS1} displays the CTEQ6.1M and EPS09 predictions (with fDSS fragmentation functions \cite{de Florian:2007hc}) compared with the BRAHMS data for the absolute $h^-$ spectra,
$$
\frac{d^3 N^{\rm pp}}{d^2p_T dy} \stackrel{\rm min. bias}{=} \frac{1}{\sigma_{NN}^{\rm inelastic}} \frac{d^3\sigma^{\rm pp}}{d^2p_T dy}
\quad ; \quad
\frac{d^3 N^{\rm dAu}}{d^2p_T dy} \stackrel{\rm min. bias}{=} \frac{{\langle N_{\rm coll}\rangle}}{\sigma_{NN}^{\rm inelastic}} \frac{\frac{1}{2A}d^3\sigma^{\rm dAu}}{d^2p_T dy}.
\label{eq:BRAHMScrosssection}
$$


In the $\eta=2.2$ panel, the measured p+p and d+Au spectra are both in good agreement with the NLO pQCD. However, in the most forward $\eta=3.2$ bin there is a systematic and significant discrepancy present between the measured and computed $p_T$ spectrum for the p+p collisions, although the d+Au spectrum in the same rapidity bin is better reproduced. This observation helps to throw some light on the inconsistency revealed by the EPS08-analysis: whether the factorization breaks down already for the p+p collisions and the agreement in d+Au is just by a chance, or there are some difficulties with the BRAHMS data in the forwardmost rapidities, where the measurements are very challenging.


\subsection{Scale-breaking effects}

\begin{figure}[p]
\center
\includegraphics[scale=0.55]{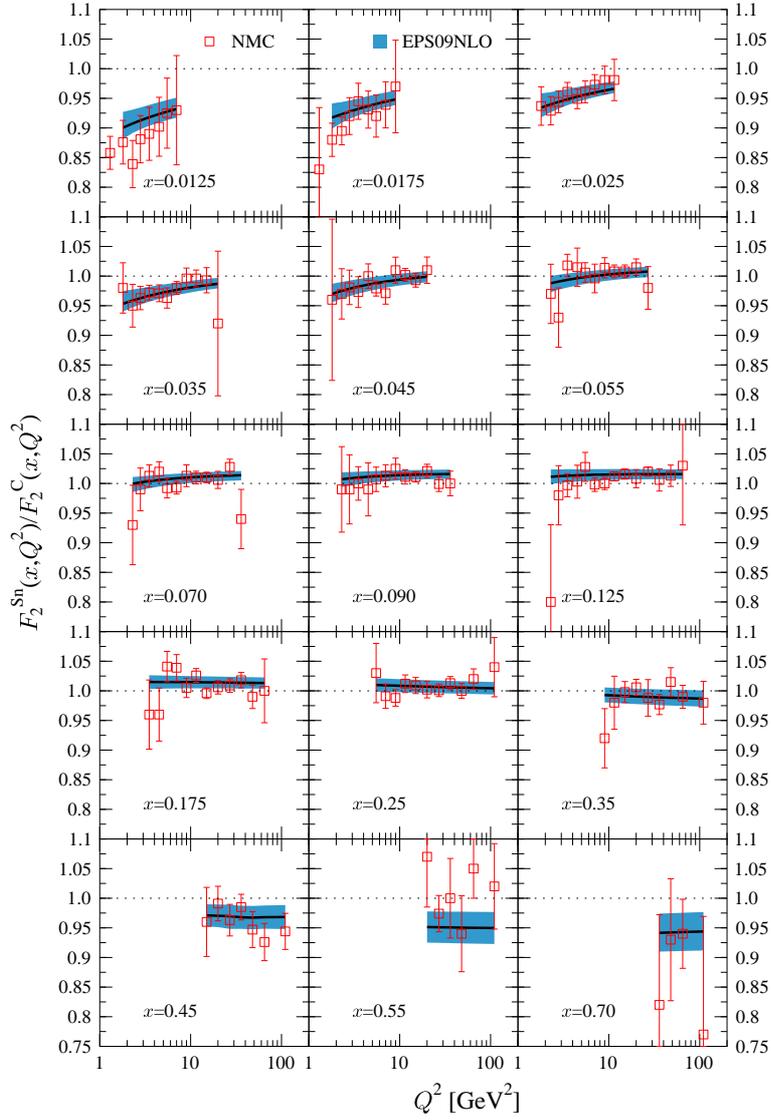}
\caption[]{The calculated NLO scale evolution (solid black lines and error bands) of the ratio $F_2^{\mathrm{Sn}}/F_2^{\mathrm{C}}$ compared with the NMC data \cite{Arneodo:1996ru} for  several fixed values of $x$.}
\label{Fig:RF2_slopes1}
\end{figure}

\begin{figure}[p]
\center
\includegraphics[scale=0.55]{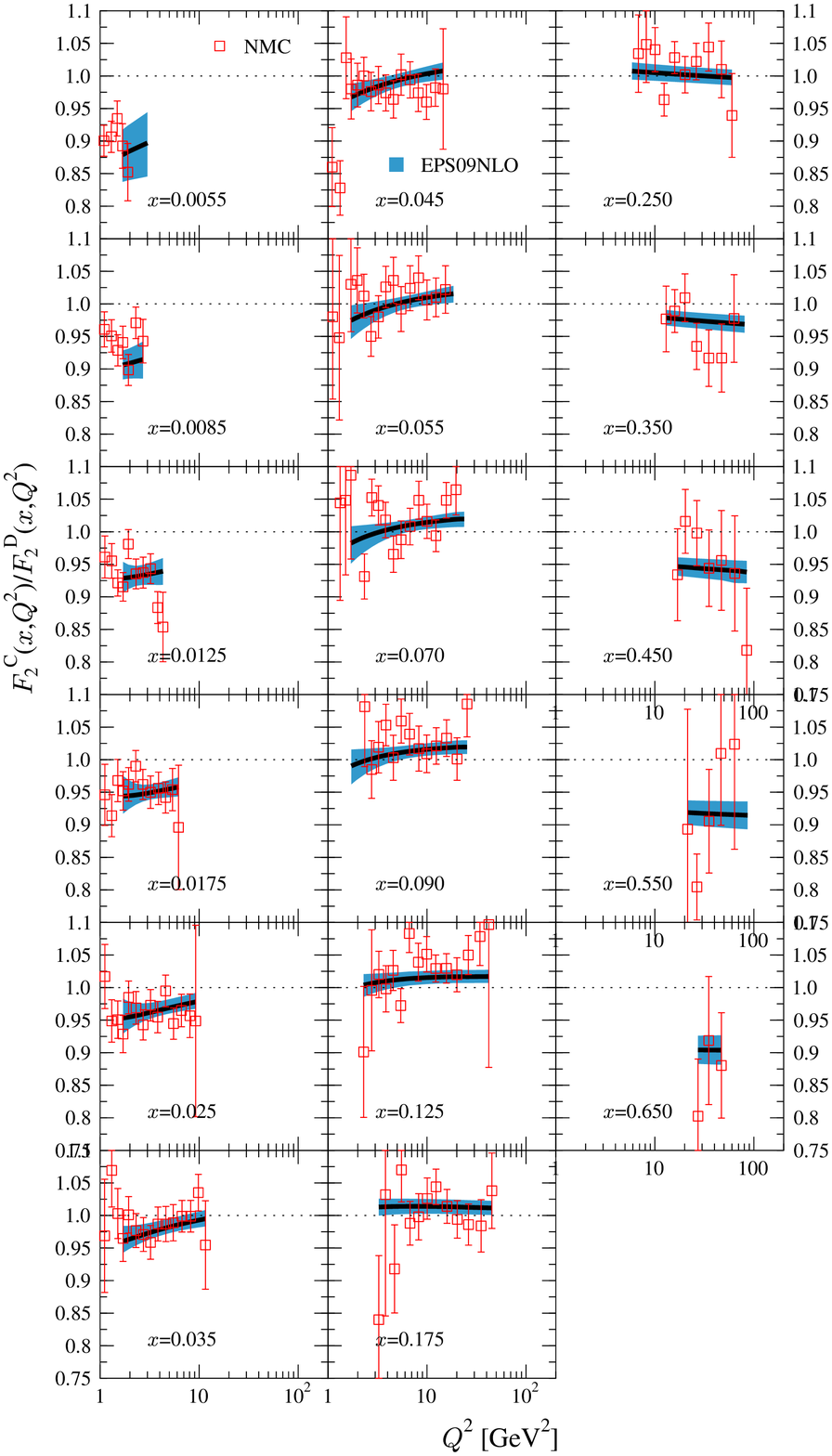}
\caption[]{The calculated NLO scale evolution (solid black lines and error bands) of the ratio $F_2^{\mathrm{C}}/F_2^{\mathrm{D}}$ compared with the NMC data \cite{Arneodo:1996ru} for  several fixed values of $x$.}
\label{Fig:RF2_slopes2}
\end{figure}

As the theory goal of the global nuclear PDF studies is largely to test the QCD factorization --- to find deviations from the DGLAP dynamics --- the scale-breaking effects deserve special attention. Such effects are cleanly visible e.g. in the E886 Drell-Yan data in Fig.~\ref{Fig:RDY} where tendency of diminishing nuclear effects toward larger invariant mass $M^2$ is observed, although the experimental uncertainties are admittedly large. The $x$-binned DIS data versus $Q^2$, shown in Figs.~\ref{Fig:RF2_slopes1} and \ref{Fig:RF2_slopes2}, also reveal some general features: At small $x$ the $Q^2$-slopes look positive, while toward larger $x$, the slopes gradually die out and become even a bit negative. In both cases, the EPS09NLO reproduces such features supporting the validity of the factorization.

\section{Comparison between the NLO works}

\begin{figure}[p]
\center
\includegraphics[scale=0.55]{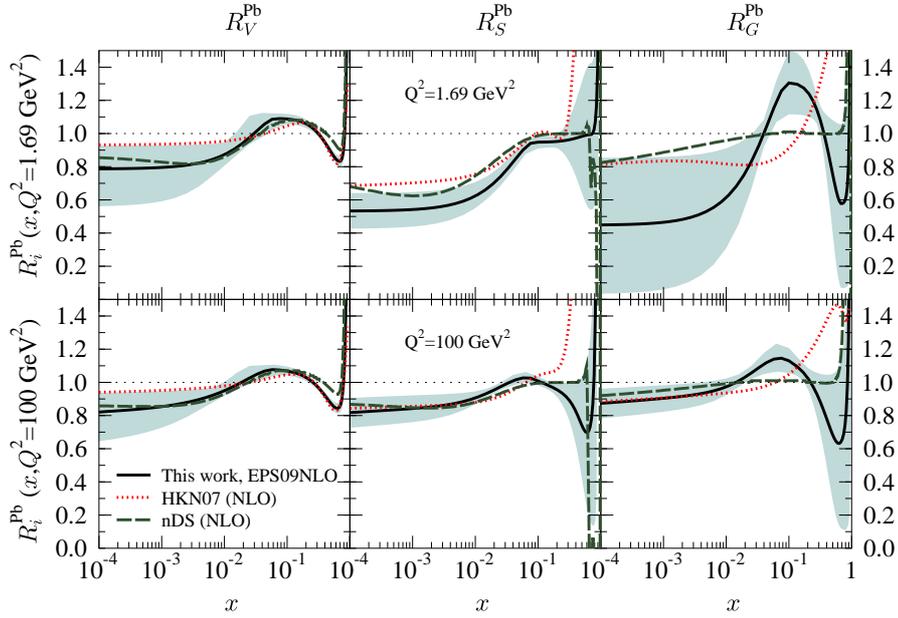}
\caption[]{Comparison of the nuclear modifications for Lead at the scales $Q^2 = 1.69 \, {\rm GeV}^2$ and $Q^2 = 100 \, {\rm GeV}^2$ from the NLO global DGLAP analyses HKN07~\cite{Hirai:2007sx}, nDS~\cite{deFlorian:2003qf}, and EPS09NLO \cite{EPS09-paperi}.}
\label{Fig:NLOcomp}
\end{figure}
\begin{figure}[p]
\centering
\includegraphics[width=20pc]{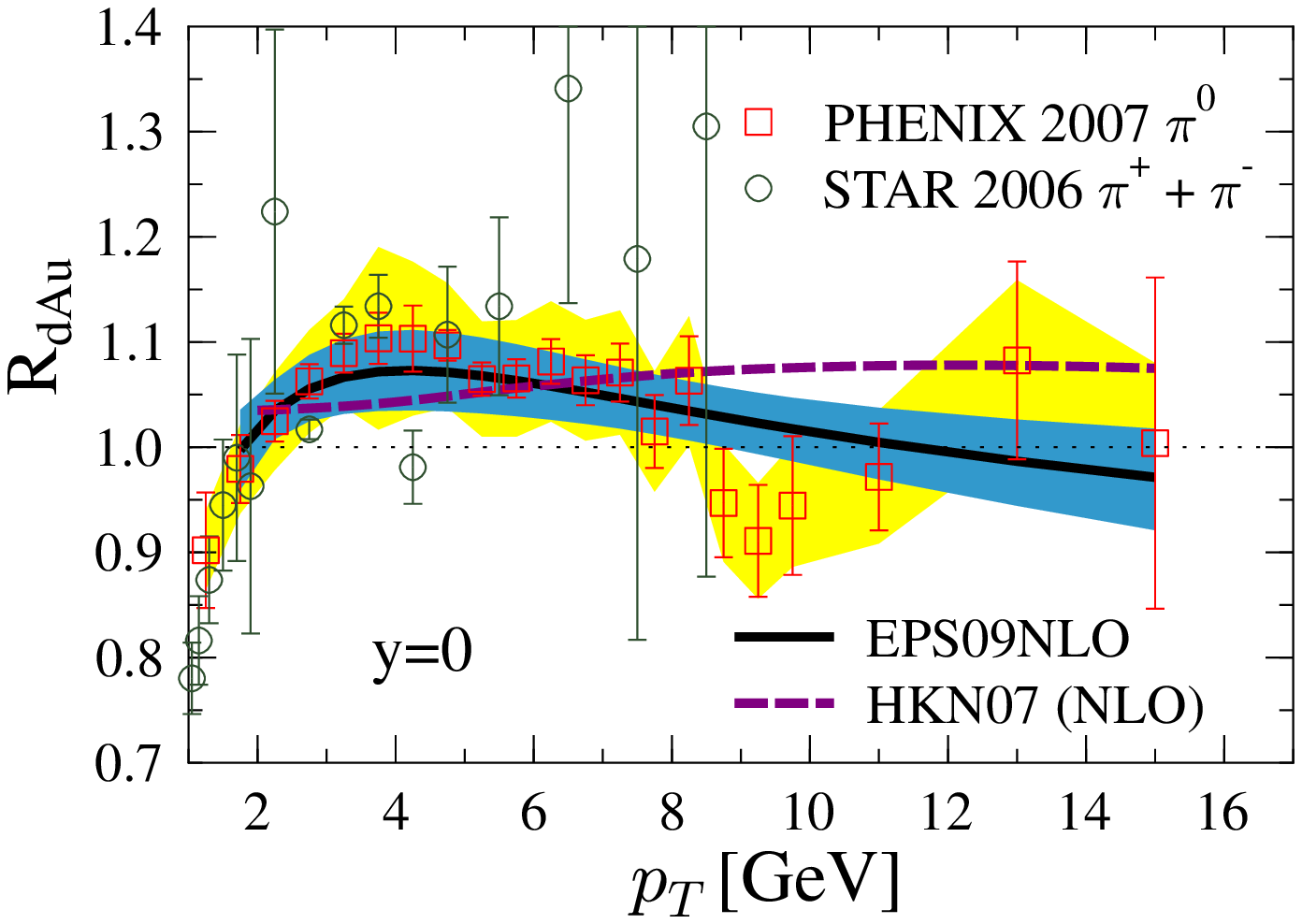}
\caption[]{{Same as Fig.~\ref{Fig:PHENIX}, but also prediction from HKN07 \cite{Hirai:2007sx} is shown.}}
\label{Fig:PHENIX_comp}
\end{figure}

\vspace{0.5cm}
Besides the EPS09NLO, there are two other NLO-level nuclear PDF sets available, and Fig.~\ref{Fig:NLOcomp} compares the nuclear modifications for Lead as they are predicted by these three sets. This comparison is shown at two scales, at $Q^2_0=1.69 \, {\rm GeV}^2$ and at $Q^2=100 \, {\rm GeV}^2$. The significant discrepancies between EPS09NLO and others --- that is, curves being outside the blue EPS09NLO error bands --- are found from the sea quark and gluon sectors. At low $x$, the differences rapidly shrink when the scale $Q^2$ is increased, but at high-$x$ region notable discrepancies persist. Most of the differences are explainable by the assumed behaviours of the fit functions, for which nDS and HKN07 allowed less freedom. The use of more rigid form of the fit function was due to lack of gluon constraints (they did not employ the RHIC pion data which is included in EPS09 \cite{EPS09-paperi}), and giving the gluons more freedom would probably not have improved their fits but instead led to a poorer convergence with parameters drifting to their limits. To clarify the consequences of distinctly different behaviour especially between EPS09NLO and HKN07, the Fig.~\ref{Fig:PHENIX_comp} shows the pion $R_{\rm dAu}$ computed also with HKN07. Despite the rather large data uncertainties, the qualitatively contradicting sign of the $p_T$-slope in $R_{\rm dAu}$ is a promising finding as more data with better precision should be soon able to decisively discriminate between different scenarios.

\section{Conclusions and future prospects}

At the end, the conclusion that can be drawn from the global QCD analyses for nuclear PDFs performed so far is that the QCD factorization conjecture seems to work very well in the explored kinematical region and for the presently analysed data types. In order to find evidence of discrepancies --- especially nucleus-enhanced power corrections in the parton evolution \cite{Mueller:1985wy} or in the cross-sections \cite{Qiu:2003vd} --- the scope of the global analysis should be therefore extended. In addition to enlarging the kinematical coverage and the variety of included processes, also an extended theoretical treatment of the parton dynamics (by including power-suppressed terms) should be of importance. 
The next steps at the horizon toward reaching smaller $x$ and higher $Q^2$, could possibly be realized by running the LHC also in the p+Pb mode. In farther future, electron-ion colliders like the planned eRHIC \cite{Deshpande:2005wd} or LHeC \cite{Dainton:2006wd}, would also penetrate deeper in the ($x,Q^2$)-plane. About new processes to include, direct photon data from RHIC d+Au and Au-Au collisions should be available shortly and there are also data from deeply inelastic neutrino-Iron scattering from the Fermilab NuTeV collaboration \cite{Tzanov:2005kr} available. The neutrino data has recently been claimed \cite{Schienbein:2007fs} to point toward different nuclear effects than what are obtained e.g. in EPS09. If true, it would indicate that there are process-dependent effects present, casting doubts on the factorization. This issue certainly deserves a further clarification.

\newpage
It is a sort of a peculiar situation that almost all global free-proton fits include data with nuclear (deuterium or heavier) targets\footnote{For example, the NuTeV data is one of the main sources for constraining the strange sea asymmetry.} --- all groups with their own favorite way to ``correct'' for the nuclear effects. Thus, there is a danger of circularity as the baseline PDFs which should be free of any nuclear effects, do actually somewhat depend on how the nuclear effects were corrected for ---  the free-proton and nuclear PDF analyses are not entirely independent. This observation opens the road for a future work which will combine the free and bound proton PDF analyses into a single, ``master'' global analysis.